\documentclass[%
reprint,
superscriptaddress,
nofootinbib,
 amsmath,amssymb,
 aps,
]{revtex4-2}

\usepackage{graphicx,amssymb,amsmath,amsthm,amsfonts,epsfig,epsf}
\usepackage[linktocpage]{hyperref}
\usepackage[usenames]{color}
\usepackage[dvipsnames]{xcolor}
\usepackage{epstopdf}
\usepackage{bm}
\usepackage{dcolumn}
\usepackage[latin1]{inputenc}
\usepackage{latexsym}
\usepackage{rotating}
\usepackage{hyperref}
\usepackage{color}
\usepackage{longtable}

\usepackage{enumerate}
\usepackage{tensor,multirow}
\usepackage{url}

\usepackage{slashed}

\definecolor{chmagenta}{rgb}{0.54, 0.17, 0.88}

\newcommand{\new}[1]{#1}
\providecommand\aj{Astronomical Journal}

\begin{document}
\title{Inferring Interference: Identifying a Perturbing Tertiary\\ with Eccentric Gravitational Wave Burst Timing}

\author{Isobel Romero-Shaw}
\affiliation{School of Physics and Astronomy, Monash University, Clayton VIC 3800, Australia}
\affiliation{OzGrav: The ARC Centre of Excellence for Gravitational Wave Discovery, Clayton VIC 3800, Australia}
\affiliation{Department of Applied Mathematics and Theoretical Physics, Cambridge CB3 0WA, United Kingdom}
\affiliation{Kavli Institute for Cosmology Cambridge, Madingley Road Cambridge CB3 0HA, United Kingdom}

\author{Nicholas Loutrel}
\affiliation{Dipartimento di Fisica, ``Sapienza'' Universit\`a di Roma \& Sezione INFN Roma1, Piazzale Aldo Moro 5, 
00185, Roma, Italy}


\author{Michael Zevin}
\affiliation{Kavli Institute for Cosmological Physics, The University of Chicago, 5640 South Ellis Avenue, Chicago, Illinois 60637, USA}
\affiliation{Enrico Fermi Institute, The University of Chicago, 933 East 56th Street, Chicago, Illinois 60637, USA}

\begin{abstract}
Binary black holes may form and merge dynamically. These binaries are likely to become bound with high eccentricities, resulting in a burst of gravitational radiation at their point of closest approach. When such a binary is perturbed by a third body, the evolution of the orbit is affected, and gravitational-wave burst times are altered. The bursts times therefore encode information about the tertiary. In order to extract this information, we require a prescription for the relationship between the tertiary properties and the gravitational-wave burst times. In this paper, we demonstrate a toy model for the burst times of a secular three-body system. We show how Bayesian inference can be employed to deduce the tertiary properties when the bursts are detected by next-generation ground-based gravitational-wave detectors. We study the bursts from an eccentric binary with a total mass of $60$~M$_\odot$ orbiting an $6 \times 10^{8}$~M$_\odot$ supermassive black hole. When we assume no knowledge of the eccentric binary, we are unable to tightly constrain the existence or properties of the tertiary, and we recover biased posterior probability distributions for the parameters of the eccentric binary. However, when the properties of the binary are already well-known---as is likely if the late inspiral and merger are also detected---we are able to more accurately infer the mass of the perturber, $m_3$, and its distance from the binary, $R$. When we assume measurement precision on the binary parameters consistent with expectations for next-generation gravitational-wave detectors, we can be greater than $90\%$ confident that the binary is perturbed. 
\new{When the orbit of the binary around the tertiary is face-on with respect to the observer, there are large statistical uncertainties on the recovered tertiary properties ($m_3$, $R_3$, and orbital phase descriptors $\omega_0$ and $V_{3,0}$) due to correlations between these parameters in the simple toy model.
However, if the orbit is tilted away from face-on, these uncertainties can be substantially reduced.}
\new{Future models allowing for non-secular evolution may further decrease measurement uncertainties by breaking more correlations between binary and tertiary parameters.}
\end{abstract}

\maketitle

\section{Introduction}
\label{sec:intro}
The third gravitational-wave transient catalogue \citep[GWTC-3;][]{LVK:2021:GWTC3} of the LIGO-Virgo-KAGRA collaboration (LVK) contains 90 signals of likely astrophysical origin ($\geq 50\%$ probability).
The vast majority of these signals are likely to originate from binary black hole (BBH) mergers.
GW190521~\cite{Abbott:2020:GW190521} is one of the more intriguing BBH systems observed. 
Due to its high total mass, the signal is short in duration in the LVK frequency band, allowing multiple population models to explain the system's origin \citep{LVK:2020:ImplicationsGW190521}. 
While it is impossible to fully determine this given the available data, one plausible explanation is that the event corresponds to a highly eccentric merger within the disk of an active galactic nucleus~\cite{Romero-Shaw:2020:GW190521, Gayathri:2022:GW190521, Gayathri:2021xwb, Gamba:2021:hyperbolic, Ford:2021:LoudVsQuiet, Samsing:2020:AGN}. 

A variety of dynamical processes, such as GW captures from a hyperbolic flyby of two BHs or resonating encounters between more than two BHs, could produce GW190521-like events.
Dynamical environments are thought capable of producing mergers with higher masses, more misaligned component spins, and higher eccentricities than can be formed through isolated evolution \citep[e.g.,][]{Samsing:2014:BinarySingle, SamsigRamirezRuiz:2017:HighlyEccentric, Rodriguez18a, Rodriguez:2018pss, Samsing18, Zevin:2019:BinaryBinary, Fragione:2020:GW190521StarClusersHierarcical, Vitale15, Stevenson15bqa, Gerosa17, Randall19, Bavera2020, Mapelli:2020:review, Sedda:2020:fingerprints, Zevin:2020:channels, Fragione:2022:NSCs}.
However, systems from \textit{different} dynamical environments---for example, globular clusters, active galactic nuclei (AGN), or nuclear clusters---are challenging to distinguish in the LVK band.
One of the key differences in source properties predicted for different dynamical environments is the overall distribution of eccentricities in the population  \citep[e.g.,][]{Gondan17, Samsing:2017xmd, Rodriguez:2018pss, Tagawa:2020jnc, Samsing:2020tda, Zevin:2021:seleccentricity, GondanKocsis2021, Romero-Shaw:2022:FourEccentricMergers}, although most of these eccentricities are below the sensitivity of current detectors \citep[e.g.,][]{Lower:2018:eccentricity, Romero-Shaw:2019:GWTC-1-ecc, Romero-Shaw:2021:GWTC-2-ecc}.

Due to their improved sensitivity, particularly at low frequencies, future ground-based detectors like the Einstein Telescope \citep[ET;][]{Maggiore:2020:ET} and Cosmic Explorer \citep{Evans:2021:CosmicExplorer} will be sensitive to a wider range of eccentricities \citep{Lower:2018:eccentricity} and more distant mergers.
As a result, they will give us greater power to identify the preferred formation environment of an ensemble of events \citep{Romero-Shaw:2021:GCs, Romero-Shaw:2022:FourEccentricMergers}.
However, distinguishing different dynamical BBH formation environments using population eccentricity distributions requires robust predictions of these distribution, which do not exist for all environments. 
For example, there are significant variations in the eccentricity distributions expected to arise in AGN under different assumptions about the properties and dynamics of the AGN itself \citep{Samsing:2020tda, Tagawa:2020jnc}.
Regardless of the uncertainty over the whole population, all dynamical formation processes are capable of producing extremely high-eccentricity ($e \gtrsim 0.9$ at detection) sources.

Instead of attempting to demonstrate inference of the formation channels of an entire population, we instead focus on the challenge of pinpointing the environment in which a single binary merged.
If formed with sufficiently high orbital eccentricity, BBHs formed through dynamical encounters will experience a burst phase, wherein GW bursts are emitted during each pericenter passage~\cite{Turner,Loutrel:2020jfx,Loutrel:2019kky,Nagar:2021:EccentricWaveform}. 
The arrival times of the bursts become analogous to the GW phase of quasi-circular binaries. 
As a result, perturbations of the binary dynamics will be imprinted in the burst arrival times, and even small deviations from the standard two-body vacuum prediction of arrival times can reduce the signal-to-noise ratio (SNR) and/or bias parameter estimation~\cite{Loutrel:2019kky}. 
One case of perturbation, which is relevant in dynamical environments and has been widely studied, is that caused by a nearby third body~\cite[e.g][]{Kozai,Lidov,Naoz:2012bx,Naoz:2016,Silsbee16,Antonini17,LiuLai17,Randall18,Liu19,Li:2022:AGNBBH,Gondan:2022:AGN,Deme:2022:IMBH}.

While the obvious culprit for an interfering tertiary when considering binaries in AGN disks is the central supermassive black hole (SMBH), triple systems with complicated dynamics can be formed either in situ or through chaotic few-body interactions~\cite{Samsing:2017xmd,Samsing:2018isx,Samsing:2018ykz,Rodriguez:2018pss,Martinez:2020}. In a dynamical setting, a wide binary is disrupted by a tertiary, causing the total system to undergo chaotic interactions. 
The system will eventually settle into a quasi-stable configuration, wherein an inspiraling eccentric binary will either merge before the tertiary can disrupt it again (see Fig.~1 of~\cite{Samsing:2017xmd}) or enter another phase of chaotic evolution. 
The influence of the tertiary on this inspiraling binary can be multi-faceted, not only inducing oscillations in the orbital eccentricity and inclination angle through tidal interactions, but also generating a time and frequency shift in the detected burst due to motion of the binary's center of mass (COM)~\cite{YuChen:2021:SMBHinference,Johan-doppler}. These perturbations to the binary's motion will be encoded in the emitted GWs, and can in principle be measured from the burst arrival times in the high-eccentricity limit, and the GW Fourier phase in the low-eccentricity limit~\cite{Chandramouli:2021kts,Xuan:2022qkw,Toubiana:2020drf}. 

Future \emph{space-based} detectors, which will be able to observe the evolution of a binary as it orbits a more massive tertiary for several years, may achieve precise measurements of tertiary properties by measuring the phase shift and/or Doppler shift due to the binary's COM motion \citep{Inayoshi:2017:LISAThreeBody,RandallXianyu:2019:BBHProbeTertiaryWithLISA}, and/or by tracking the tertiary-induced precession of the binary's orbital plane \citep{YuChen:2021:SMBHinference}. 
Previously, it was shown that existing gravitational-wave detectors are theoretically sensitive enough to discern the influence of a tertiary on a merging quasi-circular binary, particularly low-mass systems like binary neutron stars \citep{Meiron:2017:LISALIGOThreeBody}.
Here, we investigate the ability of future \emph{ground-based} detectors to identify an interfering tertiary purely from its effect on the orbit of a nearby eccentric BBH. In light of the importance of tertiaries to the relatively short-term evolution of eccentric BBHs, we perform a proof-of-concept study aimed at answering the following questions: 
\begin{enumerate}
    \item \textit{Can we infer the properties of an interfering tertiary from eccentric GW burst arrival times?}
    \item \textit{Can we constrain the parameter space of perturbing tertiaries in the absence of such effects?}
\end{enumerate}
In Sec.~\ref{sec:demo}, we demonstrate inference of an SMBH perturbing a nearby highly eccentric BBH using next-generation ground-based detectors.
In Sec.~\ref{sec:model}, we derive the toy model that represents a three-body system producing a sequence of GW bursts from the periastron passages of the eccentric BBH. The final timing model, including re-summed 2.5PN radiation reaction effects and restricted -3PN tertiary perturbations, is given in Eqs.~\eqref{eq:p-next}-\eqref{eq:time-offset}. 
\new{The additional offset induced due to the COM motion of the binary is derived in Sec.~\ref{sec:com} and given in Eq.~\ref{eq:time-doppler}.}
In Sec.~\ref{sec:pe}, we establish and demonstrate a Bayesian framework for recovering the parameters of the system from its burst signal.
When we assume that all properties of the inspiralling eccentric binary are well-known prior to the burst timing analysis, we can exclude the possibility that the binary is unperturbed at greater than $90\%$ confidence.
In Sec.~\ref{sec:disc}, we discuss our results and outline our intended future extensions to the work presented here. In this work, we use geometric units with $G=c=1$.

\section{Methodology \& Analysis}
\label{sec:demo}

In a real data analysis scenario, there would be two complications to contend with. The first is the detection of eccentric burst signals. While no searches specifically targeting highly eccentric sources are currently performed, burst search pipelines such as cWB~\cite{Klimenko:cWB:2008} or Omicron \citep{Robinet:2020:Omicron} and proposed methods, such as power stacking~\cite{Tai:2014bfa} or \texttt{BayesWave}-inspired signal-based prior searches~\cite{Cornish:2014kda,Cornish:2020dwh,CheeseboroBaker:2021:EccentricBurstSearch}, could theoretically detect such unmodeled signals. However, these methods do possess limitations. Burst searches are inefficient for detecting repeated burst signals, power stacking is sub-optimal compared to matched filtering, and the signal-based prior searches proposed in~\cite{CheeseboroBaker:2021:EccentricBurstSearch} require a timing model consistent with signals that would exist in nature.
In this work, we do not concern ourselves with the specific method by which the bursts are detected. We assume that a sufficient algorithm exists that could detect these signals, and focus on the problem of inference from eccentric bursts signals.
    
The second issue is that, in order to perform parameter estimation and infer a tertiary's parameters, we require a model of the tertiary's effect on the burst arrival times. Timing models have been developed for the vacuum two-body problem in GR~\cite{Loutrel:2014vja,Loutrel:2017fgu,Loutrel:2019kky,Arredondo:2021rdt} using the post-Newtonian (PN) approximation~\cite{PoissonWill}. While a model has been developed to 3PN order, it is only accurate to leading order in high eccentricity~\cite{Loutrel:2017fgu}. Further, models based on naive PN expansions may not be sufficiently accurate for more compact binaries, requiring a re-summation of a suitable PN timing model that must be calibrated against a signal that one might realistically expect to exist in nature~\cite{Arredondo:2021rdt}. Unfortunately, there is currently a dearth of numerical relativity simulations of highly eccentric BBHs. Re-summed models must be calibrated by other means, which will introduce instrinsic modeling error that may produce biases in parameter estimation.

In keeping with this work being a proof-of-concept study, we develop an extension of the re-summed timing model of~\cite{Arredondo:2021rdt} that incorporates the simplest three-body effects. 
This toy model focuses on hierarchical triple systems, wherein the third body moves along a circular outer orbit with radius $R \gg r_{12}$, with $r_{12}$ being the separation of the inner binary. 
Using this assumption, we expand the force induced on the inner binary by the tertiary in $r_{12}/R \ll 1$, and consider only the leading order effects. 
In doing so, the problem maps to a perturbed Kepler problem, which can be solved using the method of osculating orbits~\cite{PoissonWill,LincolnWill,Mora:2003wt,Will:2016pgm,Konigsdorffer:2006zt,Loutrel:2018ydu,Pound:2007th}. 
We further assume that the orbital angular momenta of the inner and outer orbits are aligned, and thus we do not consider any evolution of the mutual inclination angle between the two orbits. 
\new{This leads us to focus on systems in which such a configuration may be preferred, namely BBH embedded in an accretion disk within an AGN, wherein gas torques are expected to align the orbital angular momentum with the disk \citep[e.g.,][]{McKernan:2018:AGNLIGO, McKernan:2020:AGNLIGO2, Secunda:2019:AGNMigration}.}
By integrating between subsequent pericenter passages, we obtain a new timing model that includes the re-summed 2.5PN radiation reaction effects computed in~\cite{Arredondo:2021rdt}, with restricted tidal quadrupole effects of a tertiary. The latter enter the timing model at -3PN order, since the relative correction scales as $v^{-6}$, with $v$ the orbital velocity.

\subsection{Burst Timing with a Tertiary}
\label{sec:model}
Consider a highly eccentric binary being perturbed by a third body of mass $m_{3}$ on an outer orbit such that $r_{12} \ll R$, with $R$ being the separation of the third body from the inner binary's center of mass and $r_{12}$ being the radial separation of the inner binary. For simplicity, we assume that the mutual inclination of the inner and outer orbits is $\iota =0$, and that the third body moves on a circular orbit, i.e. $R = \text{constant}$. We proceed to calculate the corrections to the eccentric burst timing using the methods detailed in~\cite{Loutrel:2017fgu,Loutrel:2019kky,Arredondo:2021rdt}. \new{The response of the binary and its GW emission to the tertiary can be split into two distinct effects: (1) the tidal force induced on the inner orbit, which modifies the generation of the waves, and (2) the motion of the binary's COM around the common barycenter of the total system, which modifies the propogation of the GWs. We detail how to compute each of these below.}

\subsubsection{Tidal Effects and Generation of GWs}
\label{sec:tides}

\begin{figure}
    \centering
    \includegraphics[width=\columnwidth]{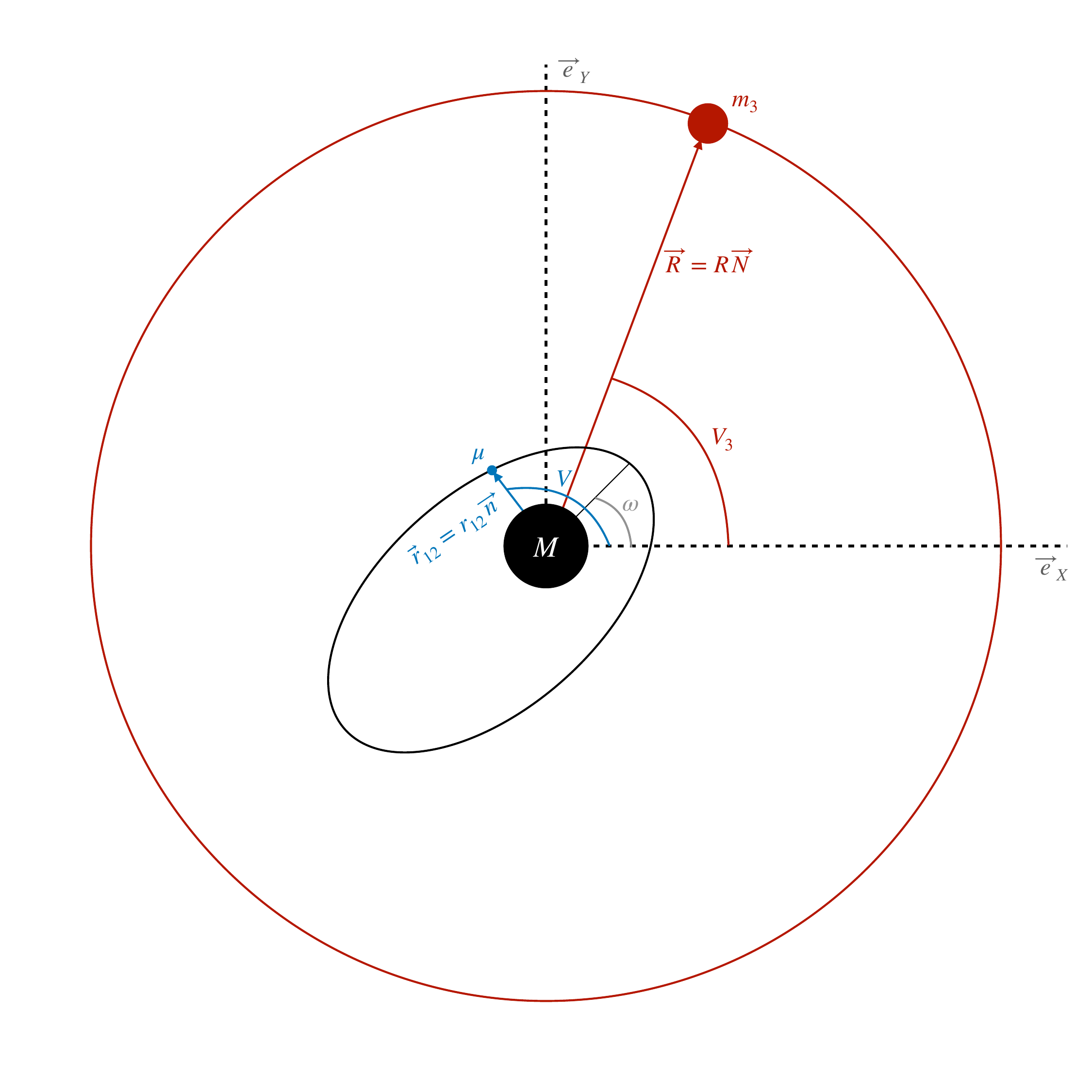}
    \caption{Schematic of the three body system under consideration. The inner binary is parameterized by orbital separation $r_{12}$, true anomaly $V$, reduced mass $\mu$, and total mass $M$. The inner orbit is misaligned relative to the fixed (XY)-frame by the angle $\omega$, corresponding to the longitude of pericenter. The tertiary has mass $m_{3}$, and moves in a circular orbit with radius $R \gg r_{12}$ and phase $V_{3}$.}
    \label{fig:orbit}
\end{figure}

In an effective one body frame, the inner binary is characterized by a small compact object with mass $\mu = m_{1} m_{2}/M$ orbiting around a larger compact object of mass $M=m_{1}+m_{2}$, with $(m_{1},m_{2})$ the masses of the individual objects, and $(\mu,M)$ the reduced and total mass, respectively. The radial separation of the inner binary obeys
\begin{equation}
    \label{eq:r12}
    r_{12} = \frac{p}{1+e\cos V}\,,
\end{equation}
where $p$ is the semi-latus rectum of the orbit, $e$ is the orbital eccentricity, and $V = \phi-\omega$ is the true anomaly with $\phi$ the orbital phase and $\omega$ the longitude of pericenter. The frame of the inner binary is defined by the unit vectors
\begin{align}
    \label{eq:nvec}
    \vec{n} &= \left[\cos(V+\omega), \sin(V+\omega), 0\right]\,,
    \\
    \label{eq:lambdavec}
    \vec{\lambda} &= \left[-\sin(V+\omega), \cos(V+\omega), 0\right]\,,
    \\
    \label{eq:eZvec}
    \vec{e}_{Z} &= \left[0,0,1\right]\,.
\end{align}
The motion of the tertiary is described by the unit vector
\begin{equation}
    \label{eq:Nvec}
    \vec{N} = \left[\cos V_{3}, \sin V_{3}, 0\right]
\end{equation}
where $V_{3} = \Omega_{3} t$ is the phase of the outer orbit, with $\Omega_{3}$ the angular frequency of the orbit, and $t$ being the time coordinate. Fig.~\ref{fig:orbit} gives a schematic of this setup.
The equations of motion of the inner binary are~\cite{Kozai,Lidov,Naoz:2016,PoissonWill}
\begin{equation}
    \label{eq:eom}
    \vec{a}_{12} = -\frac{M}{r^{2}_{12}} \vec{n} - \frac{m_{3} r_{12}}{R^{3}} \left[\vec{n} - 3\left(\vec{n}\cdot\vec{N}\right) \vec{N} + {\cal{O}}(r_{12}/R)\right]\,,
\end{equation}
where we are working in an expansion in $r_{12}/R \ll 1$. 

The equations of motion of the inner binary in Eq.~\eqref{eq:eom} are a classic example of a perturbed Kepler problem. The typical method of solving such a system of equations is the \textit{method of osculating orbits}~\cite{LincolnWill,Mora:2003wt,Will:2016pgm,Konigsdorffer:2006zt,Loutrel:2018ydu,Pound:2007th,PoissonWill}, whereby the orbit takes the standard Keplerian parameterization in Eq.~\eqref{eq:r12}, but the otherwise constant orbital elements $(p,e,\omega)$ are promoted to be functions of time. To obtain the evolution equations for the orbital elements, known as the \textit{osculating equations}\footnote{These are also sometimes referred to as the \textit{Lagrange planetary equations}.}, one decomposes the perturbing force $\vec{f}$ into components within the orbital basis defined by Eqs.~\eqref{eq:nvec}-\eqref{eq:lambdavec}. Defining ${\cal{R}} = \vec{f} \cdot \vec{n}$, ${\cal{S}}=\vec{f}\cdot\vec{\lambda}$, and ${\cal{W}} = \vec{f}\cdot\vec{e}_{Z}$, the osculating equations for $(p,e,\omega,t)$ are given, for example, in Eqs.~(3.69a)-(3.70) in~\cite{PoissonWill}. In our scenario, the perturbing force is given by the force of the tertiary on the inner binary, specifically, the term proportional to $m_{3}$ in Eq.~\eqref{eq:eom}. Following the decomposition of the perturbing force, the osculating equations are
\allowdisplaybreaks[4]
\begin{align}
    \label{eq:dpdV}
    \frac{dp}{dV} &= - 3 p A_{3} \gamma(V)^{-4} \sin\left[2 (V-V_{3} + \omega)\right]\,,
    \\
    \frac{de}{dV} &= \frac{A_{3}}{4} \gamma(V)^{-4} \sum_{k=0}^{2} \left[C_{e}^{(k)} \cos(kV) + S_{e}^{(k)} \sin(kV)\right]\,,
    \\
    \frac{d\omega}{dV} &= -\frac{A_{3}}{4e} \gamma(V)^{-4} \sum_{k=0}^{3} \left[C_{\omega}^{(k)} \cos(kV) + S_{\omega}^{(k)} \sin(kV)\right]\,,
    \\
    \label{eq:dtdV}
    \frac{dt}{dV} &= \left(\frac{p^{3}}{M}\right)^{1/2} \gamma(V)^{-2} \Bigg\{1 - \frac{A_{3}}{4e} \gamma(V)^{-4} \sum_{k=0}^{3} \left[C_{t}^{(k)} \cos(kV) 
    \right.
    \nonumber \\
    &\left.
    + S_{t}^{(k)} \sin(kV)\right]\Bigg\}
\end{align}
where
\begin{align}
    A_{3} = \frac{m_{3} p^{3}}{M R^{3}}\,, \qquad \gamma(V) = 1 + e \cos V\,,
\end{align}
and the harmonic coefficients $[C_{a}^{(k)}, S_{a}^{(k)}]$ are given in Appendix~\ref{app:harm}.

The osculating equations posses two timescales, characterized by $V$ which is short, and $V_{3}$ which is long. The general solution of these equations can then be solved via the methods of multiple scale analysis~\cite{PoissonWill, Bender, Loutrel:2018ydu, Mora:2003wt}, but the timing model can be computed by a simpler procedure. To obtain the timing model, one has to apply an approximation that accurately takes into account the dynamics of the binary under radiation reaction. For highly eccentric binaries, the gravitational waves are emitted in short bursts during pericenter passage, causing the orbital elements to be effectively constant throughout the orbit, and changing rapidly around pericenter in a nearly step-like manner. This implies that the changes to the orbital elements $\delta^{a} = [p, e, \omega]$, can be calculated by
\begin{align}
    \label{eq:master}
    \delta^{a}_{i} - \delta^{a}_{i-1} = \int_{0}^{2\pi} dV \left(\frac{d\delta^{a}}{dV}\right)_{\delta_{a} = \delta^{a}_{i-1}}\,.
\end{align}
The mappings for the orbital elements are then
\allowdisplaybreaks[4]
\begin{align}
    p_{i} &= p_{i-1} \left[1 + 15\pi \frac{m_{3}}{M} \left(\frac{p_{i-1}}{R}\right)^{3} \frac{e_{i-1}^{2} \sin[2(V_{3}-\omega_{i-1})]}{(1-e_{i-1}^{2})^{7/2}}\right]\,,
    \\
    e_{i} &= e_{i-1} - \frac{15\pi}{2} \frac{m_{3}}{M} \left(\frac{p_{i-1}}{R}\right)^{3} \frac{e_{i-1}\sin[2(V_{3}-\omega_{i-1})]}{(1-e^{2}_{i-1})^{5/2}}\,,
    \\
    \omega_{i} &= \omega_{i-1} + \frac{3\pi}{2} \frac{m_{3}}{M} \left(\frac{p_{i-1}}{R}\right)^{3}\frac{1 + 5 \cos[2(V_{3}-\omega_{i-1})]}{(1-e_{i-1}^{2})^{5/2}}\,.
\end{align}
A similar procedure produces the changes to the time between consecutive pericenter passages:
\begin{widetext}
\begin{equation}
    t_{i}-t_{i-1} = \frac{2\pi}{M^{1/2}}\left(\frac{p_{i-1}}{1-e_{i-1}^{2}}\right)^{3/2} \left[1 + \frac{m_{3}}{M} \left(\frac{p_{i-1}}{R}\right)^{3} \frac{5(4+3e_{i-1}^{2}) + (96+51e_{i-1}^{2}) \cos[2(V_{3}-\omega_{i-1})]}{16(1-e_{i-1}^{2})^{3}}\right]\,.
\end{equation}
\end{widetext}
The burst frequency, namely the peak frequency of the waveform, will still be given by the usual approximate formula, for example Eq.~(20) in~\cite{Loutrel:2014vja}.

\begin{figure*}[hbt!]
    \centering
    \includegraphics[width=\textwidth, trim={4cm 0cm 4cm 0cm}, clip]{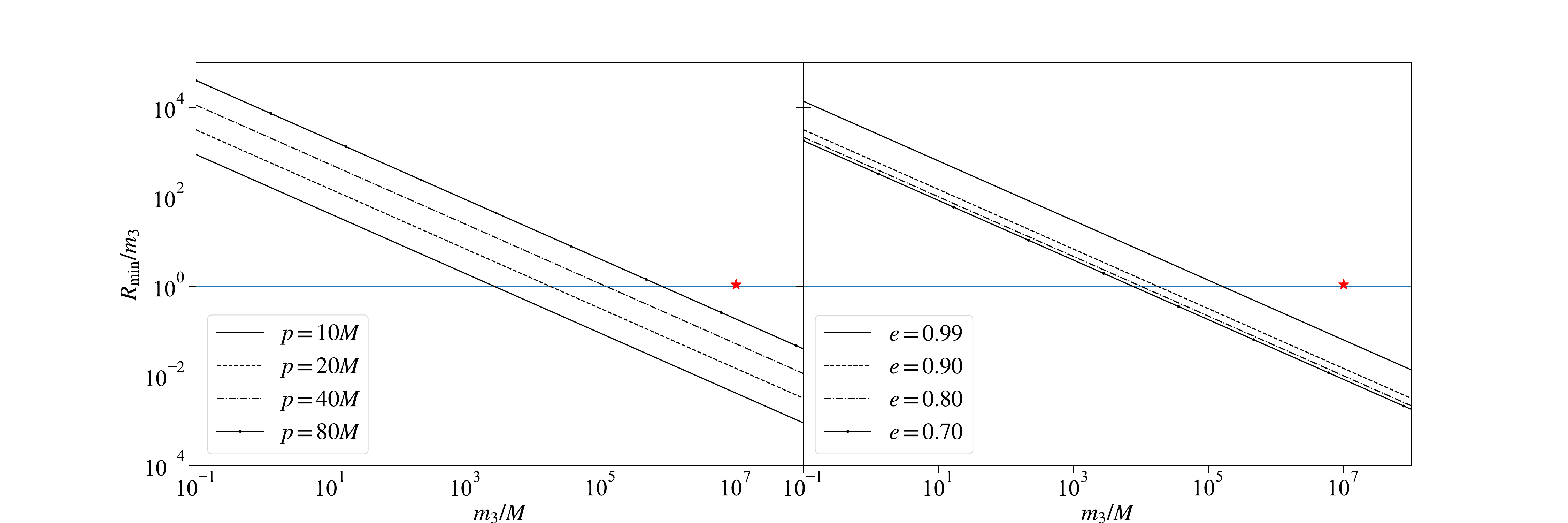}
    \caption{\new{The validity condition in Eq.~\eqref{eq:R-cond} as a function of $\eta_{3}=m_{3}/M$ for $p=[10,20,40,60]M$ with $e=0.9$ (left), and $e=[0.99,0.9,0.8,0.7]$ with $p=20M$ (right). The horizontal line corresponds to $R_{\rm min} = m_{3}$, corresponding to the radius of the innermost stable circular orbit for a maximally-spinning black hole. For any given values of $(p,e)$, values of $R_{\rm min}$ above these lines satisfy the condition in Eq.~\eqref{eq:R-cond}. The system studied in Sec.~\ref{sec:pe} is marked by the red star in each plot. We study a system close to the validity condition as this results in the largest burst detection time perturbations.}}
    \label{fig:val}
\end{figure*}

The model described above does not include the 2.5PN radiation reaction effects, but can be very easily extended to include them. The most up-to-date 2.5PN model is given in~\cite{Arredondo:2021rdt}. To add the three body terms to this, we assume the validity conditions
\begin{equation}
    \frac{m_{3}}{M} \left(\frac{p}{R}\right)^{3} (1-e^{2})^{-3} \ll \eta
    \left(\frac{M}{p}\right)^{5/2} (1-e^{2})^{-1}
\end{equation}
for any $[p,e]$ in the inspiral, and with $\eta = \mu/M$ the symmetric mass ratio of the inner binary. The above condition ensures that the three body effects are small compared to the 2.5PN effects, which is a fair assumption as long as the inner binary inspirals before the third body can disrupt it again. Rearranging terms, we have
\begin{align}
    \label{eq:R-cond}
    R &\gg R_{\rm min}\,,
    \nonumber \\
    R_{\rm min} &= \left[\frac{(1+e)^{13/2}}{(1-e)^{2}} \frac{\eta_{3}}{\eta v_{p}^{11}}\right]^{1/3} M
    \nonumber\\
    &\approx 1.3\times 10^{-3} {\rm AU} \left[\frac{f(e)}{95.6}\right] \left(\frac{\eta}{1/4}\right)^{-1/3} \left(\frac{\eta_{3}}{1/2}\right)^{1/3}
    \nonumber \\
    &\times \left(\frac{v_{p}}{1/3}\right)^{-11/3} \left(\frac{M}{20 M_{\odot}}\right)\,,
\end{align}
where $f(e)=(1+e)^{13/6}/(1-e)^{2/3}$,  $\eta_{3} = m_{3}/M$, $v_{p} = (1+e)(M/p)^{1/2}$ is the pericenter velocity, $M_{\odot}$ is the solar mass, and we have chosen $e=0.99$ to get the approximate equality for $R_{\rm min}$. Fig.~\ref{fig:val} provides a plot of Eq.~\eqref{eq:R-cond} as a function of $m_{3}/M$ for various values of the semi-latus rectum (left) and eccentricity (right). Generally, values above the relevant lines correspond to systems where Eq.~\eqref{eq:R-cond} is valid. Assuming this condition is satisfied, the \new{resulting} timing model including 2.5PN and three body effects is
\begin{widetext}
\begin{align}
    \label{eq:p-next}
    p_{i} &= p_{i-1} \left[1 - \frac{128\pi}{5} \eta \left(\frac{M}{p_{i-1}}\right)^{5/2} \left(1 + \frac{7}{8} e_{i-1}^{2}\right) + 15\pi \eta_{3} C_{R}^{3} \left(\frac{M}{p_{i-1}}\right)^{-3} \frac{e_{i-1}^{2} \sin[2(V_{3}-\omega_{i-1})]}{(1-e_{i-1}^{2})^{7/2}}\right]\,,
    \\
    \label{eq:e-next}
    e_{i} &= e_{i-1} \left[1 - \frac{608\pi}{15} \eta \left(\frac{M}{p_{i-1}}\right)^{5/2} \left(1 + \frac{121}{304} e_{i-1}^{2}\right) - \frac{15\pi}{2} 
    \eta_{3} C_{R}^{3} \left(\frac{M}{p_{i-1}}\right)^{-3} \frac{\sin[2(V_{3}-\omega_{i-1})]}{(1-e^{2}_{i-1})^{5/2}}\right]
    \\
    \label{eq:w-next}
    \omega_{i} &= \omega_{i-1} + \frac{3\pi}{2} \eta_{3} C_{R}^{3} \left(\frac{M}{p_{i-1}}\right)^{-3}\frac{1 + 5 \cos[2(V_{3}-\omega_{i-1})]}{(1-e_{i-1}^{2})^{5/2}}
    \\
    t_{i} - t_{i-1} &= \frac{2\pi}{M^{1/2}} \left[\frac{p_{i-1} + \eta \frac{M^{5/2}}{p_{i-1}^{3/2}} A(e_{i-1},p_{i-1},\eta)}{1-e_{i-1}^{2} + \eta \left(\frac{M}{p_{i-1}}\right)^{5/2} B(e_{i-1})}\right]^{3/2}
    \nonumber \\
    &\times \left[1 + \eta_{3} C_{R}^{3} \left(\frac{M}{p_{i-1}}\right)^{-3}\frac{5(4+3e_{i-1}^{2}) + (96+51e_{i-1}^{2}) \cos[2(V_{3}-\omega_{i-1})]}{16(1-e_{i-1}^{2})^{3}}\right]
    \label{eq:time-offset}
\end{align}
\end{widetext}
where $C_{R} = M/R$ is the dimensionless compactness of the outer orbit. The functions $A(e,p,\eta)$ and $B(e)$ in Eq.~\eqref{eq:time-offset} are given in Eqs.~(63)-(66d) in~\cite{Arredondo:2021rdt}. Note that there is no 2.5PN correction to the longitude of pericenter, and as a result, was not included in the model of~\cite{Arredondo:2021rdt}.

The timing model in Eqs.~\eqref{eq:p-next}-\eqref{eq:time-offset} provides a simplified model of the dynamics of a hierarchical triple system. The assumptions that have gone into the model are as follows:
\begin{itemize}
    \item The eccentricity of the inner binary is close to unity $(e\sim1)$, which provides the assumption that the evolution of the binary is nearly step-like, and allows us to apply Eq.~\eqref{eq:master}. For a standard two-body inspiral, the emission of GWs acts to circularize the binary, and as a result this assumption breaks down if the binary does not merge with appreciably large eccentricity. The action of the third body can either increase or decrease the eccentricity of the inner binary depending on the phase of the outer orbit. Since we are treating the action of the tertiary as a perturbation of the standard two body dynamics, any changes in the eccentricity (as well as the other orbital elements) due to the tertiary are small.
    \item The orbit of the third body is circular and is co-planar with the orbit of the inner binary. We have made these assumptions to simplify the analysis carried out herein. These assumption can be readily relaxed by incorporating the inclination angle $\iota$ and orbital eccentricity of the third body $e_{3}$ in Eqs.~\eqref{eq:dpdV}-\eqref{eq:dtdV}.
    \item We have neglected higher PN order effects in the dynamics of the inner binary, and as a result, the timing model of Eqs.~\eqref{eq:p-next}-\eqref{eq:time-offset} only include the Newtonian and 2.5PN order effects for the dynamics of the inner binary. Higher PN order effects in both the conservative (orbital) and dissipative (radiation reaction) dynamics of the binary are not, in general, negligible. However, the purpose of this study is to quantify whether three body effects can be measured from eccentric burst timing, and thus we have made use of a simplified model that includes only leading PN order conservative and dissipative effects. This assumption can be relaxed by repeating the calculations herein with the higher PN order effects of the inner binary in Eq.~\eqref{eq:eom}, for example, \cite{Loutrel:2017fgu}.
    \item We have made an assumption that the outer orbit is much larger than the inner orbit $R \gg r_{12}$, and have thus neglected the higher PN order effects generated by the third body. Such an assumption must be made in order to preserve the perturbative scheme needed to construct the timing model. Further, if this assumption is broken, the dynamics of the full triple system become highly complex, and possibly chaotic, rendering the methods herein difficult to perform. Under the assumption of $R \gg r_{12}$, the higher PN order effects of the third body are suppressed by powers of $(M+m_{3})/R$, which must be small in order for the assumptions about the PN expansion to hold. Thus, we do not consider them here. However, this assumption can also be relaxed by incorporating the higher PN order effects of the third body in Eq.~\eqref{eq:eom}, as long as the separation of the scales holds.
\end{itemize}
As long as the above assumptions are maintained, as well as Eq.~\eqref{eq:R-cond}, the timing model presented here is an accurate description of the dynamics of the triple system.

As a last comment, we have so far neglected to make any determination of $V_{3}$ in the model. As a matter of simplicity, one could take $V_{3} = \text{constant}$, since the condition $r_{12} \ll R$ implies $T_{\rm orb} = (t_{i}-t_{i-1})_{C_{R}=0} \ll T_{3}$, with $T_{3}$ being the period of the outer orbit. However, to incorporate the motion of the third body, this is included into the model in the following fashion. Since we assume that the third body moves on a circular orbit, the phase of the outer orbit evolves according to
\begin{equation}
    V_{3} = \int_{t_{i-1}}^{t_{i}} dt \; \Omega_{3}
\end{equation}
where
\begin{equation}
    \Omega_{3} = \left(\frac{M + m_{3}}{R^{3}}\right)^{1/2}
\end{equation}
Strictly speaking, $\Omega_{3}$ is a function of time through back-reaction effects from the inner orbit, as well as the gravitational wave emission of the outer orbit. However, because $V_{3}$ is already suppressed within the timing model, these effects can be neglected, and the integral reduces to
\begin{equation}
    \label{eq:V3}
    V_{3} = \Omega_{3} \left(t_{i}-t_{i-1}\right)_{C_{R}=0}\,.
\end{equation}
%

\subsubsection{\new{Center of Mass Motion and Propogation of Gravitational Waves}}
\label{sec:com}

\new{
If the binary is stationary with respect to the observer, then the burst arrival times are trivially given by Eqs.~\eqref{eq:p-next}-\eqref{eq:time-offset}. Such a scenario is possible if the mass of the tertiary $m_{3}$ is small compared to the total mass of the binary $M$, and thus the binary's COM would not experience significant motion. However, this need not necessarily be true for astrophysically relevant systems. Further, even if the motion of the binary's COM is slow, observations over sufficiently long timescales could make the effect observable. We here provide an explicit calculation of the effects of the binary's COM motion on the burst arrival times.
}

\new{
In the binary's COM frame, the bursts are emitted at times $t_{i}^{e}$, which are given by Eq.~\eqref{eq:time-offset}. However, the GWs arrive at the detector frame at times $t^{o}_{i}$, given by
\begin{align}
    t^{o}_{i} &= t^{e}_{i} - \vec{d}_{i} \cdot \vec{n}_{o}\,,
\end{align}
which should be recognized as the retarded time (in units where $c=1$) for a source at distance $\vec{d}_{i}$ from the detector, and with $\vec{n}_{o}$ the line of sight between the source and detector. If $d_{i}$ is constant, the correction to $t^{e}_{i}$ becomes degenerate with an overall time offset, and thus cannot be directly measured. In our case, $\vec{d}_{i}$ is not constant and is given by
\begin{equation}
    \vec{d}_{i} = \vec{d}_{\rm bary} + \vec{r}_{{\rm com}}(t_{i}^{e})\,,
\end{equation}
where $\vec{d}_{\rm bary}$ is the distance between the detector and the total barycenter of the triple system, and $\vec{r}_{\rm com}$ is the distance between the total barycenter and the binary's COM. We assume that $\vec{d}_{\rm bary}$ is constant, and as a result, can be neglected due to the aforementioned degeneracy. Thus, the observed arrival times of the burst are
\begin{equation}
    t^{o}_{i} = t^{e}_{i} - \vec{r}_{\rm com}(t^{e}_{i}) \cdot \vec{n}_{o}\,.
\end{equation}
}

\begin{figure*}[hbt!]
    \centering
    \includegraphics[width=\textwidth, trim={1cm 8cm 1cm 8cm}, clip]{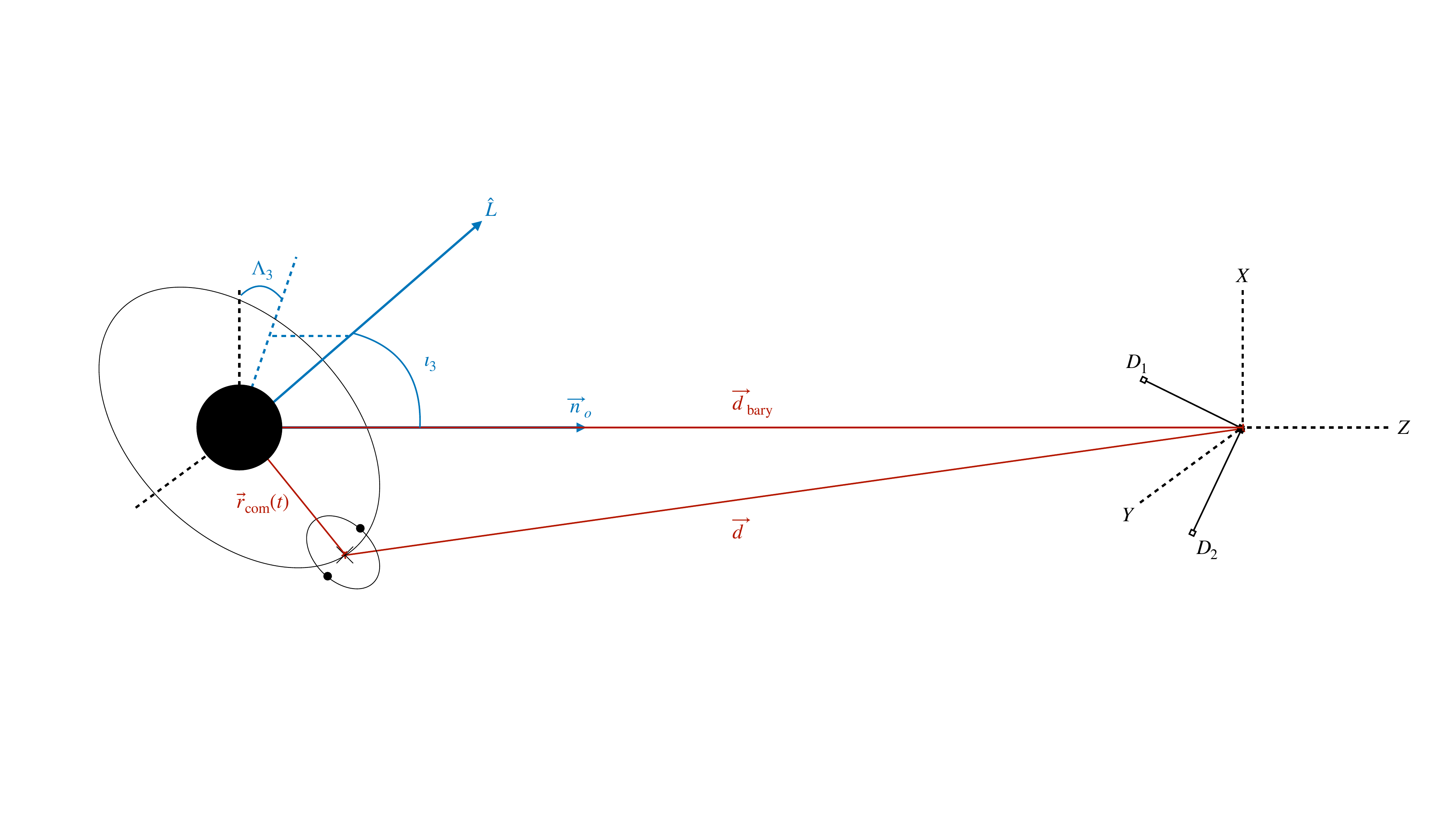}
    \caption{\new{Schematic of the COM motion of a binary in a triple system. The normal to the orbital plane $\hat{L}$ is oriented with respect to the line of sight $\vec{n}_{o}$ by angles $(\iota_{3},\Lambda_{3})$. The total barycenter of the system $d_{\rm bary}$ is assumed to be stationary with respect to the detector, which is specified by arms $D_{1,2}$. The distance from the binary to the source $\vec{d}$ is time dependent through the binary's COM position $\vec{r}_{\rm com}(t)$, modifying the arrival times of the bursts at the detector.}}
    \label{fig:com}
\end{figure*}
\new{
Fig.~\ref{fig:com} provides a schematic of the triple system's dynamics in the detector frame. The relative separation between the tertiary and the binary's COM is trivially given by $\vec{R} = R \vec{N}$, where $\vec{N}$ is given in Eq.~\eqref{eq:Nvec}. Thus, in the barycenter frame, the separation between the barycenter and the binary's COM is
\begin{equation}
    \vec{r}_{\rm com} = \vec{R} = R \left[\cos V_{3}, \sin V_{3},0 \right]\,,
\end{equation}
Recall that we are assuming that the outer orbit is circular, and thus, $R$ is constant and the only time evolution in this quantity comes from $V_{3}$. To determine $\vec{n}_{o}$, we assume the mutual plane of the inner and outer orbits can be arbitrarily oriented with respect to an observer. Since we are assuming the outer orbit is circular, the normal vector of the orbital plane will be determined by two angles, the inclination angle $\iota_{3}$ and the line of nodes $\Lambda_{3}$. Thus, in the barycenter frame
\begin{equation}
    \vec{n}_{o} = \left[\sin\iota_{3} \cos\Lambda_{3}, \sin\iota_{3} \sin\Lambda_{3}, \cos\iota_{3}\right],
\end{equation}
giving us our final expression for the burst arrival times
\begin{equation}
    \label{eq:time-doppler}
    t^{o}_{i} = t^{e}_{i} - R \sin\iota_{3} \cos\left[V_{3}(t^{e}_{i}) + \Lambda_{3}\right]
\end{equation}
where $V_{3}(t^{e}_{i})$ is given by Eq.~\eqref{eq:V3}. 
}

\new{
There are a few things to note about Eq.~\eqref{eq:time-doppler} before considering inference of the full timing model. First, note that the Doppler shift is simply the light travel time across the outer orbit, and thus, more important for systems with larger $R$ (recall that we have set $c=1$). An example of such a system is a stellar mass BBH orbiting around a SMBH, like the one we study in the next section. However, there is a subtlety with this point, namely the fact that the Doppler shift depends on the phase of the outer orbit $V_{3}$. If $R$ is sufficiently large that $V_{3}$ does no evolve significantly over the inspiral of the binary, than the Doppler shift is effectively constant and, as a result, degenerate with an overall time shift of the full series of bursts. Thus, the Doppler effect is most important for systems with larger $R$, but not sufficiently large that $V_{3}$ does not significantly evolve on the inspiral timescale. We consider this point in more detail in Sec.~\ref{sec:max-R}.
}

\begin{figure}
    \centering
    \includegraphics[width=\columnwidth]{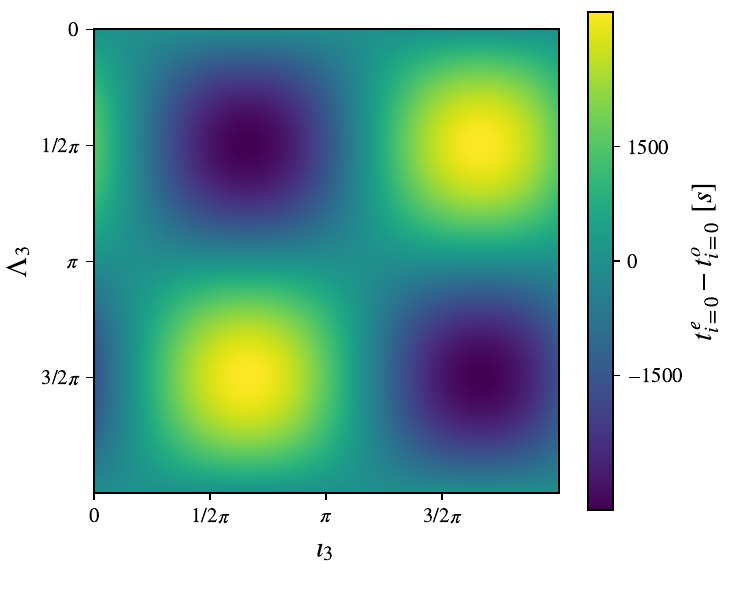}
    \caption{\new{The time offset between the emitted burst time $t_i^e$ and the detected burst time $t_i^o$ due to variations in the orbital tilt angles $\iota_3$ and $\Lambda_3$. All other parameters match the triple system described in Sec.~\ref{sec:pe}.}}
    \label{fig:burst_time_offset_due_to_tilt_only}
\end{figure}

\new{
Second, a number of approximations have gone into our computation that simplify the expression of the Dopper corrections, specifically we have assumed that the outer orbit is circular, and is not inclined relative to the inner orbit. These can be relaxed, but would require a more thorough investigation of the tidal effects, since these become significantly more complicated. Note also that this depends on the time at which the bursts are generated through $V_{3}$, and thus relies on the burst timing model of the previous section. Any considerations of the validity of the model in Eqs.~\eqref{eq:p-next}-\eqref{eq:time-offset} also then apply to Eq.~\eqref{eq:time-doppler}. Third, while the main timing model in Eqs.~\eqref{eq:p-next}-\eqref{eq:time-offset} are purely a function of parameter combinations $\frac{m_3}{R^3}$ and $2(V_3 - \omega_{i-1})$, Eq.~\eqref{eq:time-doppler} breaks these degeneracies; thus, for systems with non-zero $(\iota_{3},\Lambda_{3})$ (hereafter referred to as ``tilted'' systems), we should expect a more accurate recovery of individual parameters via parameter estimation. Finally, the Doppler effect introduces two new parameters into the model, namely $(\iota_{3},\Lambda_{3})$. The influence of varying $\iota_3$ and $\Lambda_3$ on the initial burst, expressed as the difference between the emitted time $t_i^e$ and observed time $t_i^o$, is shown in Fig.~\ref{fig:burst_time_offset_due_to_tilt_only}.
}

\subsection{Inference of a perturbing tertiary}
\label{sec:pe}

We test our ability to recover the properties of the tertiary by modelling the GW signal from a highly-eccentric binary with a total mass of $M = 60$~M$_\odot$, orbiting a SMBH of $1 \times 10^{7} M$ ($6 \times 10^{8}$~M$_\odot$) at a distance of $1.1 \times 10^{7} M$ ($6.55$~au). 
\new{This system sits close to the validity limit given in Eq.~\ref{eq:R-cond}, and its position relative to this limit is shown in Fig.~\ref{fig:val}. We choose to study a system at close to the limit of validity as this results in the largest and most detectable burst offset times; as $R$ increases for a fixed $m_3$, the perturbation the third body induces in the motion of the binary is reduced, as illustrated in Fig.~\ref{fig:varying_burst_offsets_R_iota_Lambda}.}
We study the data using hierarchical Bayesian inference, treating each burst as an individual ``event'' and the full set of bursts as a ``population'' of individual events. 
\new{In Sec.~\ref{sec:bayes}, we describe our Bayesian analysis framework. In Sec.~\ref{sec:max-R}, we estimate the maximum $R$ for which this framework is sensitive to the influence of a tertiary. In Sec.~\ref{sec:settings}, we provide our injection and analysis settings. We provide inferred tertiary properties for known inner binaries with zero and non-zero orbital tilt angles in Sections \ref{sec:results-subset} \& \ref{sec:results-subset-tilted}, respectively. We provide inferred properties of triple systems with known and unknown orbital tilt angles in Sec.~\ref{sec:results-all} \& \ref{sec:results-all-tilted}, respectively. In the latter two sections, we investigate several priors on inner binary parameters that represent different levels of confidence in downstream parameter estimation of the inner binary properties. In Sec.~\ref{sec:R-var}, we briefly discuss correlations between $R$ and other system properties.}

\begin{figure*}
    \centering
    \includegraphics[width=0.9\textwidth, trim = {1.5cm 0cm 0cm 0cm}, clip]{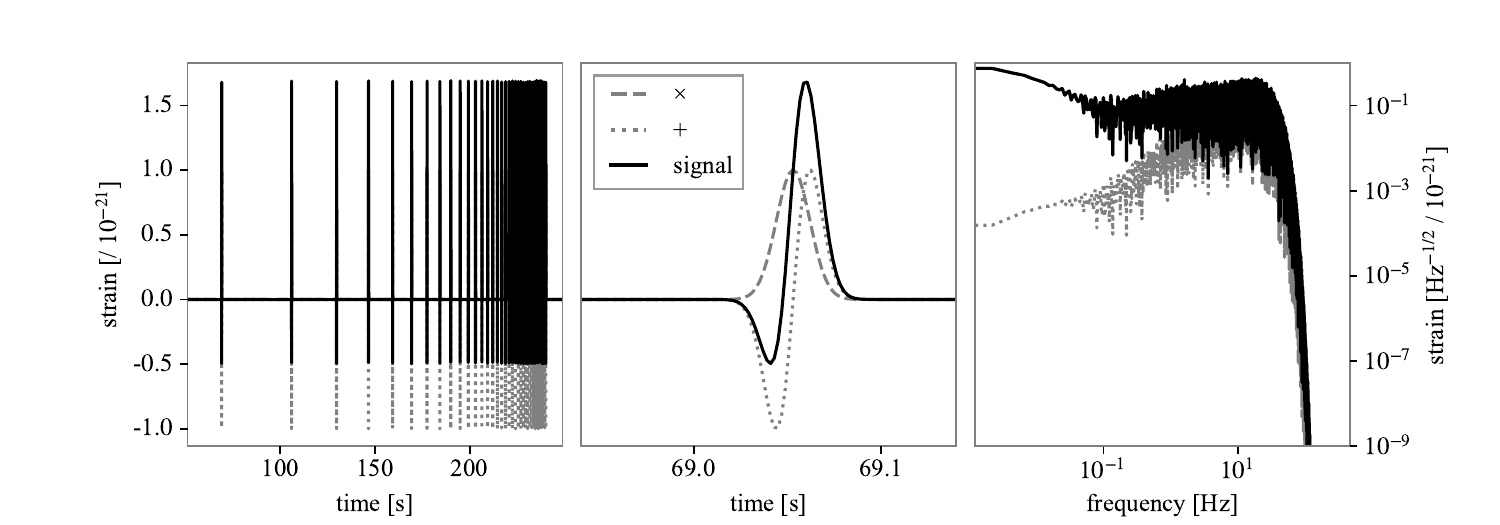}
    \caption{The full injected signal from the perturbed system in the time domain (left), a single burst in the time domain (middle) and the full signal in the frequency domain (right). The plus and cross polarisations of the signal are plotted in grey and the total signal is plotted in black.}
    \label{fig:injected_waveform}
\end{figure*}

\subsubsection{Bayesian framework}
\label{sec:bayes}

Using the timing model derived in Section~\ref{sec:model}, we aim to infer the existence and identity of a tertiary perturber from the GW signal of a highly eccentric binary. 
We use a two-step inference procedure inspired by pulsar timing \cite[e.g.][]{Edwards:2006zg,Manchester:2017:PulsarTiming,Verbiest:2021:PulsarTiming} and hierarchical Bayesian inference methods \cite[e.g.,][]{ThraneTalbot:BayesianInference:2019,LVK:2021:Population3}.

Pulsars are rotating neutron stars that emit extremely regular electromagnetic pulses, which are observed with radio telescopes on Earth. 
The pulse time of arrival (TOA) is measured by averaging over, or ``folding'', thousands of data segments of length equal to one pulse period, fitting the shape of the pulse to a standard pulse profile, and measuring the time offset ($\Delta$T) of the pulse peak from the start of the data segment \citep{Verbiest:2021:PulsarTiming}.
This gives the pulse TOA averaged over thousands of pulses, plus an associated uncertainty due to measurement noise.
In the absence of interference, every measurement should have a constant $\Delta$T.
However, multiple phenomena---e.g., pulsar spin-down, pulsar proper motion, the Earth's orbit around the Sun, low-frequency gravitational waves---can cause timing ``residuals'': differences between the expected TOA and the observed TOA.
Pulse TOA datasets are therefore fit using timing models that account for these phenomena.
These timing models contain no information about the shape of the pulse itself.
We take inspiration from pulsar timing methods in two ways: (i) by finding the time of each burst independently of the others, and (ii) by removing waveform phenomenology from the timing model analysis.

When simulating the parameter estimation pipeline, we make two key assumptions.
Firstly, we assume that the approximate location of each burst in the time-domain data is flagged by a pipeline like, e.g., Omicron \citep{Robinet:2020:Omicron} or \texttt{cWB} \cite{Klimenko:cWB:2008, Drago:cWB:2020}. 
This gives us a set of distinct data segments within which to search for bursts. 
Burst searches are expected to be more efficient for finding highly-eccentric sources than templated searches because existing template banks neglect orbital eccentricity, leading high-eccentricity sources to be missed and for sources with non-vanishing eccentricity to have reduced signal-to-noise ratios \citep[e.g.,][]{MartelPoisson:1999:EccentricSignalLoss, BrownZimmerman:2010:EccentricSignalLoss, RamosBuades:2020:EccentricSearchEfficiency, Zevin:2021:seleccentricity}.\footnote{One possible example of this effect in action is the potentially-eccentric event GW190521 \citep{Romero-Shaw:2020:GW190521, Gamba:2021:GW190521, Gayathri:2022:GW190521}, which was flagged by burst searches with a much lower false-alarm rate than by templated searches \citep{Abbott:2020:GW190521}.}
Proposed burst searches relying on the expected burst time evolution of a non-perturbed binary \citep{CheeseboroBaker:2021:EccentricBurstSearch} may be sufficient if the binary is only slightly perturbed by the tertiary.
Templated searches also rely on accurate reproductions of the signals of well-modelled systems.
For closely-interacting triples, the early burst times will be offset by a large amount from those expected from a non-perturbed binary and may evolve chaotically, so a completely uninformed burst search will likely be more applicable. 
While approximate models for eccentric LIGO-Virgo-KAGRA sources exist and/or are under development \citep[e.g.,][]{Tiwari:2019:EccentricWaveform, Islam:2021:EccentricWaveform, Nagar:2021:EccentricWaveform, SetyawatiOhme:2021:EccentricWaveform, Liu:2022:EccentricWaveform, RamosBuades:2022:EccentricWaveform, Cho:2022:EccentricWaveform,Moore:2019xkm}, these become unreliable at higher eccentricities and the lower frequencies required for next-generation detector sources, and neglect other physical effects such as spin-induced precession \citep{Romero-Shaw:2020:GW190521,Romero-Shaw:2022:EccOrPrecc}. Furthermore, no models exist for full gravitational-wave signals from eccentric binaries perturbed by a tertiary.

Our second key assumption is that each burst is unambiguously detected, i.e., we do not miss any bursts. For next-generation detectors, we can expect full eccentric signals to have SNRs comparable to their quasi-circular counterparts \citep{Chen:2021:2G3GEccentricProspects}; however, the aforementioned lack of full waveforms for perturbed eccentric binaries means that each burst will need to be detected individually for the source to be correctly interpreted. In Advanced LIGO, the \new{matched-filter} SNR of a single burst from a $10 + 10$~M$_\odot$ binary $100$~Mpc away is expected to be $\sim1$--$20$ when the binary is $\sim1$--$1000$~s from merger \citep{KocsisLevin:2012:BurstsGN, Loutrel:2020jfx}. In next-generation detectors, we anticipate that these SNRs will increase by orders of magnitude \citep{Abbott:2017:Exploring3G}. In the injection recovery studies below, each burst has a \new{matched-filter} SNR of $\approx 80$.

We represent the array of $N$ true burst times as $t_{\mathrm{b}}$, and each individual burst time within this set as $t_{\mathrm{b},i}$. 
The first step in the recovery of the tertiary parameters is to calculate a probability distribution for each individual burst arrival time, $p(t_{\mathrm{b},i} | d_i)$, where $d_i$ is the data stretch containing the burst at time $t_{\mathrm{b},i}$.
We split the data so that each burst is contained within a unique data segment. 
For this demonstration, we choose the segment start and end times to be $t_{\mathrm{start}, i} = t_{\mathrm{b},i} - \frac{1}{2}(0.125 + R_{\mathrm{start}, i})$~s and $t_{\mathrm{end}, i} = t_{\mathrm{b},i} + \frac{1}{2}(0.125 + R_{\mathrm{end}, i})$~s, respectively, where $R_{\mathrm{start}, i}$ and $R_{\mathrm{end}, i}$ are different random numbers between $0$ and $1$.
This is to emulate a situation in which several distinct data segments of different length are flagged by a burst search as likely to contain a single burst.
Using, e.g., a Markov chain Monte Carlo algorithm, one can obtain the Bayesian posterior probability distribution on each burst arrival time,
\begin{equation}
    p(t_{\mathrm{b},i} | d_i) = \frac{L(d_i | t_{\mathrm{b},i})\pi(t_{\mathrm{b},i})}{\int{d t_{\mathrm{b},i} L(d_i | t_{\mathrm{b},i})\pi(t_{\mathrm{b},i})}},
\end{equation}
where $\pi(t_{\mathrm{b},i})$ is the prior distribution (for example, a uniformly-distributed prior over $t_{\mathrm{b},i}$). $L(d_i | t_{\mathrm{b},i})$ is the Whittle likelihood,
\begin{equation}
    L(d_i | t_{\mathrm{b},i}) = \frac{1}{2 \pi S_n} \exp \left(-\frac{1}{2} \frac{|d_i-\mu(t_{\mathrm{b},i})|^{2}}{S_n}\right),
\end{equation}
where $\mu(t_{\mathrm{b},i})$ is a GW burst occurring at $t_{\mathrm{b},i}$, so that 
\begin{equation}
    h(\theta) = \sum_i^N{\mu(t_{\mathrm{b},i}(\theta))},
\end{equation}
and $S_n$ is the detector noise power spectral density (PSD), which is typically estimated by averaging over the base noise level measured at the detector close to the time of the event \citep[e.g.,][]{Abbott:2020:Noise, TalbotThrane:2020:UncertainPSD}. To quickly approximate such a probability distribution, we compute the overlap between the data and a single burst (such as that shown in the middle panel of Fig. \ref{fig:injected_waveform}) over the data segment duration, and draw $100$ samples to approximate $p(t_{\mathrm{b},i} | d_i)$ from the resulting function.\footnote{These samples are analogous to those obtained for individual events in Bayesian population inference, for example, the posterior probability distribution on chirp mass for a single binary black hole merger. In this case the number of bursts, $N$, is the size of the ``population''.}

In the second step, we combine all of these single-burst posteriors into a single dataset, $t_{\mathrm{samp}}$.
This can be fit with a hierarchical inference model with some hyper-parameters, $\Theta$, that determine the sequence of burst times.\footnote{In typical population inference studies for GW astrophysics, such a model might describe the population chirp mass distribution, with $\Theta$ describing the shape and limits of the distribution.}
In this case, the model for the burst times is the one described in Sec.~\ref{sec:model}, which depends on \new{$\Theta = [e_0, p_0, M, \eta, V_{3,0}, \omega_0, m_3, R, \iota_3, \Lambda_3]$}.
In order to compute the posterior probability distribution on the hyper-parameters (the hyper-posterior), we first need to calculate the hyper-likelihood. 
This takes the form
\begin{equation}
L(d | \Theta)=\int d t_\mathrm{b} \mathcal{L}(d | t_\mathrm{b}) \pi(t_\mathrm{b} | \Theta),
\end{equation}
where $d$ is the entire stretch of data comprising $N$ data segments $d_i$, and $\pi(t_\mathrm{b} | \Theta)$ is a new prior distribution on the burst time that is dependent on $\Theta$. 

To compute this in practice, we take our individual burst posterior probability distributions $p(t_{\mathrm{b},i} | d_i)$, and divide out the uniform prior we used to obtain that distribution in favour of a hyper-parameter-dependent prior.
For each realisation of source parameters $\Theta_j$, we generate a corresponding sequence of GW burst times, $t_{\mathrm{b}, j}(\Theta_j)$.
We then find the burst timing residuals between $t_\mathrm{samp}$ and $t_{\mathrm{b}, j}(\Theta_j)$.
A realisation of $\Theta_j$ with small residuals has a higher probability of matching the true or injected $\Theta$ than one that produces large residuals.
The likelihood for the entire series of bursts can be written as \cite[see Eq.~(32) -- (35) in][]{ThraneTalbot:BayesianInference:2019}
\begin{equation}
L(d | \Theta) =\prod_{i}^{N} \frac{\int{d t_{\mathrm{b},i} L(d_i | t_{\mathrm{b},i})\pi(t_{\mathrm{b},i})}}{n_{i}} \sum_{k}^{n_{i}} \frac{\pi(t_{\mathrm{b},i}^{k} | \Theta)}{\pi(t_{\mathrm{b},i}^{k})},
\end{equation}
where $k$ denotes a single sample from the posterior probability distribution of burst $i$, and $n_i$ is the number of posterior samples for each burst. 
Finally, the hyper-posterior is
\begin{equation}
p(\Theta | d)=\frac{L(d | \Theta) \pi(\Theta)}{\int d \Theta L(d | \Theta) \pi(\Theta)},
\end{equation}
where $\pi(\Theta)$ is the hyper-prior, which encodes our prior knowledge about the natural distributions of parameters \new{$\Theta = [e_0, p_0, M, \eta, V_{3,0}, \omega_0, m_3, R, \iota_3, \Lambda_3]$}.
We draw proposed values for $\Theta$ from these priors.
In the analyses in the following section, we use four different prior models to represent different levels of uncertainty on the properties of the inner binary; these priors are detailed in Table \ref{tab:priors}.

\subsubsection{\new{Estimation of maximum $R$ at which influence of tertiary is detectable}}
\label{sec:max-R}

\begin{figure}
    \centering
    \includegraphics[width=\columnwidth]{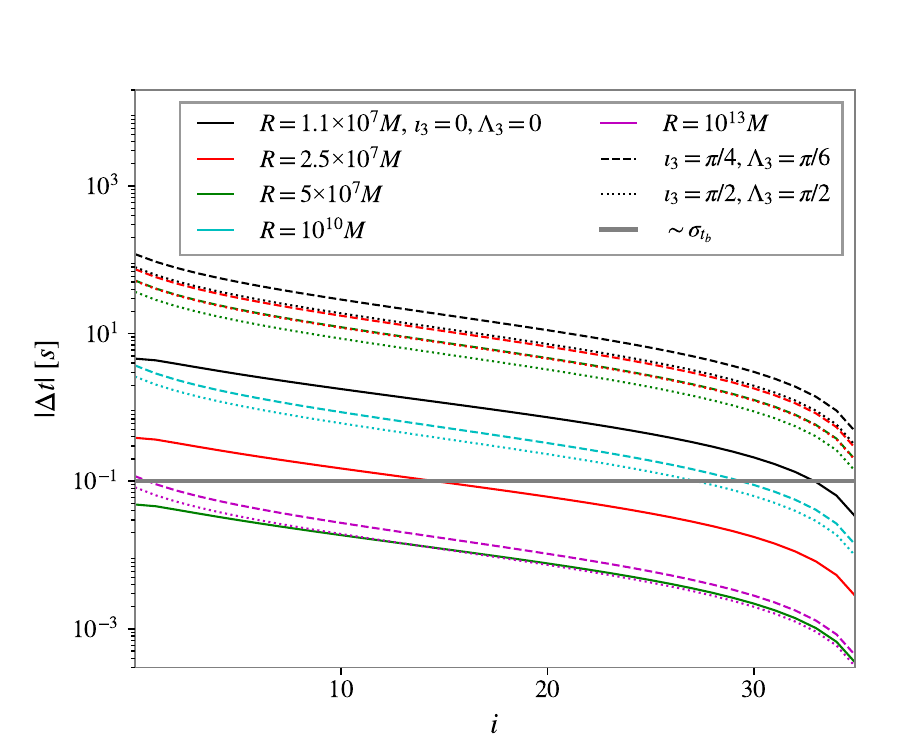}
    \caption{\new{The burst number $i$ vs the offset between that burst time and the same burst time from an unperturbed binary, for varying values of $R$, $\iota_3$ and $\Lambda_3$. Other parameters are as given in e.g. Table \ref{tab:tilted_signal_recovery}. Different values of $R$ are represented by different coloured curves. Solid curves show the burst time offsets for a face-on co-planar triple; dashed curves show burst time offsets for tilt angles $\iota_3=\frac{\pi}{4}, \Lambda_3=\frac{\pi}{6}$; and dotted curves show burst time offsets for a system with $\iota_3=\Lambda_3=\pi/2$. The horizontal grey bar shows the uncertainty on the detected burst time, which we approximately equate to the mean average standard deviation of the samples in our mock posteriors on burst arrival time. We conservatively estimate that when the maximum burst time offset is below this uncertainty, it would not be possible to confidently infer the existence of the perturber.}}
    \label{fig:varying_burst_offsets_R_iota_Lambda}
\end{figure}

\new{
We estimate the maximum $R$ for which the effect of a perturbing tertiary can be detectable, for a given value of $m_3=10^7 M$.
To produce this estimate, we compare the burst time offsets between a perturbed and non-perturbed system for a selection of $R$, $\iota_3$ and $\Lambda_3$, and compare them against our approximate uncertainty on the burst time, $\sigma_{t_b}$.
The uncertainty $\sigma_{t_b} \sim 0.1$ is the mean average of the standard deviations calculated for all mock posteriors on burst time in Sections \ref{sec:results-subset}--\ref{sec:results-all-tilted}.
This value is approximate because the samples in the posteriors are drawn at random from the overlap curves, so the exact value varies a little (between $\approx 0.06$ and $\approx 0.1$).
Since the burst timing model depends on the quantity $\frac{m_3}{R^3}$ when the co-planar system is face-on, the solid curves in Fig.~\ref{fig:varying_burst_offsets_R_iota_Lambda} are identical for different systems with the same values of $\frac{m_3}{R^3}$ (with $m_3 = 10^7 M$, $\frac{m_3}{R^3} = [7.5\times 10^{-15}, 6.4\times 10^{-16}, 8\times 10^{-17}]$ for $R = [1.1, 2.5, 5]\times 10^{7} M$). When the system is face-on, we estimate that a tertiary of mass $m_3=10^{7} M$ would become undetectable when $R \approx 5 \times 10^{7} M$. However, when the system is tilted, the effect of the tertiary may remain detectable until $R \gtrsim 10^{13} M$.
}

\subsubsection{Injection and analysis settings}
\label{sec:settings}

We create synthetic data by injecting mock signals into simulated Einstein Telescope (ET) detector noise using \texttt{bilby} \citep{bilby, Romero-Shaw:2020:Bilby}, assuming the ET-B noise PSD \citep{Hild:2008:ETB}. 
To create a mock signal, we take an array of burst times, $t_b(\Theta)$, produced by the burst timing model defined in Eqs.~\eqref{eq:p-next}-\eqref{eq:time-offset}. 
We model the plus polarisation of the signal as a Gaussian and the cross polarisation as a sine-Gaussian, both centred at the burst time $t_{\mathrm{b},i}$.
This form is a toy model of the analytic waveforms developed in \citep{Loutrel:2019kky}.
To ensure the system is in a regime in which its signal would be sufficiently ``burst-like'' (i.e., minimal GW emission between each pericenter passage) we restrict the model to $e_i > 0.7$. 
The width of the burst is $\sigma = 1.5\times10^{-4} M$, a value chosen because it produces a close match to the analytic waveforms presented in \citep{Loutrel:2019kky}.
The injected signal is plotted in both the time and frequency domains in Fig. \ref{fig:injected_waveform}. We use the \texttt{hyper} submodule of \texttt{bilby} \cite{bilby, Romero-Shaw:2020:Bilby} to perform inference.
We use sampler settings of $100$ walks, $4000$ live points, and $10$ required auto-correlation times.

\begin{figure}
    \centering
    \includegraphics[width=0.95\columnwidth]{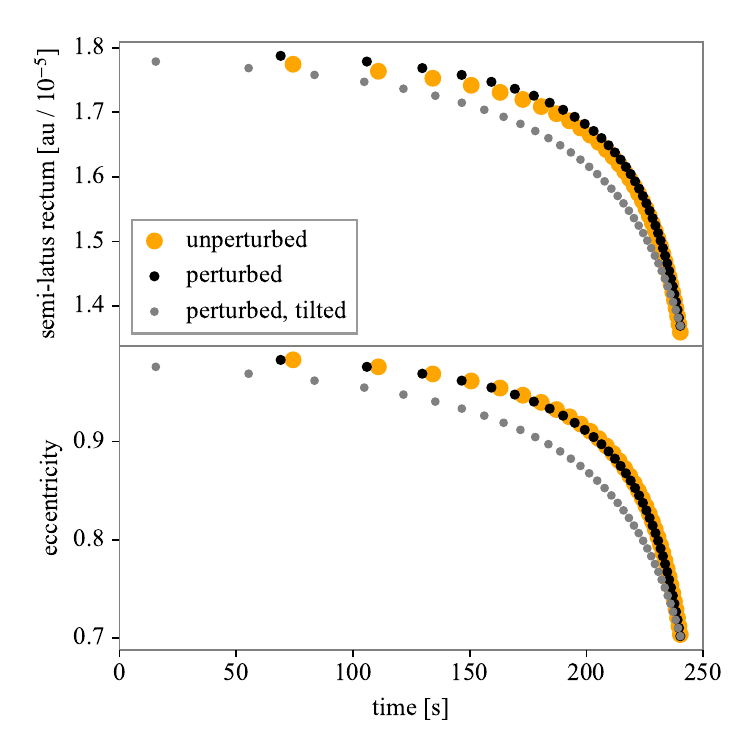}
    \caption{The evolution of the orbital eccentricity $e$ (bottom) and semi-latus rectum $p$ (top) with respect to time for a perturbed binary interacting with a third body \new{in a coplanar face-on orbit} (black)\new{, a coplanar orbit that is tilted with $\Lambda_3=\frac{\pi}{6}, \iota_3=\frac{\pi}{4}$ with respect to the observer (grey),} and an unperturbed binary (orange). The system parameters are identical to those used in our injection studies. The final bursts of each signal are time-matched so that the first-detected bursts have the greatest time offset from each other.\label{fig:perturbed-vs-unperturbed}}
\end{figure}
We inject a signal from a binary that is undergoing perturbation from a third body.
The toy model detailed in Sec.~\ref{sec:model} is only applicable in the case that the inner binary and the tertiary are co-planar, so we consider a system in which this configuration is plausible: a  $M = 60$~M$_\odot$, $\eta = 0.20$ BBH orbiting a $10^7 M$ ($6 \times 10^8$~M$_\odot$) SMBH with an accretion disk.
A maximally-spinning SMBH of this size would have an innermost \new{prograde} stable circular orbit close to its horizon radius of $\approx 5.95$~au.\footnote{Most accreting SMBH are thought have close-to-maximal spins \citep{Ananna:2020:AGNSpin}, and the radius of the innermost stable circular orbit approaches the horizon radius for maximally-spinning black holes \citep{Bardeen:1972:KerrBH}. While we ignore the influence of the spin of the SMBH on the evolution of the binary properties here, see \cite{Fang:2019:SMBHSpin} for an example of how SMBH spin may be measured via its influence on nearby binaries observed with space-based future GW detectors.} 
We position the binary $1.1 \times 10^7 M$ ($6.55$~au) from the SMBH, outside its horizon but within the radial extent of an accretion disk that could align the outer and inner orbital planes~\citep[see][and references therein]{Nasim:2022:AGNalignplane}.
The initial values of the orbital phase parameters, $\omega_0$ and $V_{3,0}$, are arbitrarily chosen to be $0$ and $\pi/3$, respectively.
The inner binary is initialised with $e_0=0.99$ and $p=30 M$ ($1.8\times10^{-5}$~au).
We scale the strain such that the \new{matched-filter} SNR for the entire signal is $\approx495$, while the \new{matched-filter} SNR of a single burst is $\approx80$.
Such SNRs are expected to be common for next-generation detectors \citep{Evans:2021:CosmicExplorer}.
We compare the burst times and evolution of orbital parameters $e$ and $p$ for this system to those of the same binary in isolation in Fig. \ref{fig:perturbed-vs-unperturbed}.

\new{In Sections \ref{sec:results-subset} \& \ref{sec:results-all}, we neglect} the burst time and frequency shift due to the binary motion around the tertiary \citep{Johan-doppler}.
The inclination of both the outer orbit and the inner orbit to the observer \new{are assumed to} be face-on \new{in these scenarios}, since in this configuration the binary COM motion does not influence the burst arrival times at the detector. 
\new{We allow the orbital tilt to vary in the analyses presented in Sections \ref{sec:results-subset-tilted} \& \ref{sec:results-all-tilted}.}

\new{In all analyses, we} assume that other extrinsic parameters (e.g., the distance to the source and its location on the sky) have already been inferred from the late inspiral and merger.
While we could in principle implement inference over these extrinsic parameters---for example, differences in the luminosity distance of the source could be explored by redshifting the waveform frequency and inversely scaling its amplitude---we do not go to these lengths for this initial proof-of-principle study. 
Thus, all of our constraints should be considered lower limits, since they do not include the additional uncertainty that would come from exploring the extrinsic parameter space.

\begin{table*}[]
    \centering
    \begin{tabular}{c||c|c|c|c|}
    Parameter &  $\delta$ + Uniform ($\delta+\mathcal{U}$) & Gaussian 1 + Uniform ($\mathcal{G}1+\mathcal{U}$) & Gaussian 2 + Uniform ($\mathcal{G}2+\mathcal{U}$) & Uniform ($\mathcal{U}$) \\
     \hline
      $M$ [M$_\odot$]  & $\delta(x-60)$ & $\mathcal{G}(\mu=60,\sigma=0.06)$ & $\mathcal{G}(\mu=60,\sigma=0.6)$ & $\mathcal{U}(\mathrm{min}=10,\mathrm{max}=400)$ \\
      $e$ & $\delta(x-0.99)$ & $\mathcal{G}(\mu=0.99,\sigma=0.99\times10^{-3})$ & $\mathcal{G}(\mu=0.99,\sigma=0.99\times10^{-2})$ & $\mathcal{U}(\mathrm{min}=0.7,\mathrm{max}=1.0)$ \\
      $p$ [$M$] & $\delta(x-30)$ & $\mathcal{G}(\mu=30,\sigma=0.03)$ & $\mathcal{G}(\mu=30,\sigma=0.3)$ & $\mathcal{U}(\mathrm{min}=10,\mathrm{max}=70)$ \\
      $\eta$  & $\delta(x-0.20)$ & $\mathcal{G}(\mu=0.20,\sigma=0.20\times10^{-3})$ & $\mathcal{G}(\mu=0.20, \sigma=0.20\times10^{-2})$ & $\mathcal{U}(\mathrm{min}=0.15,\mathrm{max}=0.25)$ \\
      $\omega_0$ & \multicolumn{4}{c|}{$\mathcal{U}(\mathrm{min}=0,\mathrm{max}=2\pi)$} \\
      $V_{3,0}$ & \multicolumn{4}{c|}{$\mathcal{U}(\mathrm{min}=0,\mathrm{max}=2\pi)$}\\
      $m_3$ [$M \times 10^7$]  & \multicolumn{4}{c|}{$\mathcal{U}(\mathrm{min}=0,\mathrm{max}=1.5)$}\\
      $R$ [$M \times 10^7$]  & \multicolumn{4}{c|}{$\mathcal{U}(\mathrm{min}=0.01,\mathrm{max}=5)$}\\
    \end{tabular}
    \caption{The prior shapes and limits that we use in this work. For $\eta$, we truncate the Gaussian priors to be within the same limits as the uniform priors. 
    \new{Extensions to priors used in the tilted case are described in Sections \ref{sec:results-subset-tilted} \& \ref{sec:results-all-tilted}.}
    }
    \label{tab:priors}
\end{table*}

\begin{table*}[]
    \centering
    \begin{tabular}{c|c||c|c|c|c|}
     Parameter & Injected & \multicolumn{4}{c|}{Recovered} \\
       &  & $\delta+\mathcal{U}$ & $\mathcal{G}1+\mathcal{U}$ & $\mathcal{G}2+\mathcal{U}$ & $\mathcal{U}$ \\
     \hline
      $M$ [M$_\odot$] & $60$ & $(60)$ & $60.00^{+0.10}_{-0.09}$ & $60.0^{+0.7}_{-0.8}$ & $60^{+9}_{-7}$ \\
      $e$  & $0.99$ & $(0.99)$ & $0.98998^{+0.00006}_{-0.00007}$ & $0.9898^{+0.0001}_{-0.0001}$ & $0.9899^{+0.0002}_{0.0002}$ \\
      $p$ [$M$] & $30$ & $(30)$ & $30.00^{+0.05}_{-0.05}$ & $30.2^{+0.2}_{-0.1}$ & $30^{+3}_{-3}$ \\
      $\eta$  & $0.20$ & $(0.20)$ & $0.2000^{+0.0003}_{-0.0003}$ & $0.199^{+0.003}_{-0.002}$ & $0.20^{+0.05}_{-0.04}$ \\
      $\omega_0$ & $0$ & $3.3^{+2.7}_{-2.9}$ & $3.2^{+2.8}_{-2.8}$ & $3.1^{+2.8}_{-2.7}$ & $3.2^{+2.7}_{-2.8}$ \\
      $V_{3,0}$ & $\pi/3$ ($1.047$) & $3.1^{+2.9}_{-2.7}$ & $3.4^{+2.6}_{-3.1}$  & $3.2^{+2.7}_{-2.8}$ & $3.2^{+2.8}_{-2.8}$ \\
      $m_3$ [$M \times 10^7$]  & $1$ & $0.9^{+0.5}_{-0.7}$ & $0.9^{+0.5}_{-0.7}$ & $0.7^{+0.7}_{-0.6}$ & $0.7^{+0.7}_{-0.7}$\\
      $R$ [$M \times 10^7$]  & $1.1$ & $1.1^{+0.2}_{-0.4}$ & $1.0^{+0.3}_{-0.4}$ & $2.9^{+1.8}_{-1.7}$ & $2.8^{+1.9}_{-1.6}$ \\
      $\mathrm{cos}\left(2(V_3-\omega_0)\right)$ & $-0.5$ & $-0.3^{+0.8}_{-0.5}$ & $0.0^{+0.9}_{-0.9}$ & $0^{+1}_{-1}$ & $0^{+1}_{-1}$ \\
      $\frac{m_3}{R^3}$ [$M^{-2} \times 10^{-15}$] & $7.5$ & $6.9^{+3.8}_{-0.7}$ & $7.6^{+8.3}_{-2.2}$ & $0.3^{+2.9}_{-0.2}$ & $0.3^{+3.0}_{-0.3}$ \\
    \end{tabular}
    \caption{Injected values compared to the median and $90\%$ credible ranges of the posterior probability distribution recovered for the perturbed system when analysed using different priors on the inner binary parameters. \new{For these analyses, we keep $\iota_3=\Lambda_3=0$ fixed.}}
    \label{tab:perturbed_signal_recovery}
\end{table*}

\begin{table}[]
    \centering
    \begin{tabular}{c|c||c|c}
     Parameter & Injected & $\delta+\mathcal{U}$ & $\mathcal{U}$ \\
     \hline
      $M$ [M$_\odot$]  & $60$ & $(60)$ & $60^{+9}_{-7}$ \\
      $e$  & $0.99$ & $(0.99)$ & $0.9900^{+0.0002}_{-0.0002}$ \\
      $p$ [$M$] & $30$ & $(30)$ & $30^{+3}_{-3}$ \\
      $\eta$  & $0.20$ & $(0.20)$ & $0.20^{+0.04}_{-0.04}$ \\
      $\omega_0$ & $0$ & $3.2^{+2.8}_{-2.9}$  & $3.3^{+2.6}_{-2.9}$ \\
      $V_{3,0}$ & N/A & $3.1^{+2.9}_{-3.0}$ & $3.1^{+2.8}_{-2.8}$ \\
      $m_3$ [$M \times 10^7$]  & $0$ & $< 1.2$ & $0.7^{+0.7}_{-0.7}$ \\
      $R$ [$M \times 10^7$]  & N/A & $> 2.4$ & $2.8^{+1.9}_{-1.6}$ \\
      $\mathrm{cos}\left(2(V_3-\omega_0)\right)$ & N/A & $0^{+1}_{-1}$ & $0^{+1}_{-1}$ \\
      $\frac{m_3}{R^3}$ [$M^{-2} \times 10^{-15}$] & $0$ & $< 0.3$ & $0.3^{+3.0}_{-0.3}$ \\
    \end{tabular}
    \caption{Injected and recovered parameters for the isolated binary. Where limits are given, they are quoted at the $90\%$ credible level. Upper and lower error bounds correspond to the limits of the $90\%$ credible interval around the median.}
    \label{tab:unperturbed_signal_recovery}
\end{table}

We perform analyses using four different priors, detailed in Table \ref{tab:priors}.
In the maximally optimistic case, we assume we have perfect knowledge of all parameters of the inner binary, and analyse the data to recover the tertiary parameters only.
This is represented by the ``Delta + Uniform'' ($\delta + \mathcal{U}$) prior, in which we fix the values of the inner binary as delta functions and set uniform priors on the tertiary parameters.
We present results under this maximally optimistic assumption in Sec.~\ref{sec:results-subset} \new{for the face-on case and Sec.~\ref{sec:results-subset-tilted} for the tilted case}.
In more realistic cases, we relax the assumption of perfect knowledge, and aim to infer the intrinsic properties of both the inner binary and the tertiary simultaneously.
We use a range of priors for these analyses.
Two have informative Gaussian priors on the inner binary parameters ($\mathcal{G}1+\mathcal{U}$ and $\mathcal{G}2+\mathcal{U}$), representative of the case where information about the binary had been inferred from its later inspiral and merger.
In the maximally pessimistic case, we set uniform priors on all parameters ($\mathcal{U}$).
This represents an unfortunate situation in which, for example, the later inspiral and merger are not detected.
We present results under these less optimistic assumptions in Sec.~\ref{sec:results-all} \new{for the face-on case and Sec.~\ref{sec:results-all-tilted} for the tilted case; we consider only an extension of the informative $\mathcal{G}1+\mathcal{U}$ in the latter scenario}.
The injected and recovered parameters are displayed in \new{Tables \ref{tab:perturbed_signal_recovery} \& \ref{tab:tilted_signal_recovery}}.

We also check that we recover sensible limits on tertiary parameters in the case that the signal comes from an unperturbed BBH.
To do this, we inject a signal from the same eccentric binary as before, but remove the influence of the tertiary. 
We analyse this signal under both the maximally optimistic ($\delta + \mathcal{U}$) and maximally pessimistic ($\mathcal{U}$) priors, and report injected and recovered parameters  in Table \ref{tab:unperturbed_signal_recovery}.

\subsubsection{Measuring tertiary parameters when inner binary parameters are known \new{(face-on case)}}
\label{sec:results-subset}

\begin{figure*}[!htbp]
    \includegraphics[width=1.2\columnwidth]{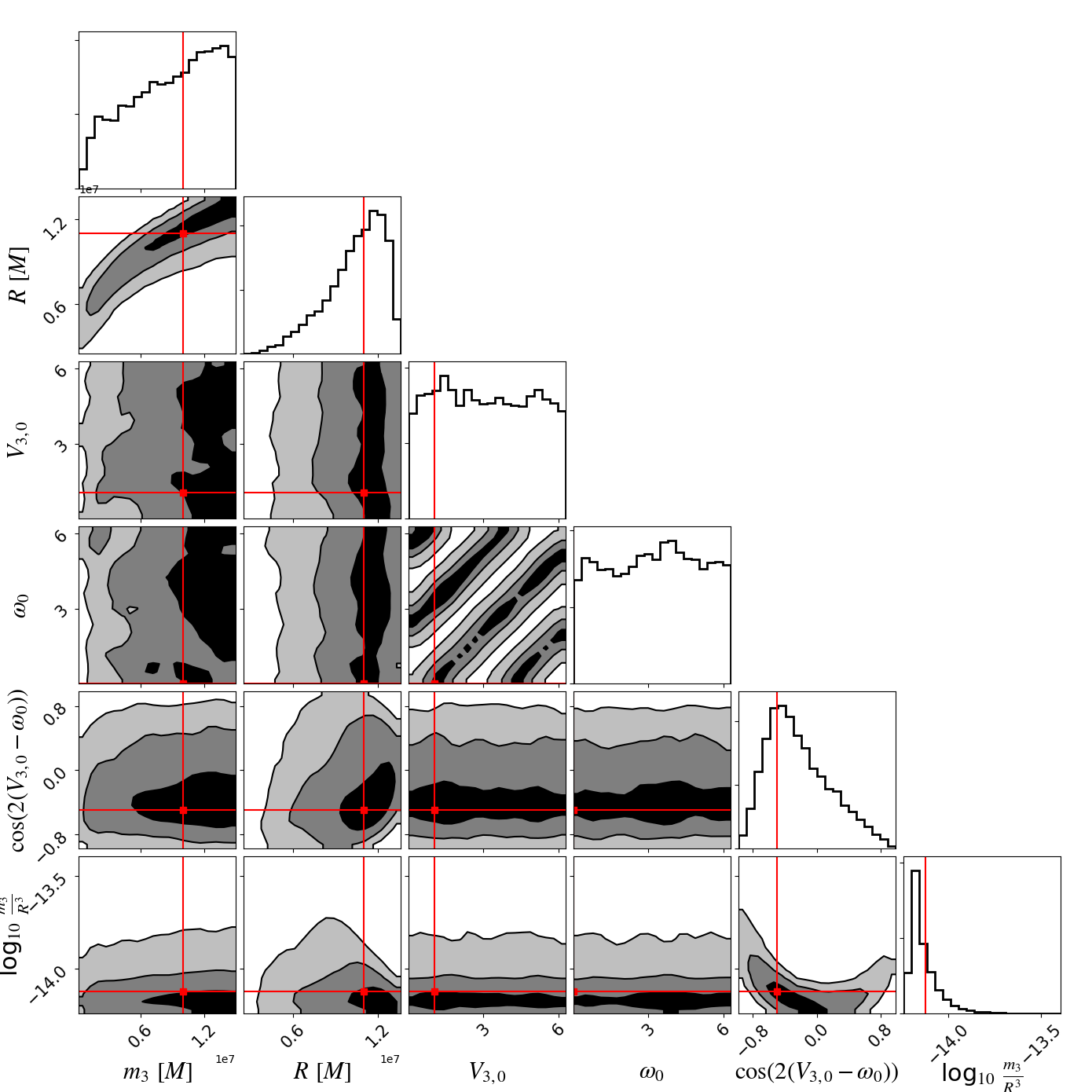}
    \includegraphics[width=1.2\columnwidth]{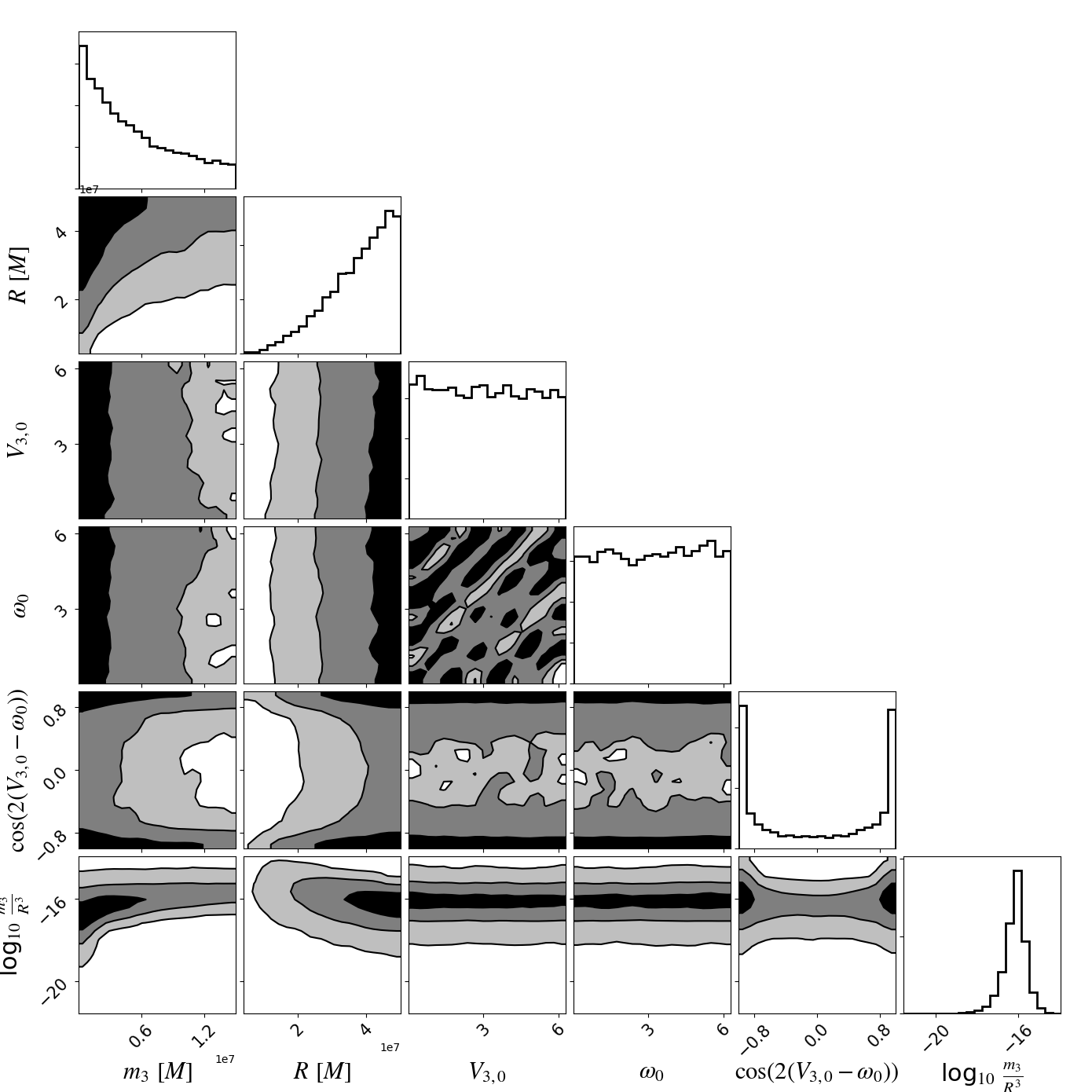}
    \caption{Posterior probability density distributions on physical three-body parameters $m_3$, $R$, $V_{3,0}$ and $\omega_0$, as well as the parameter combinations that directly influence the signal, $\mathrm{cos}\left(2(V_{3,0}-\omega_0)\right)$ and $\frac{m_3}{R^3}$, for the perturbed system (top) and the unperturbed system (bottom). 
    The injected values of the sampled quantities are indicated on the plot with red lines, and the darkest-shaded, medium-shaded and lightest-shaded regions correspond to the $1\sigma$, $2\sigma$ and $3\sigma$ credible intervals. 
    Numerical values of the median and $90\%$ credible intervals of the distributions are provided in Tables \ref{tab:perturbed_signal_recovery} \& \ref{tab:unperturbed_signal_recovery}.}
    \label{fig:subsets}
\end{figure*}

Firstly, we recover only the mass and location of the tertiary, assuming that we have perfect measurements of the parameters of the inner binary ($M$, $e_0$, $p_0$, $\eta$) through analysis of the late inspiral and merger\new{, and that the orbit of the co-planar triple is face-on ($\iota_3=\Lambda_3=0$), so there is no Doppler shift}.
Employing uniform priors on all parameters, we sample over $m_3$, $R$, $V_3$ and $\omega_0$. 
The results of this analysis are contained in Table \ref{tab:perturbed_signal_recovery}, and the posterior probability distributions obtained are shown in the top panel of Fig.~\ref{fig:subsets}. 
We obtain informative posteriors on $m_3$ and $R$ that peak close to the injected values, recovering $m_3=0.9^{+0.5}_{-0.7} \times 10^{7} M$ and $R=1.1^{+0.2}_{-0.4} \times 10^{7} M$, where the errors are the limits of the $90\%$ credible interval around the median.
In this optimistic case, we are able to exclude the possibility that the binary is unperturbed at greater than $90\%$ confidence.
While the 1D posteriors we obtain for $V_{3,0}$ and $\omega_0$ are relatively uninformative, the 2D posterior on these parameters shows clearly the expected correlations and periodicity, and reveals a peak at the injected values.

The 2D posterior on $m_3$ and $R$ shows a strong correlation between these two parameters that follows a line of constant $\frac{m_3}{R^3}$, as anticipated due to the dependence of the burst timing model on this quantity \new{when the orbit of the co-planar triple is face-on}.
Similarly, the correlations between $\omega_0$ and $V_{3,0}$ demonstrate the dependence of the burst timing model on a combination of these two parameters: the angle $2(V_{3,0} - \omega_0)$.
To demonstrate that these quantities are well-measured, we also plot the posterior distributions on $\frac{m_3}{R^3}$ and cos$\left(2(V_{3,0} - \omega_0)\right)$ in Fig.~\ref{fig:subsets} and report the recovered median and $90\%$ credible intervals in Table \ref{tab:perturbed_signal_recovery}.

Secondly, we repeat the analysis with a signal from an identical eccentric binary that is unperturbed.
We obtain posterior probability distributions on $m_3$ and $R$ that indicate either a nonexistent, negligibly small or infinitely distant tertiary. We show the posterior probability distributions obtained for the unperturbed binary in the bottom panel of Fig.~\ref{fig:subsets}.
The 1D posterior probability distributions for $\omega_0$ and $V_{3,0}$ are uninformative, although there is some structure in the 2D posterior, with preferred values found where $V_{3,0}$ and $\omega_0$ are an integer value of $\frac{\pi}{2}$ offset.
The constraints obtained on the system parameters are provided in Table \ref{tab:unperturbed_signal_recovery}.
We note in particular that the maximum $90\%$ credible value of $\frac{m_3}{R^3}$ is below the lower limit of the $90\%$ credible region recovered in the perturbed case, which would indicate strongly that this system is effectively isolated.

\subsubsection{\new{Measuring tertiary and inner binary parameters simultaneously \new{(known face-on case)}}}
\label{sec:results-all}

\begin{figure*}[!htbp]
    \centering
    \includegraphics[width=\textwidth]{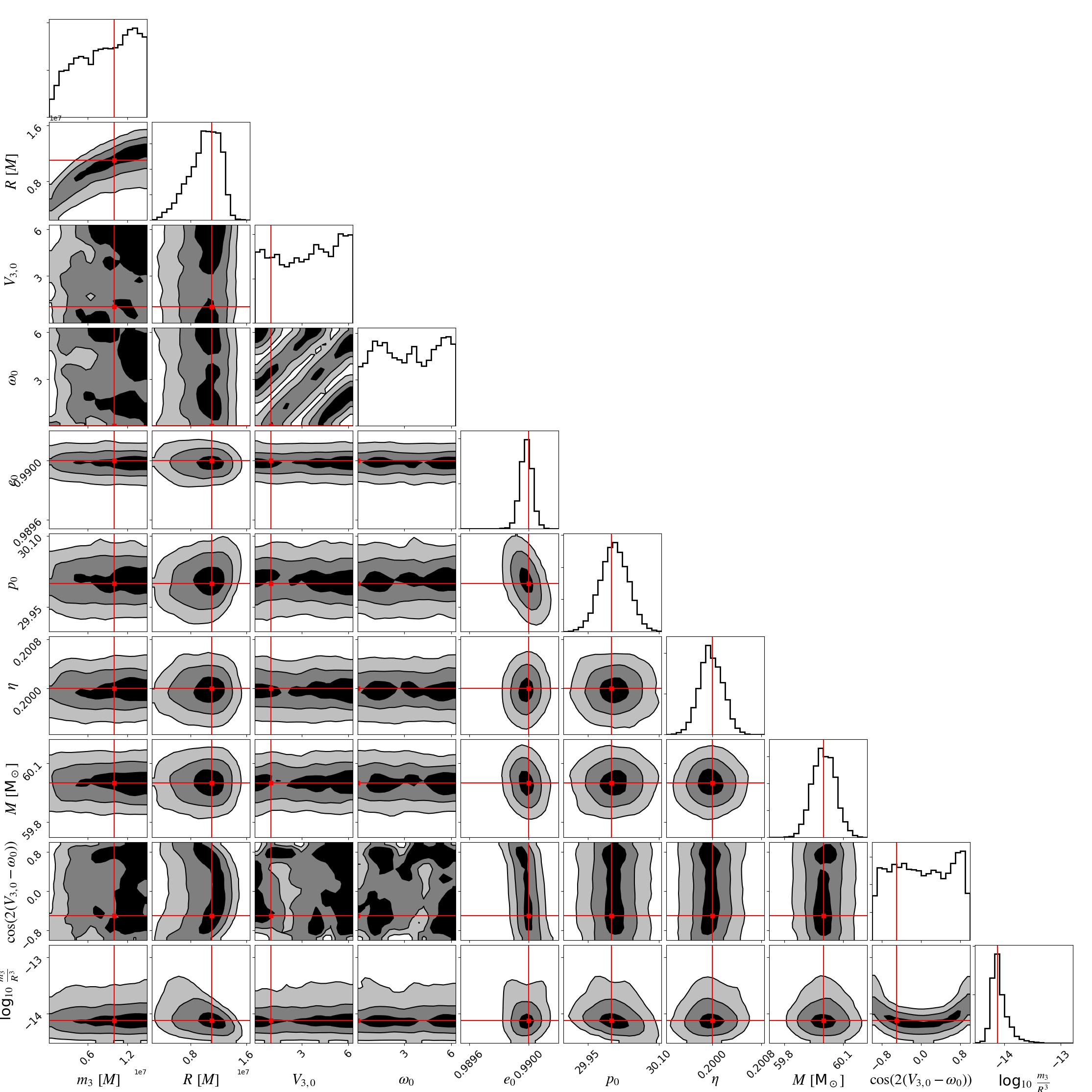}
    \caption{Posterior probability distributions on the extrinsic parameters of a three-body system, obtained through analysis of the burst signal from a perturbed binary using the $\mathcal{G}1+\mathcal{U}$ prior. The injected values of the sampled quantities are indicated on the plot with red lines, and the darkest-shaded, medium-shaded and lightest-shaded regions correspond to the $1\sigma$, $2\sigma$ and $3\sigma$ credible intervals.}
    \label{fig:all_parameters_triple_G1}
\end{figure*}

\begin{figure*}[!htbp]
    \centering
    \includegraphics[width=\textwidth]{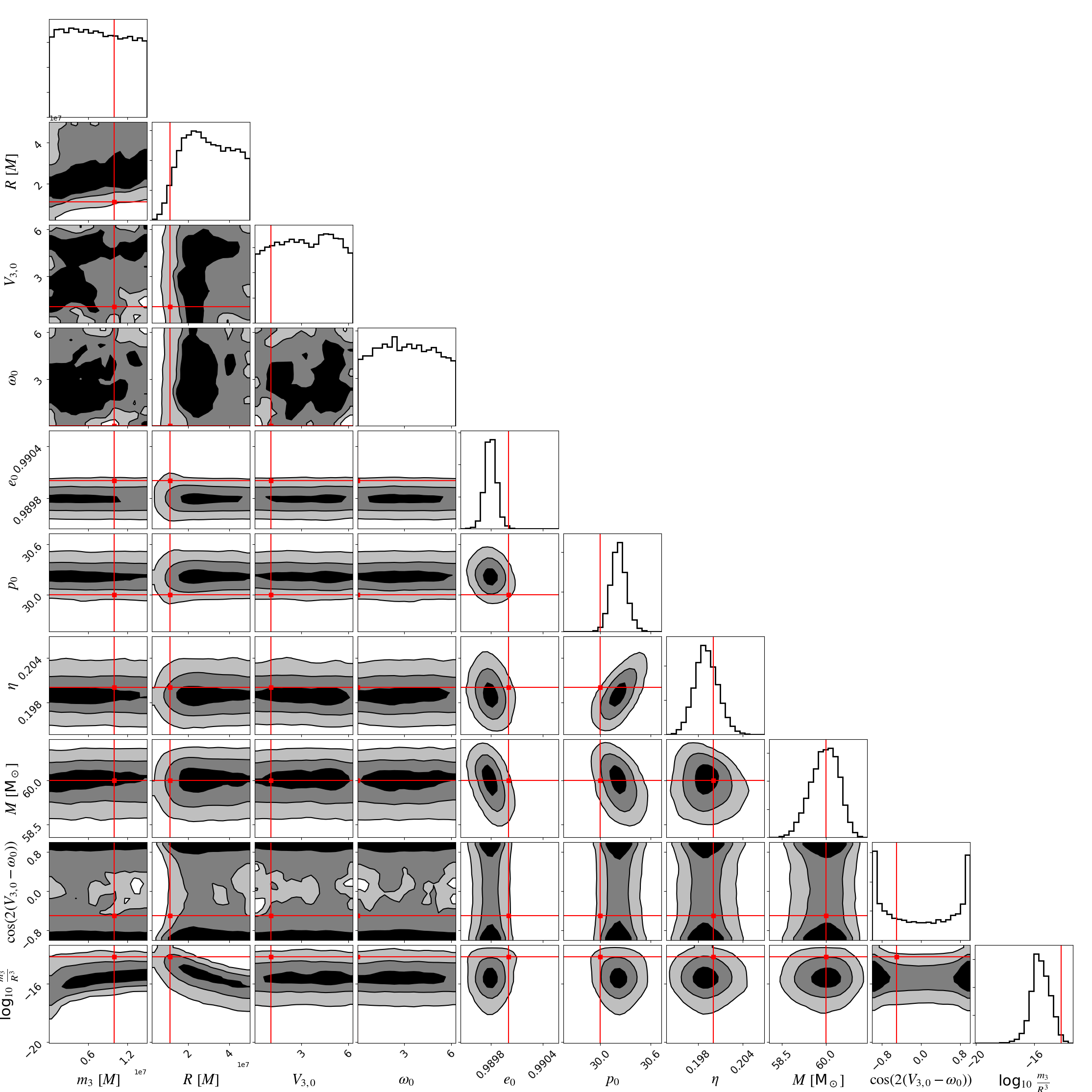}
    \caption{Posterior probability distributions on the extrinsic parameters of a three-body system, obtained through analysis of the burst signal from a perturbed binary  using the $\mathcal{G}2+\mathcal{U}$ prior. As in previous corner plots, the injected values of the sampled quantities are indicated on the plot with red lines, and the darkest-shaded, medium-shaded and lightest-shaded regions correspond to the $1\sigma$, $2\sigma$ and $3\sigma$ credible intervals.}
    \label{fig:all_parameters_triple_G2}
\end{figure*}

\begin{figure*}[!htbp]
    \centering
    \includegraphics[width=\textwidth]{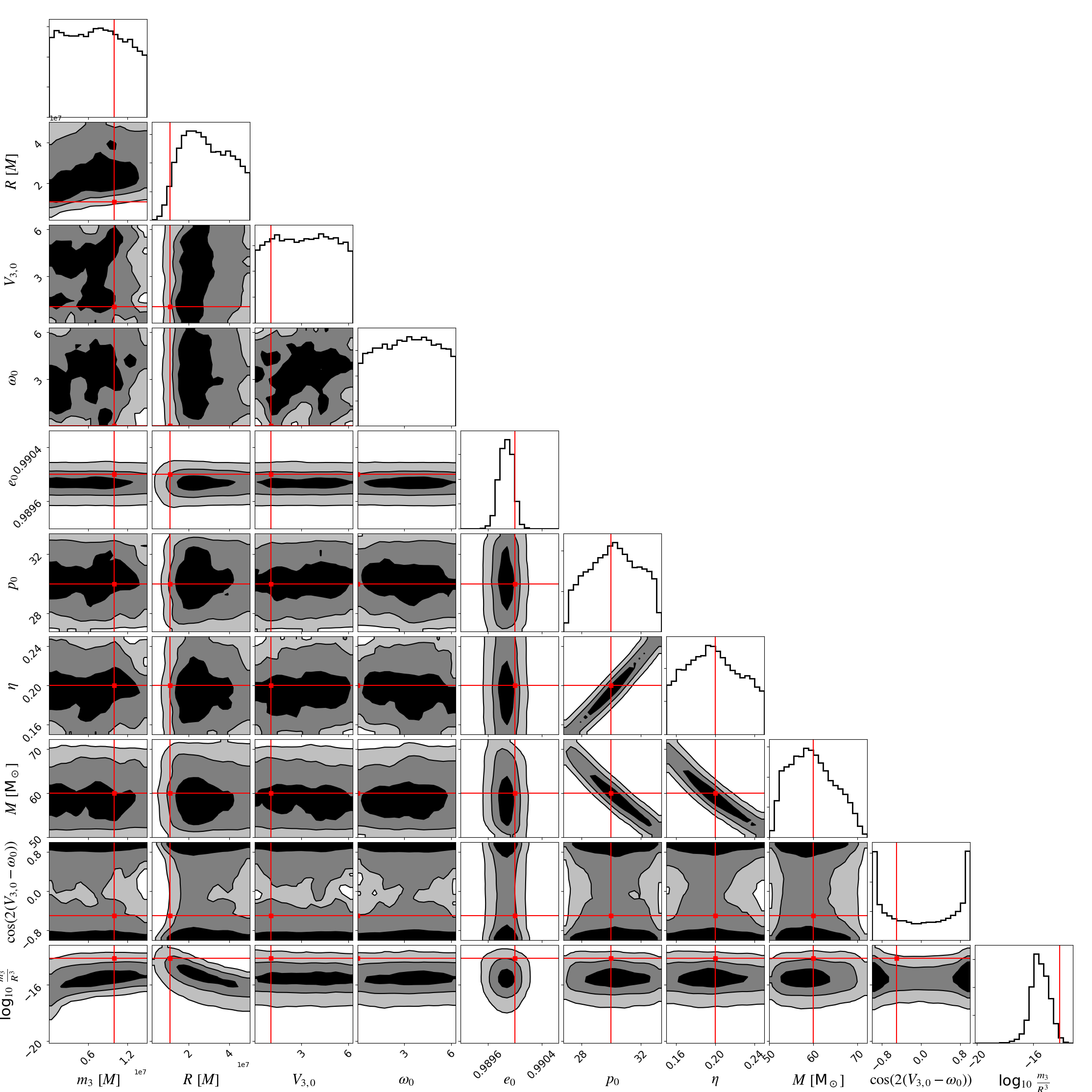}
    \caption{Posterior probability distributions on the extrinsic parameters of a three-body system, obtained through analysis of the burst signal from a perturbed binary using the $\mathcal{U}$ prior. As in previous corner plots, the injected values of the sampled quantities are indicated on the plot with red lines, and the darkest-shaded, medium-shaded and lightest-shaded regions correspond to the $1\sigma$, $2\sigma$ and $3\sigma$ credible intervals.}
    \label{fig:all_parameters_triple}
\end{figure*}

\begin{figure*}[!htbp]
    \centering
    \includegraphics[width=\textwidth]{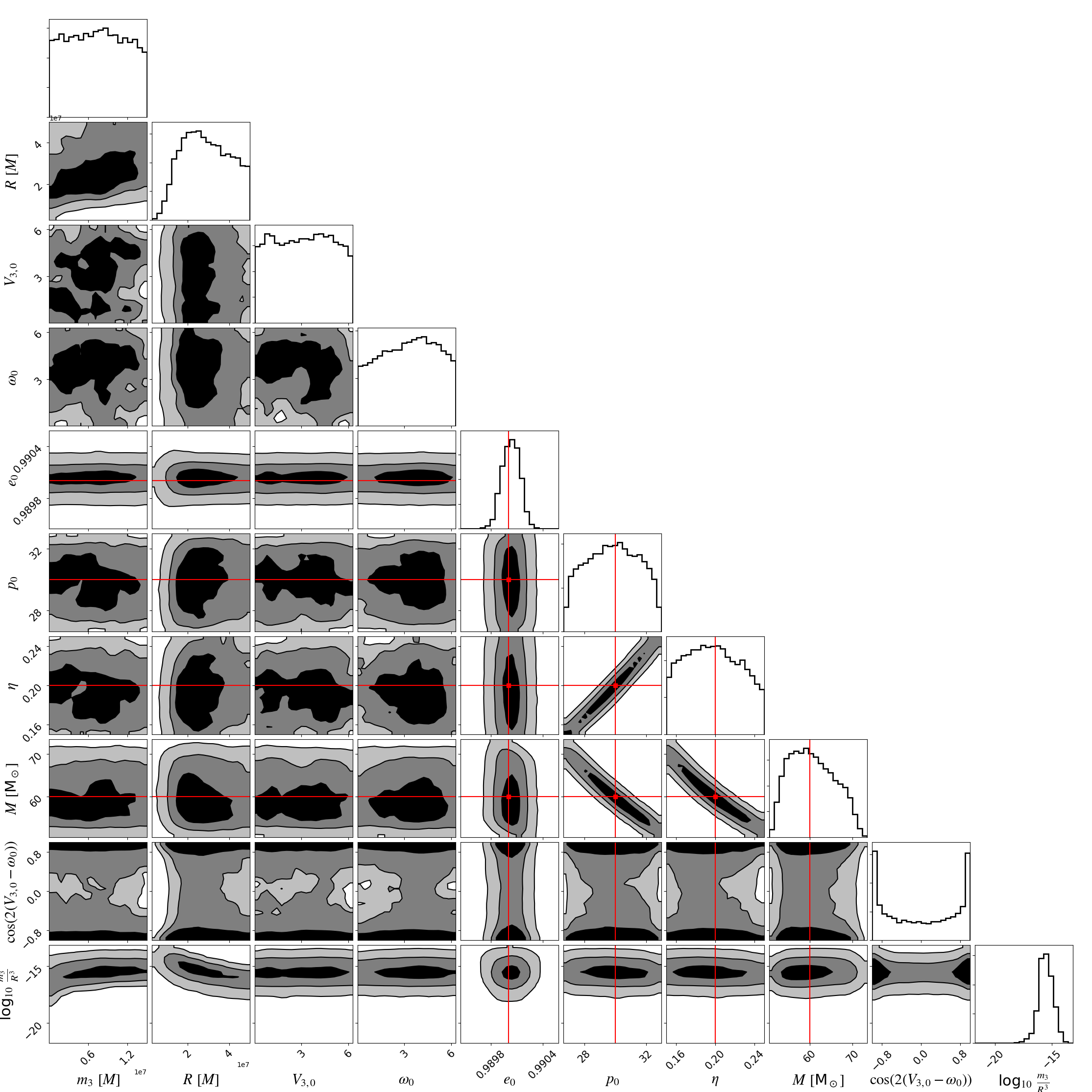}
    \caption{Posterior probability distributions on the extrinsic parameters of a three-body system, obtained through analysis of the burst signal from an unperturbed binary using the $\mathcal{U}$ prior. As in previous corner plots, the injected values of the sampled quantities are indicated on the plot with red lines, and the darkest-shaded, medium-shaded and lightest-shaded regions correspond to the $1\sigma$, $2\sigma$ and $3\sigma$ credible intervals.}
    \label{fig:all_parameters_binary}
\end{figure*}

\new{We now loosen our priors on the inner binary parameters, allowing greater uncertainty to enter our inference of the tertiary parameters.}
The priors we use are detailed in Table \ref{tab:priors} \new{for the case of the face-on ($\iota_3=\Lambda_3=0$) triple, which we study first}.
To represent scenarios in which we have some uncertain measurements on the intrinsic inner binary parameters, we use Gaussian priors centered at the injected value $\mu$ with some uncertainty represented by $\sigma$.
We retain wide uniform priors on tertiary parameters.
We use two such priors---one tighter ($\mathcal{G}1+\mathcal{U}$), and one more relaxed ($\mathcal{G}2+\mathcal{U}$).
Both sets of priors have widths within the range of measurement precision expected for BBH detected with ET \citep{Cho:2022:AccuracyET, Pieroni:2022:AccuracyETCE}.
The posterior probability distributions recovered under each of these priors are shown in Figs.~\ref{fig:all_parameters_triple_G1} \& \ref{fig:all_parameters_triple_G2}, respectively.
The median values and $90\%$ credible interval recovered under each of these priors are reported in Table \ref{tab:perturbed_signal_recovery}.
With the more restricted prior, $\mathcal{G}1+\mathcal{U}$, we achieve comparable measurement precision and accuracy to the $\delta + \mathcal{U}$ prior, recovering  $m_3=0.9^{+0.5}_{-0.7} \times 10^{7} M$ and $R=1.0^{+0.3}_{-0.4} \times 10^{7} M$.
In the case of the less restricted prior, the posterior probability distributions for the inner binary properties are noticeably biased, preferring a lower eccentricity, lower total mass, wider separation, and more extreme mass ratio than injected.

In the final case, we relax our assumption that we have any prior knowledge about the eccentric inner binary by using the $\mathcal{U}$ prior, in which uniform priors are set on all parameters.
In this case, the parameters recovered for the perturbed binary and unperturbed binary are near-identical, implying that GW burst timing offsets---at least in the SMBH scenario that we consider---can be well-explained by many binary and triple configurations.
The posterior probability distributions recovered for both the perturbed and unperturbed binary are shown in Figs.~\ref{fig:all_parameters_triple} \& \ref{fig:all_parameters_binary}, respectively.
The median and $90\%$ credible interval values recovered for all parameters, reported in Tables \ref{tab:perturbed_signal_recovery} \& \ref{tab:unperturbed_signal_recovery}, are almost the same between the perturbed and unperturbed, with the median of the eccentricity posterior slightly lower than the injected value for the perturbed binary.
These results highlight that, within the limits of our toy model, accurately constraining the inner binary properties is crucial if the tertiary properties are to be studied using the burst timing method. 
The correlations in the tertiary parameters are less well-established when all parameters are sampled over, as can be expected due to the increased uncertainty.
However, strong correlations between the well-measured inner binary parameters can be seen in Figs.~\ref{fig:all_parameters_triple} \& \ref{fig:all_parameters_binary}.

\subsubsection{Measuring tertiary parameters when inner binary parameters are known \new{(tilted case)}}
\label{sec:results-subset-tilted}

\begin{table*}[]
    \centering
    \begin{tabular}{|c|c||c|c|||c||c|}
     Parameter & Injected & \multicolumn{2}{c|||}{Recovered} & Injected & Recovered \\
       &  & $\delta^\tau+\mathcal{U}$ & $\mathcal{G}1^\tau+\mathcal{U}$ & & $\mathcal{G}1^\tau+\mathcal{U}$ \\
     \hline
      $M$ [M$_\odot$] & $60$ & $(60)$ & $60.0^{+0.1}_{-0.1}$ & $60$ & $60.0^{+0.1}_{-0.1}$ \\
      $e$ & $0.99$ & $(0.99)$ & $0.9900^{+0.0005}_{-0.0004}$ & $0.99$ & $0.9900^{+0.0004}_{-0.0003}$ \\
      $p$ [$M$] & $30$ & $(30)$ & $30.01^{+0.05}_{-0.05}$ &  $30$ & $30.00^{+0.05}_{-0.05}$ \\
      $\eta$ & $0.20$ & $(0.20)$ & $0.2000^{+0.0004}_{-0.0003}$ & $0.20$ & $0.2000^{+0.0003}_{-0.0003}$ \\
      $\iota_3$ & $\frac{\pi}{4}$ ($0.79$) & $\left(\frac{\pi}{4}\right)$ & $0.79^{+0.02}_{-0.02}$ & $0$ & $0.000^{+0.008}_{-0.007}$ \\
      $\Lambda_3$ & $\frac{\pi}{6}$ ($0.52$) & $\left(\frac{\pi}{6}\right)$ & $0.52^{+0.02}_{-0.02}$ & $0$ & $0.00^{+0.02}_{0.02}$ \\
      $\omega_0$ & $0$ & $3.4^{+2.8}_{-3.2}$ &$3.1^{+3.0}_{-3.0}$ & $0$ & $2.7^{+3.0}_{-2.5}$ \\
      $V_{3,0}$ & $\pi/3$ ($1.047$) & $1.047^{+0.002}_{-0.001}$ & $1.047^{+0.02}_{-0.02}$ & $\pi/3$ ($1.047$) & $2.5^{+3.1}_{-1.8}$ \\
      $m_3$ [$M \times 10^7$] & $1$ & $0.9^{+0.3}_{-0.4}$ & $0.9^{+0.4}_{-0.4}$ & $1$ & $0.8^{+0.7}_{-0.7}$ \\
      $R$ [$M \times 10^7$] & $1.1$ & $1.0^{+0.3}_{-0.5}$ & $1.0^{+0.4}_{-0.5}$ & $1.1$ & $0.9^{+0.4}_{-0.5}$ \\
      $\mathrm{cos}\left(2(V_3-\omega_0)\right)$ & $-0.5$ & $-0.1^{+1.0}_{-0.8}$ & $0.00^{+0.9}_{-0.9}$ & $-0.5$ & $0.1^{+1.0}_{-0.9}$ \\
      $\frac{m_3}{R^3}$ [$M^{-2} \times 10^{-15}$] & $7.5$ & $8.6^{+22}_{-3.2}$ & $9.0^{+23}_{-3.9}$ & $7.5$ & $8.7^{+27}_{-3.6}$ \\
    \end{tabular}
    \caption{\new{Injected and recovered parameters for a perturbed binary orbiting a tertiary on an orbit that is tilted away from face-on (second to fourth columns) and face-on (fifth to seventh columns), where the tilt angles are now sampled over. Where limits are given, they are quoted at the $90\%$ credible level. Upper and lower error bounds correspond to the limits of the $90\%$ credible interval around the median. Details of the extended priors with $\tau$ superscripts are given in the text.}}
    \label{tab:tilted_signal_recovery}
\end{table*}

\new{We now relax the requirement that $\iota_3=\Lambda_3=0$, incorporating the Doppler effect due to the COM motion of the binary as formulated in Sec.~\ref{sec:com}. As in Sec.~\ref{sec:results-subset}, we fix the inner binary parameters, now including the orbital tilt angles. We infer only the tertiary parameters. While it is of course unrealistic to assume that we would know whether a detected binary was co-planar with any outer orbit, this analysis demonstrates clearly the benefit of including well-constrained orbital tilt parameters. Recovered properties are shown in the third column of Table \ref{tab:tilted_signal_recovery}.}

\begin{figure*}
    \centering
    \includegraphics[width=1.2\columnwidth]{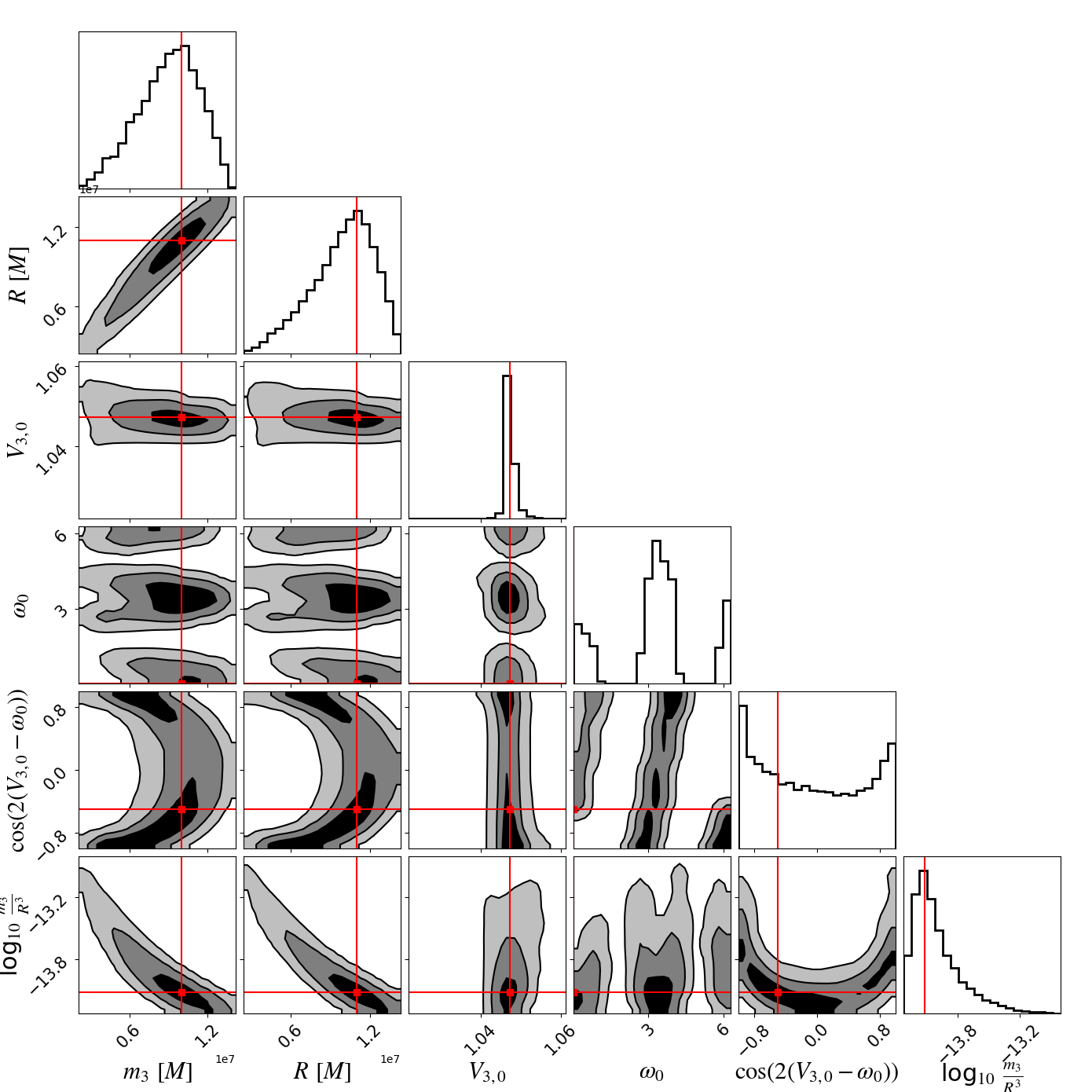}
    \caption{\new{Posterior probability distributions on three-body parameters for a perturbed signal identical to that studied in Sec.~\ref{sec:pe}, but tilted at an angle $\iota_3=\frac{\pi}{4}$, $\Lambda_3=\frac{\pi}{6}$. The injected tertiary mass is much more confidently recovered when the orbital plane is tilted at an angle to the observer because, while the burst times themselves are a function of $\frac{m_3}{R^3}$, the \textit{detected} time offset due to the outer orbital tilt depends only on $R$. The parameters $\omega_3$ and $V_{3,0}$ also become well-measured because the same time offset depends on $V_3$ but not $\omega_3$. }}
    \label{fig:tilted_tertiary_only}
\end{figure*}

\new{Comparing the results plotted in the top panel of Fig.~\ref{fig:perturbed-vs-unperturbed} to those plotted in Fig.~\ref{fig:tilted_tertiary_only}, it is clear that when the outer orbit is tilted with respect to the observer, $m_3$ can be more precisely constrained. This is expected, since the difference between the emitted and detected burst times (calculated in Eq.~\ref{eq:time-doppler}) does not depend on $m_3$, while the burst time itself (calculated in Eq.~\ref{eq:time-offset}) is a function of $\frac{m_3}{R^3}$. Therefore, the time lag induced by the binary's COM motion breaks the degeneracy between $m_3$ and $R^3$, allowing $m_3$ to be more well-constrained. Similarly, Eq.~\ref{eq:time-doppler} is a function of $V_3$, but does not include $\omega_0$, thereby introducing another way to distinguish two correlated parameters.}

\new{Comparing the median and $90\%$ credible intervals recovered for the tilted system vs the face-on system (third columns of Tables \ref{tab:tilted_signal_recovery} \& \ref{tab:perturbed_signal_recovery} respectively), it can be seen that $V_{3, 0}$ is recovered much more precisely when the system is tilted away from face-on. This parameter is measured accurately with $90\%$ error bars of order $10^{-3}$ when the system is tilted, while the same quantity is poorly-measured when the system is face on, with $90\%$ error bars spanning the majority of the prior.}

\subsubsection{\new{Measuring tertiary and binary parameters simultaneously (uncertain tilt case)}}
\label{sec:results-all-tilted}

\begin{figure*}[!htbp]
    \centering
    \includegraphics[width=\textwidth]{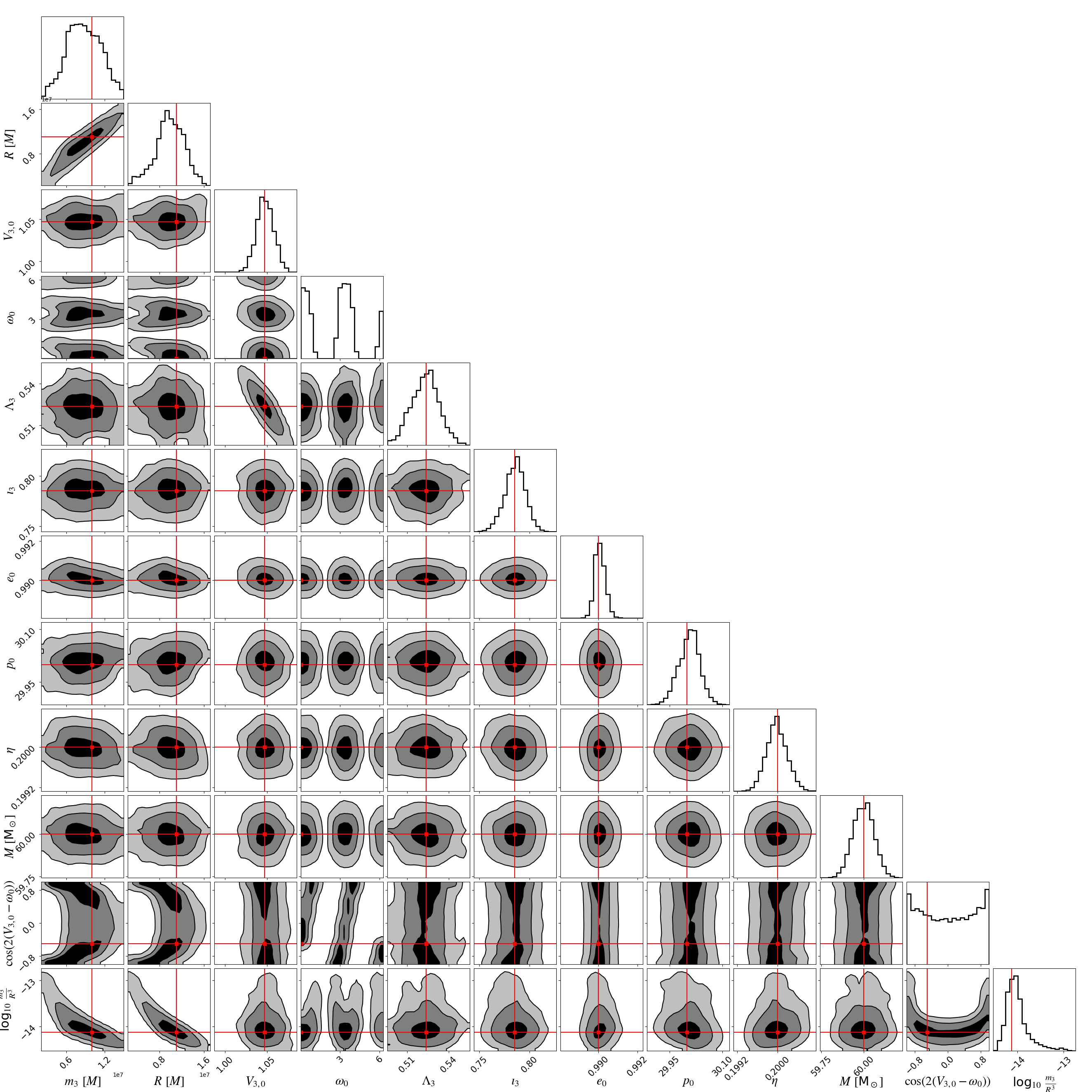}
    \caption{\new{Posterior probability distributions on the extrinsic parameters of a tilted triple system, obtained through analysis of the burst signal from a perturbed binary using the narrow $\mathcal{G}1^\tau+\mathcal{U}$ prior. The additional uncertainty from the extra two tilt parameters is more than counteracted by the degeneracy-breaking effect of the tilted system, which enables tighter constraints on $m_3$ and $R$. The parameters $V_3$ and $\omega_3$ are again easier to measure when the tilt angles $\iota_3$ and $\Lambda_3$ are constrained.}}
    \label{fig:all_parameters_triple_G1_extended_tilted}
\end{figure*}

\begin{figure*}[!htbp]
    \centering
    \includegraphics[width=\textwidth]{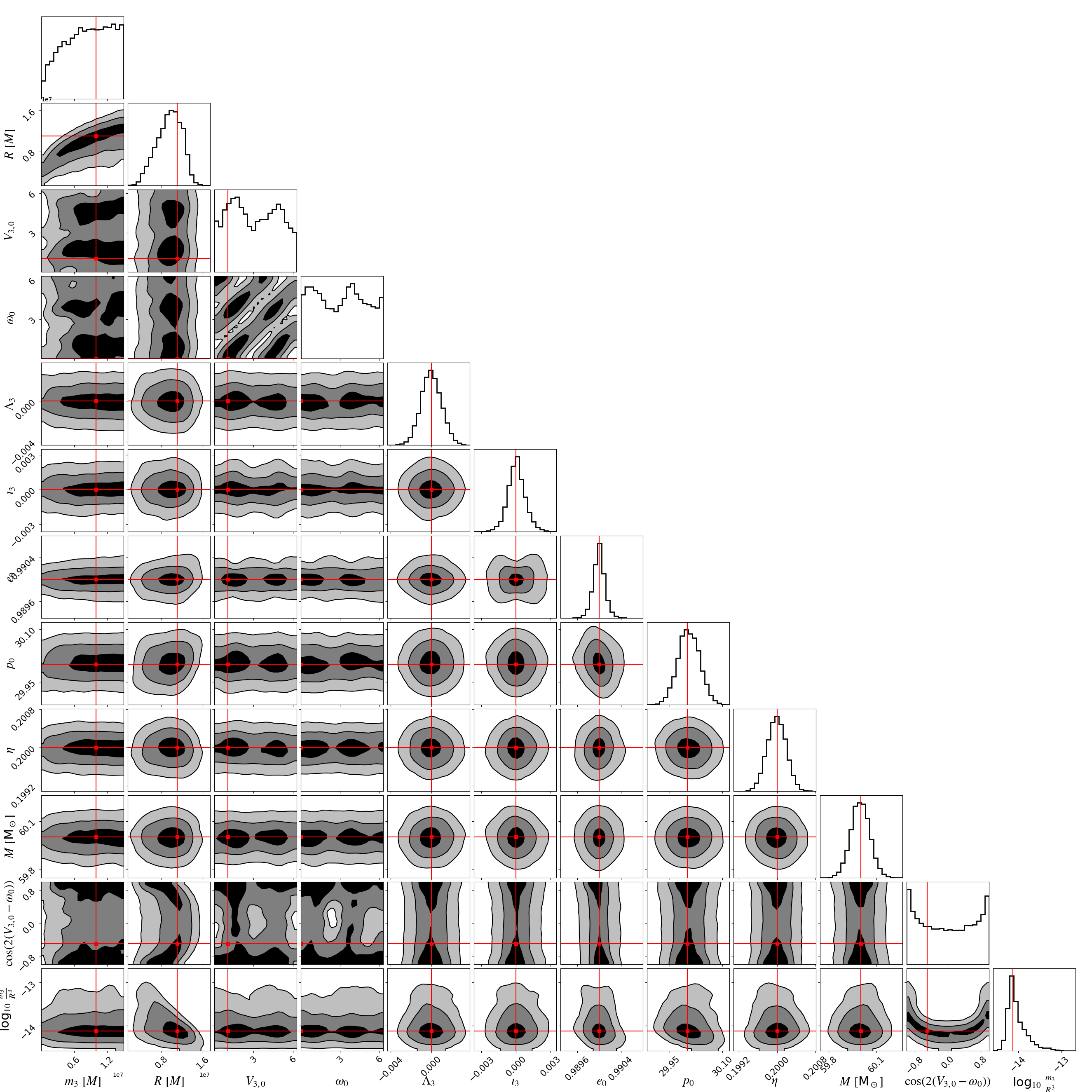}
    \caption{\new{Posterior probability distributions on the extrinsic parameters of a three-body system, obtained through analysis of the burst signal from a face-on triple system, using the narrow $\mathcal{G}1^\tau+\mathcal{U}$ prior, showing the additional uncertainty coming from the extra two tilt parameters $\iota_3$ and $\Lambda$. Again, it can be seen that the parameters $V_3$ and $\omega_3$ become much easier to measure when the tilt angles $\iota_3$ and $\Lambda_3$ are well-measured. However, when the system is face-on rather than tilted, the benefit is reduced.}}
    \label{fig:all_parameters_triple_G1_extended}
\end{figure*}

\new{We again allow tilt angles $\iota_3$ and $\Lambda_3$ to be non-zero, and now explore the influence of allowing some uncertainty on their values. We use an extension of our Gaussian 1 + Uniform ($\mathcal{G}1^\tau + \mathcal{U}$) prior in which the prior on both $\iota_3$ and $\Lambda_3$ is a Gaussian at the injected value with $\sigma=1\times10^{-2}$. We study two systems: one with $\iota_3=\frac{\pi}{4}$, $\Lambda_3=\frac{\pi}{6}$, and one with $\iota_3=\Lambda_3=0$. The posteriors obtained via these analyses are shown in Figures \ref{fig:all_parameters_triple_G1_extended_tilted} \& \ref{fig:all_parameters_triple_G1_extended}. Median and $90\%$ credibility ranges recovered are reported in Table \ref{tab:tilted_signal_recovery} (fourth and sixth columns).}

\new{Comparing the constraints obtained on the tilted system vs the non-tilted system when $\iota_3$ and $\Lambda_3$ are sampled over demonstrates the advantages of non-zero orbital tilt. 
The tertiary mass $m_3$ is more precisely constrained when the co-planar orbit is tilted away from face-on, and posteriors on the parameters $w_0$ and $V_{3,0}$ are both more strongly peaked.
The marginal posterior on tilt angle parameter $\iota_3$, however, is less well-measured when $\iota_3 \neq 0$; this is likely because $\iota_3=0$ produces a more unique set of burst offset times for a given system than $\iota_3 \neq 0$.
Comparing both of these results to those of the analysis with the $\mathcal{G}1 + \mathcal{U}$ prior and $\iota_3=\Lambda_{3,0}=0$ fixed, it is clear that sampling over varying orbital tilt angles enables better measurement of orbital phase parameters $w_0$ and $V_{3,0}$, even when the orbit is face-on.}

\subsubsection{Correlations between $R$ and inner binary parameters}
\label{sec:R-var}

\new{$R$ is one of the best-measured parameters for both tilted and untilted systems.}
Our narrowest set of priors, $\mathcal{G}1+\mathcal{U}$, have standard deviations of $0.1\%$ of the injected values.
The wider set, $\mathcal{G}2+\mathcal{U}$, have standard deviations exactly an order of magnitude higher.
Both sets are well within the measurement precision expected for the Einstein Telescope \citep{Cho:2022:AccuracyET, Pieroni:2022:AccuracyETCE}.
The wider set of priors enables correlations between $R$ and the inner binary parameters to be seen: \new{in Fig.~\ref{fig:all_parameters_triple_G2}, the posteriors suggest that the} perturbed signal injected \new{may be} confused for a binary with a more distant, less influential tertiary and a more eccentric, wider, lower-mass and less equal-mass inner binary. 
The narrower set of priors enables the injected values of the parameters to be correctly recovered, giving an idea of the prior measurement accuracy required to reliably identify a perturbing SMBH from the eccentric GW burst times of a nearby stellar-mass BBH.

\section{Discussion}
\label{sec:disc}

For systems where the effects due to a tertiary are small, the results of our analysis give us confidence that when we have tight constraints on the parameters of the eccentric binary (represented by the $\delta + {\cal{U}}$ and ${\cal{G}}1 + {\cal{U}}$ priors), we can infer the existence of a perturbing tertiary using next-generation ground-based GW detectors and already-existing hierarchical inference techniques.
The statistical uncertainties on measurements of tertiary parameters are large, even in the most optimistic case, due to how these parameters enter the burst timing model: they \new{predominantly} appear in the combinations $\frac{m_3}{R^3}$ \new{and $\cos (2(V_{3,0}-\omega_0))$, so these are the quantities that strongly influence the signal.}
\new{When the co-planar system is tilted, the degeneracies between these parameters are broken, enabling tighter constraints on tertiary parameters.}

When the binary parameters are not tightly restricted (as emulated by the ${\cal{G}}2 + {\cal{U}}$ prior) or completely unknown (in the case of the ${\cal{U}}$ prior), the existence of a tertiary \new{is unlikely to confidently} be inferred \new{if the system exhibits no orbital tilt}, due to covariances among the tertiary's parameters and those of the inner binary \new{in the case that $\iota_3 = \Lambda_3 = 0$}. 
In order to measure the properties of a perturbing SMBH from its effects on the orbit of a nearby eccentric BBH, achieving the low measurement errors on the binary parameters expected from next-generation GW detectors \citep{Cho:2022:AccuracyET,Pieroni:2022:AccuracyETCE} is crucial.
Therefore, our ability to infer the existence and properties of a perturbing tertiary from its influence on the GW burst signal of an eccentric binary depends both on measurement constraints on binary parameters and the model itself. 
While we are limited here to the toy model in Eqs.~\eqref{eq:p-next}-\eqref{eq:time-doppler}, which requires the effects of the tertiary to be small compared to radiation reaction effects, future developments to this model to allow for non-secular evolution may enable tighter constraints on tertiary parameters.

Nonetheless, the timing model of eccentric GW bursts in Eqs.~\eqref{eq:p-next}-\eqref{eq:time-doppler} provides a first approximation of the effects due to a third body, especially in the case when the angular momentum of the outer orbit is aligned with that of the inner orbit. Such a scenario may be most relevant to binaries formed within AGN disks, and where the perturbing tertiary is the central supermassive BH. However, the tidal perturbations due to a tertiary on a binary formed during binary-single interactions will induce rich dynamics on the binary~\cite{Kozai,Lidov}. Both the orbital plane and orbital eccentricity can undergo significant oscillations, especially if the tidal perturbations become of the same order or stronger than PN corrections to the binary's dynamics~\cite{Naoz:2016,Naoz:2012bx}. 
We leave the extension of the model presented here to other scenarios for future work.

The parameter estimation method used here also stands as a simplified example of the more complicated and computationally-intensive requirements of a real analysis.
The crude GW burst signal model described in Sec.~\ref{sec:pe} can be replaced with something like the waveform described in \cite{Loutrel:2019kky}, and the initial posterior probability distribution on each burst time obtained through a parameterised Bayesian inference approach using this model.
It may be possible to create a hybrid waveform for injection by combining the burst-like section of the signal with a numerical relativity simulation of a GW from an eccentric binary's late inspiral and merger (the chirp), when the effect of any tertiary will be minimal.
This would facilitate simulation of the full inference pipeline: from detection of both the chirp section and the burst section of the signal, to the preliminary inference of the binary parameters through analysis of the chirp, to using the resultant posteriors on binary parameters as priors during analysis of the burst section of the signal.
The extrinsic parameters of the eccentric binary can be implemented into this signal model and subsequently inferred.
We plan to expand the method developed here in these directions in future work.
\section{Acknowledgements}
\new{The authors thank their anonymous referees for their comments, which improved the manuscript.}
The authors would like to thank Johan Samsing for his involvement in this project and many fruitful discussions. They would also like to thank Rowena Nathan, Valentina Di Marco, Eric Thrane, and Paul Lasky for their tutorial on pulsar timing methods, and Daniel Reardon for his talk about pulsar timing at Monash University. I.M.R-S acknowledges support received from the Herchel Smith Postdoctoral Fellowship Fund and the Australian Research Council (ARC) Centre of Excellence CE170100004. N.L. acknowledges financial support provided under the European Union's H2020 ERC, Starting Grant agreement no.~DarkGRA--757480. N.L. also acknowledges support under the MIUR PRIN and FARE programmes (GW-NEXT, CUP: B84I20000100001), and from the Amaldi Research Center funded by the MIUR program ``Dipartimento di Eccellenza'' (CUP: B81I18001170001).
Support for M.Z. was provided by NASA through the NASA Hubble Fellowship grant HST-HF2-51474.001-A awarded by the Space Telescope Science Institute, which is operated by the Association of Universities for Research in Astronomy, Incorporated, under NASA contract NAS5-26555. The authors are grateful for computational resources provided by the LIGO Laboratory and supported by National Science Foundation Grants PHY-0757058 and PHY-0823459.
This work was performed in part at the Aspen Center for Physics, which is supported by NSF grant PHY-1607611. 

\appendix

\section{\label{app:harm} Harmonic Coefficients of the Osculating Equations}

The osculating equations associated with the tidal perturbing force due to a third body are given in Eqs.~\eqref{eq:dpdV}-\eqref{eq:dtdV}. The harmonic coefficients $[C_{a}^{(k)}, S_{a}^{(k)}]$ are
\allowdisplaybreaks[4]
\begin{align}
    C_{e}^{(0)} &= 3 e \sin[2(V_{3}-\omega)]
    \\
    C_{1}^{(1)} &= 9 \sin[2 (V_{3}-\omega)]
    \\
    C_{2}^{(1)} &= 9 e \sin[2 (V_{3}-\omega)]
    \\
    C_{3}^{(1)} &= 3 \sin[2(V_{3} - \omega)]
    \\
    S_{1}^{(1)} &= 2 - 9 \cos[2(V_{3}-\omega)]
    \\
    S_{2}^{(1)} &= e - 9 e \cos[2(V_{3}-\omega)]
    \\
    S_{3}^{(1)} &= -3 \cos[2(V_{3}-\omega)]
    \\
    C_{\omega}^{(0)} &= -e \left\{1 + e \cos[2(V_{3}-\omega)]\right\}
    \\
    C_{\omega}^{(1)} &= -2 - 9 \cos[2(V_{3}-\omega)]
    \\
    C_{\omega}^{(2)} &= -e \left\{1 + e \cos[2(V_{3}-\omega)]\right\}
    \\
    C_{\omega}^{(3)} &= 3 \cos[2(V_{3}-\omega)]
    \\
    S_{\omega}^{(1)} &= -9 \sin[2(V_{3}-\omega)]
    \\
    S_{\omega}^{(2)} &= - 3 e \sin[2(V_{3}-\omega)]
    \\
    S_{\omega}^{(3)} &= 3 \sin[2(V_{3}-\omega)]
    \\
    C_{t}^{(0)} &= -e \left\{1 + e \cos[2(V_{3}-\omega)]\right\}
    \\
    C_{t}^{(1)} &= -2 - 9 \cos[2(V_{3}-\omega)]
    \\
    C_{t}^{(2)} &= -e \left\{1 + e \cos[2(V_{3}-\omega)]\right\}
    \\
    C_{t}^{(3)} &= 3 \cos[2(V_{3}-\omega)]
    \\
    S_{t}^{(1)} &= - 9 \sin[2(V_{3}-\omega)]
    \\
    S_{t}^{(2)} &= -3 e \sin[2(V_{3}-\omega)]
    \\
    S_{t}^{(3)} &= 3 \sin[2(V_{3}-\omega)]
\end{align}

\section{\label{app:stab} Stability Considerations for BBHs Orbiting SMBHs}

The binary black hole system under consideration in Sec.~\ref{sec:pe} consists of an equal mass binary with $M=60M_{\odot}$, orbiting around a SMBH with mass $m_{3}=60\times10^{7}M_{\odot}$ and orbital radius $R_{3} = 1.1 m_{3}$. As noted previously, the value of the orbital radius is close to the ISCO associated with an extremal (or near-extremal) Kerr BH. There are two considerations that arise as a result of this:
\begin{enumerate}
    \item How long can a single BH exist in this environment before coalescing with the SMBH?
    \item Is a binary at this radius stable under the influence of the tertiary (the SMBH in this case)?
\end{enumerate}
The first question relates to whether a single BH can exist in such a close orbit long enough to form a binary with another BH. To provide an estimate of this timescale, we use the computation of the time to plunge for a near horizon inspiral into a near extremal Kerr BH, specifically~\cite{Gralla:2016qfw}
\begin{align}
    \label{eq:Tins}
    T_{\rm ins} = 0.150 \mu \left(\frac{M}{\mu}\right)^{2} \left(k - 1 - \log k\right)
\end{align}
where $\mu$ is the mass of the particle, $M$ is the masss of the SMBH, and
\begin{equation}
    k = \frac{2 \left(1 - \chi^{2}\right)}{\left(\frac{r_{0}}{r_{+}} - 1\right)^{3}}
\end{equation}
with $\chi$ the BH's dimensionless spin parameter, $r_{+}$ the horizon radius, and $r_{0}$ the initial radial separation. For our scenario $\mu = 60 M_{\odot}, M = 6\times10^{8} M_{\odot}$, and $r_{0} \approx 1.1 r_{+}$. When $\chi \rightarrow 1$, $k \ll 1$ and Eq.~\eqref{eq:Tins} is dominated by the logarithm term. Applying our values gives
\begin{equation}
    T_{\rm ins} \approx 143 \left[-9.29 - \log\delta\chi + {\cal{O}}(\delta\chi)\right] \; {\rm yrs}\,,
\end{equation}
with $\delta \chi = 1-\chi$, and we have expanded in $\delta \chi \ll 1$. For a value of $\delta \chi = 10^{-5}$, $T_{\rm ins} \approx 317$ years.

The second question is related to whether or not the tertiary (the SMBH in this case), can disrupt the binary. There are multiple methods of quantifying the stability criterion~\cite{Hill:1878a, Hill:1878b, Hill:1878c, Zare:1976a,Zare:1977,Mardling:2001}, but a useful conceptualization of this is related to Lagrange points and Hill stability~\cite{Georga:2008}. For the triple system described in Sec.~\ref{sec:pe}, the $L_{1,2}$ Lagrange points are equidistant from the binary's COM, and are equal to the radius of the Hill sphere,
\begin{equation}
    R_{\rm Hill} = \frac{R}{(3 \eta_{3})^{1/3}}\,.
\end{equation}
where $R$ is the radial separation between the tertiary and the COM of the inner binary. If the maximum radial separation of the inner binary is smaller than $R_{\rm Hill}$, then the SMBH cannot disrupt the binary. The largest separation of the inner binary is the apocenter distance $r_{a} = p/(1-e)$, while $R=R_{3}$ in this case. From the values of Table~\ref{tab:perturbed_signal_recovery}, it is straightforward to show that the binary is stable.

\bibliography{refs}

\begin{thebibliography}{125}%
\makeatletter
\providecommand \@ifxundefined [1]{%
 \@ifx{#1\undefined}
}%
\providecommand \@ifnum [1]{%
 \ifnum #1\expandafter \@firstoftwo
 \else \expandafter \@secondoftwo
 \fi
}%
\providecommand \@ifx [1]{%
 \ifx #1\expandafter \@firstoftwo
 \else \expandafter \@secondoftwo
 \fi
}%
\providecommand \natexlab [1]{#1}%
\providecommand \enquote  [1]{``#1''}%
\providecommand \bibnamefont  [1]{#1}%
\providecommand \bibfnamefont [1]{#1}%
\providecommand \citenamefont [1]{#1}%
\providecommand \href@noop [0]{\@secondoftwo}%
\providecommand \href [0]{\begingroup \@sanitize@url \@href}%
\providecommand \@href[1]{\@@startlink{#1}\@@href}%
\providecommand \@@href[1]{\endgroup#1\@@endlink}%
\providecommand \@sanitize@url [0]{\catcode `\\12\catcode `\$12\catcode
  `\&12\catcode `\#12\catcode `\^12\catcode `\_12\catcode `\%12\relax}%
\providecommand \@@startlink[1]{}%
\providecommand \@@endlink[0]{}%
\providecommand \url  [0]{\begingroup\@sanitize@url \@url }%
\providecommand \@url [1]{\endgroup\@href {#1}{\urlprefix }}%
\providecommand \urlprefix  [0]{URL }%
\providecommand \Eprint [0]{\href }%
\providecommand \doibase [0]{http://dx.doi.org/}%
\providecommand \selectlanguage [0]{\@gobble}%
\providecommand \bibinfo  [0]{\@secondoftwo}%
\providecommand \bibfield  [0]{\@secondoftwo}%
\providecommand \translation [1]{[#1]}%
\providecommand \BibitemOpen [0]{}%
\providecommand \bibitemStop [0]{}%
\providecommand \bibitemNoStop [0]{.\EOS\space}%
\providecommand \EOS [0]{\spacefactor3000\relax}%
\providecommand \BibitemShut  [1]{\csname bibitem#1\endcsname}%
\let\auto@bib@innerbib\@empty
\bibitem [{\citenamefont {Abbott}\ \emph {et~al.}(2021)\citenamefont {Abbott}
  \emph {et~al.}}]{LVK:2021:GWTC3}%
  \BibitemOpen
  \bibfield  {author} {\bibinfo {author} {\bibfnamefont {R.}~\bibnamefont
  {Abbott}} \emph {et~al.} (\bibinfo {collaboration} {LIGO Scientific, VIRGO,
  KAGRA}),\ }\href@noop {} {\  (\bibinfo {year} {2021})},\ \Eprint
  {http://arxiv.org/abs/2111.03606} {arXiv:2111.03606 [gr-qc]} \BibitemShut
  {NoStop}%
\bibitem [{\citenamefont {Abbott}\ \emph {et~al.}(2020)\citenamefont {Abbott},
  \citenamefont {Abbott}, \citenamefont {Abraham}, \citenamefont {Acernese},
  \citenamefont {Ackley}, \citenamefont {Adams}, \citenamefont {Adhikari},
  \citenamefont {Adya}, \citenamefont {Affeldt}, \citenamefont {Agathos} \emph
  {et~al.}}]{Abbott:2020:GW190521}%
  \BibitemOpen
  \bibfield  {author} {\bibinfo {author} {\bibfnamefont {R.}~\bibnamefont
  {Abbott}}, \bibinfo {author} {\bibfnamefont {T.}~\bibnamefont {Abbott}},
  \bibinfo {author} {\bibfnamefont {S.}~\bibnamefont {Abraham}}, \bibinfo
  {author} {\bibfnamefont {F.}~\bibnamefont {Acernese}}, \bibinfo {author}
  {\bibfnamefont {K.}~\bibnamefont {Ackley}}, \bibinfo {author} {\bibfnamefont
  {C.}~\bibnamefont {Adams}}, \bibinfo {author} {\bibfnamefont
  {R.}~\bibnamefont {Adhikari}}, \bibinfo {author} {\bibfnamefont
  {V.}~\bibnamefont {Adya}}, \bibinfo {author} {\bibfnamefont {C.}~\bibnamefont
  {Affeldt}}, \bibinfo {author} {\bibfnamefont {M.}~\bibnamefont {Agathos}},
  \emph {et~al.},\ }\href@noop {} {\bibfield  {journal} {\bibinfo  {journal}
  {Phys. Rev. Lett.}\ }\textbf {\bibinfo {volume} {125}},\ \bibinfo {pages}
  {101102} (\bibinfo {year} {2020})}\BibitemShut {NoStop}%
\bibitem [{\citenamefont {{Abbott}}\ \emph
  {et~al.}(2020{\natexlab{a}})\citenamefont {{Abbott}}, \citenamefont
  {{Abbott}}, \citenamefont {{Abraham}}, \citenamefont {{Acernese}},
  \citenamefont {{Ackley}}, \citenamefont {{Adams}}, \citenamefont {{Adhikari}}
  \emph {et~al.}}]{LVK:2020:ImplicationsGW190521}%
  \BibitemOpen
  \bibfield  {author} {\bibinfo {author} {\bibfnamefont {R.}~\bibnamefont
  {{Abbott}}}, \bibinfo {author} {\bibfnamefont {T.~D.}\ \bibnamefont
  {{Abbott}}}, \bibinfo {author} {\bibfnamefont {S.}~\bibnamefont {{Abraham}}},
  \bibinfo {author} {\bibfnamefont {F.}~\bibnamefont {{Acernese}}}, \bibinfo
  {author} {\bibfnamefont {K.}~\bibnamefont {{Ackley}}}, \bibinfo {author}
  {\bibfnamefont {C.}~\bibnamefont {{Adams}}}, \bibinfo {author} {\bibfnamefont
  {R.~X.}\ \bibnamefont {{Adhikari}}},  \emph {et~al.},\ }\href {\doibase
  10.3847/2041-8213/aba493} {\bibfield  {journal} {\bibinfo  {journal}
  {Astrophys. J. Lett.}\ }\textbf {\bibinfo {volume} {900}},\ \bibinfo {eid}
  {L13} (\bibinfo {year} {2020}{\natexlab{a}})},\ \Eprint
  {http://arxiv.org/abs/2009.01190} {arXiv:2009.01190 [astro-ph.HE]}
  \BibitemShut {NoStop}%
\bibitem [{\citenamefont {{Romero-Shaw}}\ \emph {et~al.}(2020)\citenamefont
  {{Romero-Shaw}}, \citenamefont {Lasky}, \citenamefont {Thrane},\ and\
  \citenamefont {Bustillo}}]{Romero-Shaw:2020:GW190521}%
  \BibitemOpen
  \bibfield  {author} {\bibinfo {author} {\bibfnamefont {I.~M.}\ \bibnamefont
  {{Romero-Shaw}}}, \bibinfo {author} {\bibfnamefont {P.~D.}\ \bibnamefont
  {Lasky}}, \bibinfo {author} {\bibfnamefont {E.}~\bibnamefont {Thrane}}, \
  and\ \bibinfo {author} {\bibfnamefont {J.~C.}\ \bibnamefont {Bustillo}},\
  }\href {\doibase 10.3847/2041-8213/abbe26} {\bibfield  {journal} {\bibinfo
  {journal} {Astrophys. J. Lett.}\ }\textbf {\bibinfo {volume} {903}},\
  \bibinfo {pages} {L5} (\bibinfo {year} {2020})},\ \Eprint
  {http://arxiv.org/abs/2009.04771} {arXiv:2009.04771 [astro-ph.HE]}
  \BibitemShut {NoStop}%
\bibitem [{\citenamefont {Gayathri}\ \emph {et~al.}(2022)\citenamefont
  {Gayathri}, \citenamefont {Healy}, \citenamefont {Lange}, \citenamefont
  {O'Brien}, \citenamefont {Szczepanczyk}, \citenamefont {Bartos},
  \citenamefont {Campanelli}, \citenamefont {Klimenko}, \citenamefont
  {Lousto},\ and\ \citenamefont {O'Shaughnessy}}]{Gayathri:2022:GW190521}%
  \BibitemOpen
  \bibfield  {author} {\bibinfo {author} {\bibfnamefont {V.}~\bibnamefont
  {Gayathri}}, \bibinfo {author} {\bibfnamefont {J.}~\bibnamefont {Healy}},
  \bibinfo {author} {\bibfnamefont {J.}~\bibnamefont {Lange}}, \bibinfo
  {author} {\bibfnamefont {B.}~\bibnamefont {O'Brien}}, \bibinfo {author}
  {\bibfnamefont {M.}~\bibnamefont {Szczepanczyk}}, \bibinfo {author}
  {\bibfnamefont {I.}~\bibnamefont {Bartos}}, \bibinfo {author} {\bibfnamefont
  {M.}~\bibnamefont {Campanelli}}, \bibinfo {author} {\bibfnamefont
  {S.}~\bibnamefont {Klimenko}}, \bibinfo {author} {\bibfnamefont {C.~O.}\
  \bibnamefont {Lousto}}, \ and\ \bibinfo {author} {\bibfnamefont
  {R.}~\bibnamefont {O'Shaughnessy}},\ }\href {\doibase
  10.1038/s41550-021-01568-w} {\bibfield  {journal} {\bibinfo  {journal}
  {Nature Astron.}\ }\textbf {\bibinfo {volume} {6}},\ \bibinfo {pages} {344}
  (\bibinfo {year} {2022})},\ \Eprint {http://arxiv.org/abs/2009.05461}
  {arXiv:2009.05461 [astro-ph.HE]} \BibitemShut {NoStop}%
\bibitem [{\citenamefont {Gayathri}\ \emph {et~al.}(2021)\citenamefont
  {Gayathri}, \citenamefont {Yang}, \citenamefont {Tagawa}, \citenamefont
  {Haiman},\ and\ \citenamefont {Bartos}}]{Gayathri:2021xwb}%
  \BibitemOpen
  \bibfield  {author} {\bibinfo {author} {\bibfnamefont {V.}~\bibnamefont
  {Gayathri}}, \bibinfo {author} {\bibfnamefont {Y.}~\bibnamefont {Yang}},
  \bibinfo {author} {\bibfnamefont {H.}~\bibnamefont {Tagawa}}, \bibinfo
  {author} {\bibfnamefont {Z.}~\bibnamefont {Haiman}}, \ and\ \bibinfo {author}
  {\bibfnamefont {I.}~\bibnamefont {Bartos}},\ }\href {\doibase
  10.3847/2041-8213/ac2cc1} {\bibfield  {journal} {\bibinfo  {journal}
  {Astrophys. J. Lett.}\ }\textbf {\bibinfo {volume} {920}},\ \bibinfo {pages}
  {L42} (\bibinfo {year} {2021})},\ \Eprint {http://arxiv.org/abs/2104.10253}
  {arXiv:2104.10253 [gr-qc]} \BibitemShut {NoStop}%
\bibitem [{\citenamefont {{Gamba}}\ \emph {et~al.}(2023)\citenamefont
  {{Gamba}}, \citenamefont {{Breschi}}, \citenamefont {{Carullo}},
  \citenamefont {{Albanesi}}, \citenamefont {{Rettegno}}, \citenamefont
  {{Bernuzzi}},\ and\ \citenamefont {{Nagar}}}]{Gamba:2021:hyperbolic}%
  \BibitemOpen
  \bibfield  {author} {\bibinfo {author} {\bibfnamefont {R.}~\bibnamefont
  {{Gamba}}}, \bibinfo {author} {\bibfnamefont {M.}~\bibnamefont {{Breschi}}},
  \bibinfo {author} {\bibfnamefont {G.}~\bibnamefont {{Carullo}}}, \bibinfo
  {author} {\bibfnamefont {S.}~\bibnamefont {{Albanesi}}}, \bibinfo {author}
  {\bibfnamefont {P.}~\bibnamefont {{Rettegno}}}, \bibinfo {author}
  {\bibfnamefont {S.}~\bibnamefont {{Bernuzzi}}}, \ and\ \bibinfo {author}
  {\bibfnamefont {A.}~\bibnamefont {{Nagar}}},\ }\href {\doibase
  10.1038/s41550-022-01813-w} {\bibfield  {journal} {\bibinfo  {journal}
  {Nature Astronomy}\ }\textbf {\bibinfo {volume} {7}},\ \bibinfo {pages} {11}
  (\bibinfo {year} {2023})},\ \Eprint {http://arxiv.org/abs/2106.05575}
  {arXiv:2106.05575 [gr-qc]} \BibitemShut {NoStop}%
\bibitem [{\citenamefont {{Ford}}\ and\ \citenamefont
  {{McKernan}}(2022)}]{Ford:2021:LoudVsQuiet}%
  \BibitemOpen
  \bibfield  {author} {\bibinfo {author} {\bibfnamefont {K.~E.~S.}\
  \bibnamefont {{Ford}}}\ and\ \bibinfo {author} {\bibfnamefont
  {B.}~\bibnamefont {{McKernan}}},\ }\href {\doibase 10.1093/mnras/stac2861}
  {\bibfield  {journal} {\bibinfo  {journal} {Mon. Not. Roy. Astron. Soc.}\
  }\textbf {\bibinfo {volume} {517}},\ \bibinfo {pages} {5827} (\bibinfo {year}
  {2022})},\ \Eprint {http://arxiv.org/abs/2109.03212} {arXiv:2109.03212
  [astro-ph.HE]} \BibitemShut {NoStop}%
\bibitem [{\citenamefont {{Samsing}}\ \emph {et~al.}(2020)\citenamefont
  {{Samsing}}, \citenamefont {{Bartos}}, \citenamefont {{D'Orazio}},
  \citenamefont {{Haiman}}, \citenamefont {{Kocsis}}, \citenamefont {{Leigh}},
  \citenamefont {{Liu}}, \citenamefont {{Pessah}},\ and\ \citenamefont
  {{Tagawa}}}]{Samsing:2020:AGN}%
  \BibitemOpen
  \bibfield  {author} {\bibinfo {author} {\bibfnamefont {J.}~\bibnamefont
  {{Samsing}}}, \bibinfo {author} {\bibfnamefont {I.}~\bibnamefont {{Bartos}}},
  \bibinfo {author} {\bibfnamefont {D.~J.}\ \bibnamefont {{D'Orazio}}},
  \bibinfo {author} {\bibfnamefont {Z.}~\bibnamefont {{Haiman}}}, \bibinfo
  {author} {\bibfnamefont {B.}~\bibnamefont {{Kocsis}}}, \bibinfo {author}
  {\bibfnamefont {N.~W.~C.}\ \bibnamefont {{Leigh}}}, \bibinfo {author}
  {\bibfnamefont {B.}~\bibnamefont {{Liu}}}, \bibinfo {author} {\bibfnamefont
  {M.~E.}\ \bibnamefont {{Pessah}}}, \ and\ \bibinfo {author} {\bibfnamefont
  {H.}~\bibnamefont {{Tagawa}}},\ }\href@noop {} {\bibfield  {journal}
  {\bibinfo  {journal} {arXiv e-prints}\ ,\ \bibinfo {eid} {arXiv:2010.09765}}
  (\bibinfo {year} {2020})},\ \Eprint {http://arxiv.org/abs/2010.09765}
  {arXiv:2010.09765 [astro-ph.HE]} \BibitemShut {NoStop}%
\bibitem [{\citenamefont {{Samsing}}\ \emph {et~al.}(2014)\citenamefont
  {{Samsing}}, \citenamefont {{MacLeod}},\ and\ \citenamefont
  {{Ramirez-Ruiz}}}]{Samsing:2014:BinarySingle}%
  \BibitemOpen
  \bibfield  {author} {\bibinfo {author} {\bibfnamefont {J.}~\bibnamefont
  {{Samsing}}}, \bibinfo {author} {\bibfnamefont {M.}~\bibnamefont
  {{MacLeod}}}, \ and\ \bibinfo {author} {\bibfnamefont {E.}~\bibnamefont
  {{Ramirez-Ruiz}}},\ }\href {\doibase 10.1088/0004-637X/784/1/71} {\bibfield
  {journal} {\bibinfo  {journal} {\apj}\ }\textbf {\bibinfo {volume} {784}},\
  \bibinfo {eid} {71} (\bibinfo {year} {2014})},\ \Eprint
  {http://arxiv.org/abs/1308.2964} {arXiv:1308.2964 [astro-ph.HE]} \BibitemShut
  {NoStop}%
\bibitem [{\citenamefont {{Samsing}}\ and\ \citenamefont
  {{Ramirez-Ruiz}}(2017)}]{SamsigRamirezRuiz:2017:HighlyEccentric}%
  \BibitemOpen
  \bibfield  {author} {\bibinfo {author} {\bibfnamefont {J.}~\bibnamefont
  {{Samsing}}}\ and\ \bibinfo {author} {\bibfnamefont {E.}~\bibnamefont
  {{Ramirez-Ruiz}}},\ }\href {\doibase 10.3847/2041-8213/aa6f0b} {\bibfield
  {journal} {\bibinfo  {journal} {Astrophys. J. Lett.}\ }\textbf {\bibinfo
  {volume} {840}},\ \bibinfo {eid} {L14} (\bibinfo {year} {2017})},\ \Eprint
  {http://arxiv.org/abs/1703.09703} {arXiv:1703.09703 [astro-ph.HE]}
  \BibitemShut {NoStop}%
\bibitem [{\citenamefont {Rodriguez}\ \emph
  {et~al.}(2018{\natexlab{a}})\citenamefont {Rodriguez}, \citenamefont
  {Amaro-Seoane}, \citenamefont {Chatterjee},\ and\ \citenamefont
  {Rasio}}]{Rodriguez18a}%
  \BibitemOpen
  \bibfield  {author} {\bibinfo {author} {\bibfnamefont {C.~L.}\ \bibnamefont
  {Rodriguez}}, \bibinfo {author} {\bibfnamefont {P.}~\bibnamefont
  {Amaro-Seoane}}, \bibinfo {author} {\bibfnamefont {S.}~\bibnamefont
  {Chatterjee}}, \ and\ \bibinfo {author} {\bibfnamefont {F.~A.}\ \bibnamefont
  {Rasio}},\ }\href {\doibase 10.1103/PhysRevLett.120.151101} {\bibfield
  {journal} {\bibinfo  {journal} {Phys. Rev. Lett.}\ }\textbf {\bibinfo
  {volume} {120}},\ \bibinfo {pages} {151101} (\bibinfo {year}
  {2018}{\natexlab{a}})},\ \Eprint {http://arxiv.org/abs/1712.04937}
  {arXiv:1712.04937} \BibitemShut {NoStop}%
\bibitem [{\citenamefont {Rodriguez}\ \emph
  {et~al.}(2018{\natexlab{b}})\citenamefont {Rodriguez}, \citenamefont
  {Amaro-Seoane}, \citenamefont {Chatterjee}, \citenamefont {Kremer},
  \citenamefont {Rasio}, \citenamefont {Samsing}, \citenamefont {Ye},\ and\
  \citenamefont {Zevin}}]{Rodriguez:2018pss}%
  \BibitemOpen
  \bibfield  {author} {\bibinfo {author} {\bibfnamefont {C.~L.}\ \bibnamefont
  {Rodriguez}}, \bibinfo {author} {\bibfnamefont {P.}~\bibnamefont
  {Amaro-Seoane}}, \bibinfo {author} {\bibfnamefont {S.}~\bibnamefont
  {Chatterjee}}, \bibinfo {author} {\bibfnamefont {K.}~\bibnamefont {Kremer}},
  \bibinfo {author} {\bibfnamefont {F.~A.}\ \bibnamefont {Rasio}}, \bibinfo
  {author} {\bibfnamefont {J.}~\bibnamefont {Samsing}}, \bibinfo {author}
  {\bibfnamefont {C.~S.}\ \bibnamefont {Ye}}, \ and\ \bibinfo {author}
  {\bibfnamefont {M.}~\bibnamefont {Zevin}},\ }\href {\doibase
  10.1103/PhysRevD.98.123005} {\bibfield  {journal} {\bibinfo  {journal} {Phys.
  Rev. D}\ }\textbf {\bibinfo {volume} {98}},\ \bibinfo {pages} {123005}
  (\bibinfo {year} {2018}{\natexlab{b}})},\ \Eprint
  {http://arxiv.org/abs/1811.04926} {arXiv:1811.04926 [astro-ph.HE]}
  \BibitemShut {NoStop}%
\bibitem [{\citenamefont {Samsing}\ \emph
  {et~al.}(2018{\natexlab{a}})\citenamefont {Samsing}, \citenamefont
  {D'Orazio}, \citenamefont {Askar},\ and\ \citenamefont {Giersz}}]{Samsing18}%
  \BibitemOpen
  \bibfield  {author} {\bibinfo {author} {\bibfnamefont {J.}~\bibnamefont
  {Samsing}}, \bibinfo {author} {\bibfnamefont {D.~J.}\ \bibnamefont
  {D'Orazio}}, \bibinfo {author} {\bibfnamefont {A.}~\bibnamefont {Askar}}, \
  and\ \bibinfo {author} {\bibfnamefont {M.}~\bibnamefont {Giersz}},\
  }\href@noop {} {\bibfield  {journal} {\bibinfo  {journal} {arXiv e-prints}\
  ,\ \bibinfo {eid} {arXiv:1802.08654}} (\bibinfo {year}
  {2018}{\natexlab{a}})},\ \Eprint {http://arxiv.org/abs/1802.08654}
  {arXiv:1802.08654 [astro-ph.HE]} \BibitemShut {NoStop}%
\bibitem [{\citenamefont {{Zevin}}\ \emph {et~al.}(2019)\citenamefont
  {{Zevin}}, \citenamefont {{Samsing}}, \citenamefont {{Rodriguez}},
  \citenamefont {{Haster}},\ and\ \citenamefont
  {{Ramirez-Ruiz}}}]{Zevin:2019:BinaryBinary}%
  \BibitemOpen
  \bibfield  {author} {\bibinfo {author} {\bibfnamefont {M.}~\bibnamefont
  {{Zevin}}}, \bibinfo {author} {\bibfnamefont {J.}~\bibnamefont {{Samsing}}},
  \bibinfo {author} {\bibfnamefont {C.}~\bibnamefont {{Rodriguez}}}, \bibinfo
  {author} {\bibfnamefont {C.-J.}\ \bibnamefont {{Haster}}}, \ and\ \bibinfo
  {author} {\bibfnamefont {E.}~\bibnamefont {{Ramirez-Ruiz}}},\ }\href
  {\doibase 10.3847/1538-4357/aaf6ec} {\bibfield  {journal} {\bibinfo
  {journal} {\apj}\ }\textbf {\bibinfo {volume} {871}},\ \bibinfo {eid} {91}
  (\bibinfo {year} {2019})},\ \Eprint {http://arxiv.org/abs/1810.00901}
  {arXiv:1810.00901 [astro-ph.HE]} \BibitemShut {NoStop}%
\bibitem [{\citenamefont {{Fragione}}\ \emph {et~al.}(2020)\citenamefont
  {{Fragione}}, \citenamefont {{Loeb}},\ and\ \citenamefont
  {{Rasio}}}]{Fragione:2020:GW190521StarClusersHierarcical}%
  \BibitemOpen
  \bibfield  {author} {\bibinfo {author} {\bibfnamefont {G.}~\bibnamefont
  {{Fragione}}}, \bibinfo {author} {\bibfnamefont {A.}~\bibnamefont {{Loeb}}},
  \ and\ \bibinfo {author} {\bibfnamefont {F.~A.}\ \bibnamefont {{Rasio}}},\
  }\href {\doibase 10.3847/2041-8213/abbc0a} {\bibfield  {journal} {\bibinfo
  {journal} {Astrophys. J. Lett.}\ }\textbf {\bibinfo {volume} {902}},\
  \bibinfo {eid} {L26} (\bibinfo {year} {2020})},\ \Eprint
  {http://arxiv.org/abs/2009.05065} {arXiv:2009.05065 [astro-ph.GA]}
  \BibitemShut {NoStop}%
\bibitem [{\citenamefont {Vitale}\ \emph {et~al.}(2017)\citenamefont {Vitale},
  \citenamefont {Lynch}, \citenamefont {Sturani},\ and\ \citenamefont
  {Graff}}]{Vitale15}%
  \BibitemOpen
  \bibfield  {author} {\bibinfo {author} {\bibfnamefont {S.}~\bibnamefont
  {Vitale}}, \bibinfo {author} {\bibfnamefont {R.}~\bibnamefont {Lynch}},
  \bibinfo {author} {\bibfnamefont {R.}~\bibnamefont {Sturani}}, \ and\
  \bibinfo {author} {\bibfnamefont {P.}~\bibnamefont {Graff}},\ }\href
  {\doibase 10.1088/1361-6382/aa552e} {\bibfield  {journal} {\bibinfo
  {journal} {Class. Quant. Grav.}\ }\textbf {\bibinfo {volume} {34}},\ \bibinfo
  {pages} {03LT01} (\bibinfo {year} {2017})},\ \Eprint
  {http://arxiv.org/abs/1503.04307} {arXiv:1503.04307 [gr-qc]} \BibitemShut
  {NoStop}%
\bibitem [{\citenamefont {Stevenson}\ \emph {et~al.}(2015)\citenamefont
  {Stevenson}, \citenamefont {Ohme},\ and\ \citenamefont
  {Fairhurst}}]{Stevenson15bqa}%
  \BibitemOpen
  \bibfield  {author} {\bibinfo {author} {\bibfnamefont {S.}~\bibnamefont
  {Stevenson}}, \bibinfo {author} {\bibfnamefont {F.}~\bibnamefont {Ohme}}, \
  and\ \bibinfo {author} {\bibfnamefont {S.}~\bibnamefont {Fairhurst}},\ }\href
  {\doibase 10.1088/0004-637X/810/1/58} {\bibfield  {journal} {\bibinfo
  {journal} {Astrophys. J.}\ }\textbf {\bibinfo {volume} {810}},\ \bibinfo
  {pages} {58} (\bibinfo {year} {2015})},\ \Eprint
  {http://arxiv.org/abs/1504.07802} {arXiv:1504.07802 [astro-ph.HE]}
  \BibitemShut {NoStop}%
\bibitem [{\citenamefont {Gerosa}\ and\ \citenamefont
  {Berti}(2017)}]{Gerosa17}%
  \BibitemOpen
  \bibfield  {author} {\bibinfo {author} {\bibfnamefont {D.}~\bibnamefont
  {Gerosa}}\ and\ \bibinfo {author} {\bibfnamefont {E.}~\bibnamefont {Berti}},\
  }\href {\doibase 10.1103/PhysRevD.95.124046} {\bibfield  {journal} {\bibinfo
  {journal} {Phys. Rev.}\ }\textbf {\bibinfo {volume} {D95}},\ \bibinfo {pages}
  {124046} (\bibinfo {year} {2017})},\ \Eprint
  {http://arxiv.org/abs/1703.06223} {arXiv:1703.06223 [gr-qc]} \BibitemShut
  {NoStop}%
\bibitem [{\citenamefont {Randall}\ and\ \citenamefont
  {Xianyu}(2019)}]{Randall19}%
  \BibitemOpen
  \bibfield  {author} {\bibinfo {author} {\bibfnamefont {L.}~\bibnamefont
  {Randall}}\ and\ \bibinfo {author} {\bibfnamefont {Z.-Z.}\ \bibnamefont
  {Xianyu}},\ }\href@noop {} {\  (\bibinfo {year} {2019})},\ \Eprint
  {http://arxiv.org/abs/1907.02283} {arXiv:1907.02283 [astro-ph.HE]}
  \BibitemShut {NoStop}%
\bibitem [{\citenamefont {{Bavera}}\ \emph {et~al.}(2020)\citenamefont
  {{Bavera}}, \citenamefont {{Fragos}}, \citenamefont {{Qin}}, \citenamefont
  {{Zapartas}}, \citenamefont {{Neijssel}}, \citenamefont {{Mandel}},
  \citenamefont {{Batta}}, \citenamefont {{Gaebel}}, \citenamefont
  {{Kimball}},\ and\ \citenamefont {{Stevenson}}}]{Bavera2020}%
  \BibitemOpen
  \bibfield  {author} {\bibinfo {author} {\bibfnamefont {S.~S.}\ \bibnamefont
  {{Bavera}}}, \bibinfo {author} {\bibfnamefont {T.}~\bibnamefont {{Fragos}}},
  \bibinfo {author} {\bibfnamefont {Y.}~\bibnamefont {{Qin}}}, \bibinfo
  {author} {\bibfnamefont {E.}~\bibnamefont {{Zapartas}}}, \bibinfo {author}
  {\bibfnamefont {C.~J.}\ \bibnamefont {{Neijssel}}}, \bibinfo {author}
  {\bibfnamefont {I.}~\bibnamefont {{Mandel}}}, \bibinfo {author}
  {\bibfnamefont {A.}~\bibnamefont {{Batta}}}, \bibinfo {author} {\bibfnamefont
  {S.~M.}\ \bibnamefont {{Gaebel}}}, \bibinfo {author} {\bibfnamefont
  {C.}~\bibnamefont {{Kimball}}}, \ and\ \bibinfo {author} {\bibfnamefont
  {S.}~\bibnamefont {{Stevenson}}},\ }\href {\doibase
  10.1051/0004-6361/201936204} {\bibfield  {journal} {\bibinfo  {journal}
  {Astron. Astrophys.}\ }\textbf {\bibinfo {volume} {635}},\ \bibinfo {eid}
  {A97} (\bibinfo {year} {2020})},\ \Eprint {http://arxiv.org/abs/1906.12257}
  {arXiv:1906.12257 [astro-ph.HE]} \BibitemShut {NoStop}%
\bibitem [{\citenamefont {{Mapelli}}(2020)}]{Mapelli:2020:review}%
  \BibitemOpen
  \bibfield  {author} {\bibinfo {author} {\bibfnamefont {M.}~\bibnamefont
  {{Mapelli}}},\ }\href {\doibase 10.3389/fspas.2020.00038} {\bibfield
  {journal} {\bibinfo  {journal} {Frontiers in Astronomy and Space Sciences}\
  }\textbf {\bibinfo {volume} {7}},\ \bibinfo {eid} {38} (\bibinfo {year}
  {2020})},\ \Eprint {http://arxiv.org/abs/2105.12455} {arXiv:2105.12455
  [astro-ph.HE]} \BibitemShut {NoStop}%
\bibitem [{\citenamefont {Sedda}\ \emph {et~al.}(2020)\citenamefont {Sedda},
  \citenamefont {Mapelli}, \citenamefont {Spera}, \citenamefont {Benacquista},\
  and\ \citenamefont {Giacobbo}}]{Sedda:2020:fingerprints}%
  \BibitemOpen
  \bibfield  {author} {\bibinfo {author} {\bibfnamefont {M.~A.}\ \bibnamefont
  {Sedda}}, \bibinfo {author} {\bibfnamefont {M.}~\bibnamefont {Mapelli}},
  \bibinfo {author} {\bibfnamefont {M.}~\bibnamefont {Spera}}, \bibinfo
  {author} {\bibfnamefont {M.}~\bibnamefont {Benacquista}}, \ and\ \bibinfo
  {author} {\bibfnamefont {N.}~\bibnamefont {Giacobbo}},\ }\href {\doibase
  10.3847/1538-4357/ab88b2} {\bibfield  {journal} {\bibinfo  {journal}
  {Astrophys. J.}\ }\textbf {\bibinfo {volume} {894}},\ \bibinfo {pages} {133}
  (\bibinfo {year} {2020})},\ \Eprint {http://arxiv.org/abs/2003.07409}
  {arXiv:2003.07409 [astro-ph.GA]} \BibitemShut {NoStop}%
\bibitem [{\citenamefont {{Zevin}}\ \emph
  {et~al.}(2021{\natexlab{a}})\citenamefont {{Zevin}}, \citenamefont
  {{Bavera}}, \citenamefont {{Berry}}, \citenamefont {{Kalogera}},
  \citenamefont {{Fragos}}, \citenamefont {{Marchant}}, \citenamefont
  {{Rodriguez}}, \citenamefont {{Antonini}}, \citenamefont {{Holz}},\ and\
  \citenamefont {{Pankow}}}]{Zevin:2020:channels}%
  \BibitemOpen
  \bibfield  {author} {\bibinfo {author} {\bibfnamefont {M.}~\bibnamefont
  {{Zevin}}}, \bibinfo {author} {\bibfnamefont {S.~S.}\ \bibnamefont
  {{Bavera}}}, \bibinfo {author} {\bibfnamefont {C.~P.~L.}\ \bibnamefont
  {{Berry}}}, \bibinfo {author} {\bibfnamefont {V.}~\bibnamefont {{Kalogera}}},
  \bibinfo {author} {\bibfnamefont {T.}~\bibnamefont {{Fragos}}}, \bibinfo
  {author} {\bibfnamefont {P.}~\bibnamefont {{Marchant}}}, \bibinfo {author}
  {\bibfnamefont {C.~L.}\ \bibnamefont {{Rodriguez}}}, \bibinfo {author}
  {\bibfnamefont {F.}~\bibnamefont {{Antonini}}}, \bibinfo {author}
  {\bibfnamefont {D.~E.}\ \bibnamefont {{Holz}}}, \ and\ \bibinfo {author}
  {\bibfnamefont {C.}~\bibnamefont {{Pankow}}},\ }\href {\doibase
  10.3847/1538-4357/abe40e} {\bibfield  {journal} {\bibinfo  {journal} {apj}\
  }\textbf {\bibinfo {volume} {910}},\ \bibinfo {eid} {152} (\bibinfo {year}
  {2021}{\natexlab{a}})},\ \Eprint {http://arxiv.org/abs/2011.10057}
  {arXiv:2011.10057 [astro-ph.HE]} \BibitemShut {NoStop}%
\bibitem [{\citenamefont {{Fragione}}\ \emph {et~al.}(2022)\citenamefont
  {{Fragione}}, \citenamefont {{Kocsis}}, \citenamefont {{Rasio}},\ and\
  \citenamefont {{Silk}}}]{Fragione:2022:NSCs}%
  \BibitemOpen
  \bibfield  {author} {\bibinfo {author} {\bibfnamefont {G.}~\bibnamefont
  {{Fragione}}}, \bibinfo {author} {\bibfnamefont {B.}~\bibnamefont
  {{Kocsis}}}, \bibinfo {author} {\bibfnamefont {F.~A.}\ \bibnamefont
  {{Rasio}}}, \ and\ \bibinfo {author} {\bibfnamefont {J.}~\bibnamefont
  {{Silk}}},\ }\href {\doibase 10.3847/1538-4357/ac5026} {\bibfield  {journal}
  {\bibinfo  {journal} {apj}\ }\textbf {\bibinfo {volume} {927}},\ \bibinfo
  {eid} {231} (\bibinfo {year} {2022})},\ \Eprint
  {http://arxiv.org/abs/2107.04639} {arXiv:2107.04639 [astro-ph.GA]}
  \BibitemShut {NoStop}%
\bibitem [{\citenamefont {Gond\'an}\ \emph {et~al.}(2018)\citenamefont
  {Gond\'an}, \citenamefont {Kocsis}, \citenamefont {Raffai},\ and\
  \citenamefont {Frei}}]{Gondan17}%
  \BibitemOpen
  \bibfield  {author} {\bibinfo {author} {\bibfnamefont {L.}~\bibnamefont
  {Gond\'an}}, \bibinfo {author} {\bibfnamefont {B.}~\bibnamefont {Kocsis}},
  \bibinfo {author} {\bibfnamefont {P.}~\bibnamefont {Raffai}}, \ and\ \bibinfo
  {author} {\bibfnamefont {Z.}~\bibnamefont {Frei}},\ }\href {\doibase
  10.3847/1538-4357/aabfee} {\bibfield  {journal} {\bibinfo  {journal}
  {Astrophys. J.}\ }\textbf {\bibinfo {volume} {860}},\ \bibinfo {pages} {5}
  (\bibinfo {year} {2018})},\ \Eprint {http://arxiv.org/abs/1711.09989}
  {arXiv:1711.09989} \BibitemShut {NoStop}%
\bibitem [{\citenamefont {Samsing}(2018)}]{Samsing:2017xmd}%
  \BibitemOpen
  \bibfield  {author} {\bibinfo {author} {\bibfnamefont {J.}~\bibnamefont
  {Samsing}},\ }\href {\doibase 10.1103/PhysRevD.97.103014} {\bibfield
  {journal} {\bibinfo  {journal} {Phys. Rev. D}\ }\textbf {\bibinfo {volume}
  {97}},\ \bibinfo {pages} {103014} (\bibinfo {year} {2018})},\ \Eprint
  {http://arxiv.org/abs/1711.07452} {arXiv:1711.07452 [astro-ph.HE]}
  \BibitemShut {NoStop}%
\bibitem [{\citenamefont {Tagawa}\ \emph {et~al.}(2021)\citenamefont {Tagawa},
  \citenamefont {Kocsis}, \citenamefont {Haiman}, \citenamefont {Bartos},
  \citenamefont {Omukai},\ and\ \citenamefont {Samsing}}]{Tagawa:2020jnc}%
  \BibitemOpen
  \bibfield  {author} {\bibinfo {author} {\bibfnamefont {H.}~\bibnamefont
  {Tagawa}}, \bibinfo {author} {\bibfnamefont {B.}~\bibnamefont {Kocsis}},
  \bibinfo {author} {\bibfnamefont {Z.}~\bibnamefont {Haiman}}, \bibinfo
  {author} {\bibfnamefont {I.}~\bibnamefont {Bartos}}, \bibinfo {author}
  {\bibfnamefont {K.}~\bibnamefont {Omukai}}, \ and\ \bibinfo {author}
  {\bibfnamefont {J.}~\bibnamefont {Samsing}},\ }\href {\doibase
  10.3847/2041-8213/abd4d3} {\bibfield  {journal} {\bibinfo  {journal}
  {Astrophys. J. Lett.}\ }\textbf {\bibinfo {volume} {907}},\ \bibinfo {pages}
  {L20} (\bibinfo {year} {2021})},\ \Eprint {http://arxiv.org/abs/2010.10526}
  {arXiv:2010.10526 [astro-ph.HE]} \BibitemShut {NoStop}%
\bibitem [{\citenamefont {{Samsing}}\ \emph {et~al.}(2022)\citenamefont
  {{Samsing}}, \citenamefont {{Bartos}}, \citenamefont {{D'Orazio}},
  \citenamefont {{Haiman}}, \citenamefont {{Kocsis}}, \citenamefont {{Leigh}},
  \citenamefont {{Liu}}, \citenamefont {{Pessah}},\ and\ \citenamefont
  {{Tagawa}}}]{Samsing:2020tda}%
  \BibitemOpen
  \bibfield  {author} {\bibinfo {author} {\bibfnamefont {J.}~\bibnamefont
  {{Samsing}}}, \bibinfo {author} {\bibfnamefont {I.}~\bibnamefont {{Bartos}}},
  \bibinfo {author} {\bibfnamefont {D.~J.}\ \bibnamefont {{D'Orazio}}},
  \bibinfo {author} {\bibfnamefont {Z.}~\bibnamefont {{Haiman}}}, \bibinfo
  {author} {\bibfnamefont {B.}~\bibnamefont {{Kocsis}}}, \bibinfo {author}
  {\bibfnamefont {N.~W.~C.}\ \bibnamefont {{Leigh}}}, \bibinfo {author}
  {\bibfnamefont {B.}~\bibnamefont {{Liu}}}, \bibinfo {author} {\bibfnamefont
  {M.~E.}\ \bibnamefont {{Pessah}}}, \ and\ \bibinfo {author} {\bibfnamefont
  {H.}~\bibnamefont {{Tagawa}}},\ }\href {\doibase 10.1038/s41586-021-04333-1}
  {\bibfield  {journal} {\bibinfo  {journal} {Nature}\ }\textbf {\bibinfo
  {volume} {603}},\ \bibinfo {pages} {237} (\bibinfo {year} {2022})},\ \Eprint
  {http://arxiv.org/abs/2010.09765} {arXiv:2010.09765 [astro-ph.HE]}
  \BibitemShut {NoStop}%
\bibitem [{\citenamefont {{Zevin}}\ \emph
  {et~al.}(2021{\natexlab{b}})\citenamefont {{Zevin}}, \citenamefont
  {{Romero-Shaw}}, \citenamefont {{Kremer}}, \citenamefont {{Thrane}},\ and\
  \citenamefont {{Lasky}}}]{Zevin:2021:seleccentricity}%
  \BibitemOpen
  \bibfield  {author} {\bibinfo {author} {\bibfnamefont {M.}~\bibnamefont
  {{Zevin}}}, \bibinfo {author} {\bibfnamefont {I.~M.}\ \bibnamefont
  {{Romero-Shaw}}}, \bibinfo {author} {\bibfnamefont {K.}~\bibnamefont
  {{Kremer}}}, \bibinfo {author} {\bibfnamefont {E.}~\bibnamefont {{Thrane}}},
  \ and\ \bibinfo {author} {\bibfnamefont {P.~D.}\ \bibnamefont {{Lasky}}},\
  }\href {\doibase 10.3847/2041-8213/ac32dc} {\bibfield  {journal} {\bibinfo
  {journal} {Astrophys. J. Lett.}\ }\textbf {\bibinfo {volume} {921}},\
  \bibinfo {eid} {L43} (\bibinfo {year} {2021}{\natexlab{b}})},\ \Eprint
  {http://arxiv.org/abs/2106.09042} {arXiv:2106.09042 [astro-ph.HE]}
  \BibitemShut {NoStop}%
\bibitem [{\citenamefont {{Gond{\'a}n}}\ and\ \citenamefont
  {{Kocsis}}(2021)}]{GondanKocsis2021}%
  \BibitemOpen
  \bibfield  {author} {\bibinfo {author} {\bibfnamefont {L.}~\bibnamefont
  {{Gond{\'a}n}}}\ and\ \bibinfo {author} {\bibfnamefont {B.}~\bibnamefont
  {{Kocsis}}},\ }\href {\doibase 10.1093/mnras/stab1722} {\bibfield  {journal}
  {\bibinfo  {journal} {Mon. Not. Roy. Astron. Soc.}\ } (\bibinfo {year}
  {2021}),\ 10.1093/mnras/stab1722},\ \Eprint {http://arxiv.org/abs/2011.02507}
  {arXiv:2011.02507 [astro-ph.HE]} \BibitemShut {NoStop}%
\bibitem [{\citenamefont {{Romero-Shaw}}\ \emph {et~al.}(2022)\citenamefont
  {{Romero-Shaw}}, \citenamefont {{Lasky}},\ and\ \citenamefont
  {{Thrane}}}]{Romero-Shaw:2022:FourEccentricMergers}%
  \BibitemOpen
  \bibfield  {author} {\bibinfo {author} {\bibfnamefont {I.}~\bibnamefont
  {{Romero-Shaw}}}, \bibinfo {author} {\bibfnamefont {P.~D.}\ \bibnamefont
  {{Lasky}}}, \ and\ \bibinfo {author} {\bibfnamefont {E.}~\bibnamefont
  {{Thrane}}},\ }\href {\doibase 10.3847/1538-4357/ac9798} {\bibfield
  {journal} {\bibinfo  {journal} {Astrophys. J. Lett.}\ }\textbf {\bibinfo
  {volume} {940}},\ \bibinfo {eid} {171} (\bibinfo {year} {2022})},\ \Eprint
  {http://arxiv.org/abs/2206.14695} {arXiv:2206.14695 [astro-ph.HE]}
  \BibitemShut {NoStop}%
\bibitem [{\citenamefont {{Lower}}\ \emph {et~al.}(2018)\citenamefont
  {{Lower}}, \citenamefont {{Thrane}}, \citenamefont {{Lasky}},\ and\
  \citenamefont {{Smith}}}]{Lower:2018:eccentricity}%
  \BibitemOpen
  \bibfield  {author} {\bibinfo {author} {\bibfnamefont {M.~E.}\ \bibnamefont
  {{Lower}}}, \bibinfo {author} {\bibfnamefont {E.}~\bibnamefont {{Thrane}}},
  \bibinfo {author} {\bibfnamefont {P.~D.}\ \bibnamefont {{Lasky}}}, \ and\
  \bibinfo {author} {\bibfnamefont {R.}~\bibnamefont {{Smith}}},\ }\href
  {\doibase 10.1103/PhysRevD.98.083028} {\bibfield  {journal} {\bibinfo
  {journal} {\prd}\ }\textbf {\bibinfo {volume} {98}},\ \bibinfo {eid} {083028}
  (\bibinfo {year} {2018})},\ \Eprint {http://arxiv.org/abs/1806.05350}
  {arXiv:1806.05350 [astro-ph.HE]} \BibitemShut {NoStop}%
\bibitem [{\citenamefont {{Romero-Shaw}}\ \emph {et~al.}(2019)\citenamefont
  {{Romero-Shaw}}, \citenamefont {{Lasky}},\ and\ \citenamefont
  {{Thrane}}}]{Romero-Shaw:2019:GWTC-1-ecc}%
  \BibitemOpen
  \bibfield  {author} {\bibinfo {author} {\bibfnamefont {I.~M.}\ \bibnamefont
  {{Romero-Shaw}}}, \bibinfo {author} {\bibfnamefont {P.~D.}\ \bibnamefont
  {{Lasky}}}, \ and\ \bibinfo {author} {\bibfnamefont {E.}~\bibnamefont
  {{Thrane}}},\ }\href {\doibase 10.1093/mnras/stz2996} {\bibfield  {journal}
  {\bibinfo  {journal} {Mon. Not. Roy. Astron. Soc.}\ }\textbf {\bibinfo
  {volume} {490}},\ \bibinfo {pages} {5210} (\bibinfo {year} {2019})},\ \Eprint
  {http://arxiv.org/abs/1909.05466} {arXiv:1909.05466 [astro-ph.HE]}
  \BibitemShut {NoStop}%
\bibitem [{\citenamefont {{Romero-Shaw}}\ \emph
  {et~al.}(2021{\natexlab{a}})\citenamefont {{Romero-Shaw}}, \citenamefont
  {{Lasky}},\ and\ \citenamefont {{Thrane}}}]{Romero-Shaw:2021:GWTC-2-ecc}%
  \BibitemOpen
  \bibfield  {author} {\bibinfo {author} {\bibfnamefont {I.}~\bibnamefont
  {{Romero-Shaw}}}, \bibinfo {author} {\bibfnamefont {P.~D.}\ \bibnamefont
  {{Lasky}}}, \ and\ \bibinfo {author} {\bibfnamefont {E.}~\bibnamefont
  {{Thrane}}},\ }\href {\doibase 10.3847/2041-8213/ac3138} {\bibfield
  {journal} {\bibinfo  {journal} {Astrophys. J. Lett.}\ }\textbf {\bibinfo
  {volume} {921}},\ \bibinfo {eid} {L31} (\bibinfo {year}
  {2021}{\natexlab{a}})},\ \Eprint {http://arxiv.org/abs/2108.01284}
  {arXiv:2108.01284 [astro-ph.HE]} \BibitemShut {NoStop}%
\bibitem [{\citenamefont {{Maggiore}}\ \emph {et~al.}(2020)\citenamefont
  {{Maggiore}}, \citenamefont {{Van Den Broeck}}, \citenamefont {{Bartolo}},
  \citenamefont {{Belgacem}}, \citenamefont {{Bertacca}}, \citenamefont
  {{Bizouard}}, \citenamefont {{Branchesi}}, \citenamefont {{Clesse}},
  \citenamefont {{Foffa}}, \citenamefont {{Garc{\'\i}a-Bellido}}, \citenamefont
  {{Grimm}}, \citenamefont {{Harms}}, \citenamefont {{Hinderer}}, \citenamefont
  {{Matarrese}}, \citenamefont {{Palomba}}, \citenamefont {{Peloso}},
  \citenamefont {{Ricciardone}},\ and\ \citenamefont
  {{Sakellariadou}}}]{Maggiore:2020:ET}%
  \BibitemOpen
  \bibfield  {author} {\bibinfo {author} {\bibfnamefont {M.}~\bibnamefont
  {{Maggiore}}}, \bibinfo {author} {\bibfnamefont {C.}~\bibnamefont {{Van Den
  Broeck}}}, \bibinfo {author} {\bibfnamefont {N.}~\bibnamefont {{Bartolo}}},
  \bibinfo {author} {\bibfnamefont {E.}~\bibnamefont {{Belgacem}}}, \bibinfo
  {author} {\bibfnamefont {D.}~\bibnamefont {{Bertacca}}}, \bibinfo {author}
  {\bibfnamefont {M.~A.}\ \bibnamefont {{Bizouard}}}, \bibinfo {author}
  {\bibfnamefont {M.}~\bibnamefont {{Branchesi}}}, \bibinfo {author}
  {\bibfnamefont {S.}~\bibnamefont {{Clesse}}}, \bibinfo {author}
  {\bibfnamefont {S.}~\bibnamefont {{Foffa}}}, \bibinfo {author} {\bibfnamefont
  {J.}~\bibnamefont {{Garc{\'\i}a-Bellido}}}, \bibinfo {author} {\bibfnamefont
  {S.}~\bibnamefont {{Grimm}}}, \bibinfo {author} {\bibfnamefont
  {J.}~\bibnamefont {{Harms}}}, \bibinfo {author} {\bibfnamefont
  {T.}~\bibnamefont {{Hinderer}}}, \bibinfo {author} {\bibfnamefont
  {S.}~\bibnamefont {{Matarrese}}}, \bibinfo {author} {\bibfnamefont
  {C.}~\bibnamefont {{Palomba}}}, \bibinfo {author} {\bibfnamefont
  {M.}~\bibnamefont {{Peloso}}}, \bibinfo {author} {\bibfnamefont
  {A.}~\bibnamefont {{Ricciardone}}}, \ and\ \bibinfo {author} {\bibfnamefont
  {M.}~\bibnamefont {{Sakellariadou}}},\ }\href {\doibase
  10.1088/1475-7516/2020/03/050} {\bibfield  {journal} {\bibinfo  {journal}
  {jcap}\ }\textbf {\bibinfo {volume} {2020}},\ \bibinfo {eid} {050} (\bibinfo
  {year} {2020})},\ \Eprint {http://arxiv.org/abs/1912.02622} {arXiv:1912.02622
  [astro-ph.CO]} \BibitemShut {NoStop}%
\bibitem [{\citenamefont {{Evans}}\ \emph {et~al.}(2021)\citenamefont
  {{Evans}}, \citenamefont {{Adhikari}}, \citenamefont {{Afle}}, \citenamefont
  {{Ballmer}}, \citenamefont {{Biscoveanu}}, \citenamefont {{Borhanian}},
  \citenamefont {{Brown}}, \citenamefont {{Chen}}, \citenamefont
  {{Eisenstein}}, \citenamefont {{Gruson}}, \citenamefont {{Gupta}},
  \citenamefont {{Hall}}, \citenamefont {{Huxford}}, \citenamefont {{Kamai}},
  \citenamefont {{Kashyap}}, \citenamefont {{Kissel}}, \citenamefont {{Kuns}},
  \citenamefont {{Landry}}, \citenamefont {{Lenon}}, \citenamefont
  {{Lovelace}}, \citenamefont {{McCuller}}, \citenamefont {{Ng}}, \citenamefont
  {{Nitz}}, \citenamefont {{Read}}, \citenamefont {{Sathyaprakash}},
  \citenamefont {{Shoemaker}}, \citenamefont {{Slagmolen}}, \citenamefont
  {{Smith}}, \citenamefont {{Srivastava}}, \citenamefont {{Sun}}, \citenamefont
  {{Vitale}},\ and\ \citenamefont {{Weiss}}}]{Evans:2021:CosmicExplorer}%
  \BibitemOpen
  \bibfield  {author} {\bibinfo {author} {\bibfnamefont {M.}~\bibnamefont
  {{Evans}}}, \bibinfo {author} {\bibfnamefont {R.~X.}\ \bibnamefont
  {{Adhikari}}}, \bibinfo {author} {\bibfnamefont {C.}~\bibnamefont {{Afle}}},
  \bibinfo {author} {\bibfnamefont {S.~W.}\ \bibnamefont {{Ballmer}}}, \bibinfo
  {author} {\bibfnamefont {S.}~\bibnamefont {{Biscoveanu}}}, \bibinfo {author}
  {\bibfnamefont {S.}~\bibnamefont {{Borhanian}}}, \bibinfo {author}
  {\bibfnamefont {D.~A.}\ \bibnamefont {{Brown}}}, \bibinfo {author}
  {\bibfnamefont {Y.}~\bibnamefont {{Chen}}}, \bibinfo {author} {\bibfnamefont
  {R.}~\bibnamefont {{Eisenstein}}}, \bibinfo {author} {\bibfnamefont
  {A.}~\bibnamefont {{Gruson}}}, \bibinfo {author} {\bibfnamefont
  {A.}~\bibnamefont {{Gupta}}}, \bibinfo {author} {\bibfnamefont {E.~D.}\
  \bibnamefont {{Hall}}}, \bibinfo {author} {\bibfnamefont {R.}~\bibnamefont
  {{Huxford}}}, \bibinfo {author} {\bibfnamefont {B.}~\bibnamefont {{Kamai}}},
  \bibinfo {author} {\bibfnamefont {R.}~\bibnamefont {{Kashyap}}}, \bibinfo
  {author} {\bibfnamefont {J.~S.}\ \bibnamefont {{Kissel}}}, \bibinfo {author}
  {\bibfnamefont {K.}~\bibnamefont {{Kuns}}}, \bibinfo {author} {\bibfnamefont
  {P.}~\bibnamefont {{Landry}}}, \bibinfo {author} {\bibfnamefont
  {A.}~\bibnamefont {{Lenon}}}, \bibinfo {author} {\bibfnamefont
  {G.}~\bibnamefont {{Lovelace}}}, \bibinfo {author} {\bibfnamefont
  {L.}~\bibnamefont {{McCuller}}}, \bibinfo {author} {\bibfnamefont {K.~K.~Y.}\
  \bibnamefont {{Ng}}}, \bibinfo {author} {\bibfnamefont {A.~H.}\ \bibnamefont
  {{Nitz}}}, \bibinfo {author} {\bibfnamefont {J.}~\bibnamefont {{Read}}},
  \bibinfo {author} {\bibfnamefont {B.~S.}\ \bibnamefont {{Sathyaprakash}}},
  \bibinfo {author} {\bibfnamefont {D.~H.}\ \bibnamefont {{Shoemaker}}},
  \bibinfo {author} {\bibfnamefont {B.~J.~J.}\ \bibnamefont {{Slagmolen}}},
  \bibinfo {author} {\bibfnamefont {J.~R.}\ \bibnamefont {{Smith}}}, \bibinfo
  {author} {\bibfnamefont {V.}~\bibnamefont {{Srivastava}}}, \bibinfo {author}
  {\bibfnamefont {L.}~\bibnamefont {{Sun}}}, \bibinfo {author} {\bibfnamefont
  {S.}~\bibnamefont {{Vitale}}}, \ and\ \bibinfo {author} {\bibfnamefont
  {R.}~\bibnamefont {{Weiss}}},\ }\href@noop {} {\bibfield  {journal} {\bibinfo
   {journal} {arXiv e-prints}\ ,\ \bibinfo {eid} {arXiv:2109.09882}} (\bibinfo
  {year} {2021})},\ \Eprint {http://arxiv.org/abs/2109.09882} {arXiv:2109.09882
  [astro-ph.IM]} \BibitemShut {NoStop}%
\bibitem [{\citenamefont {{Romero-Shaw}}\ \emph
  {et~al.}(2021{\natexlab{b}})\citenamefont {{Romero-Shaw}}, \citenamefont
  {{Kremer}}, \citenamefont {{Lasky}}, \citenamefont {{Thrane}},\ and\
  \citenamefont {{Samsing}}}]{Romero-Shaw:2021:GCs}%
  \BibitemOpen
  \bibfield  {author} {\bibinfo {author} {\bibfnamefont {I.~M.}\ \bibnamefont
  {{Romero-Shaw}}}, \bibinfo {author} {\bibfnamefont {K.}~\bibnamefont
  {{Kremer}}}, \bibinfo {author} {\bibfnamefont {P.~D.}\ \bibnamefont
  {{Lasky}}}, \bibinfo {author} {\bibfnamefont {E.}~\bibnamefont {{Thrane}}}, \
  and\ \bibinfo {author} {\bibfnamefont {J.}~\bibnamefont {{Samsing}}},\ }\href
  {\doibase 10.1093/mnras/stab1815} {\bibfield  {journal} {\bibinfo  {journal}
  {Mon. Not. Roy. Astron. Soc.}\ }\textbf {\bibinfo {volume} {506}},\ \bibinfo
  {pages} {2362} (\bibinfo {year} {2021}{\natexlab{b}})},\ \Eprint
  {http://arxiv.org/abs/2011.14541} {arXiv:2011.14541 [astro-ph.HE]}
  \BibitemShut {NoStop}%
\bibitem [{\citenamefont {{Turner}}(1977)}]{Turner}%
  \BibitemOpen
  \bibfield  {author} {\bibinfo {author} {\bibfnamefont {M.}~\bibnamefont
  {{Turner}}},\ }\href {\doibase 10.1086/155501} {\bibfield  {journal}
  {\bibinfo  {journal} {\apj}\ }\textbf {\bibinfo {volume} {216}},\ \bibinfo
  {pages} {610} (\bibinfo {year} {1977})}\BibitemShut {NoStop}%
\bibitem [{\citenamefont {Loutrel}(2021)}]{Loutrel:2020jfx}%
  \BibitemOpen
  \bibfield  {author} {\bibinfo {author} {\bibfnamefont {N.}~\bibnamefont
  {Loutrel}},\ }\href {\doibase 10.1088/1361-6382/abc3f6} {\bibfield  {journal}
  {\bibinfo  {journal} {Class. Quant. Grav.}\ }\textbf {\bibinfo {volume}
  {38}},\ \bibinfo {pages} {015005} (\bibinfo {year} {2021})},\ \Eprint
  {http://arxiv.org/abs/2003.13673} {arXiv:2003.13673 [gr-qc]} \BibitemShut
  {NoStop}%
\bibitem [{\citenamefont {Loutrel}(2020)}]{Loutrel:2019kky}%
  \BibitemOpen
  \bibfield  {author} {\bibinfo {author} {\bibfnamefont {N.}~\bibnamefont
  {Loutrel}},\ }\href {\doibase 10.1088/1361-6382/ab745f} {\bibfield  {journal}
  {\bibinfo  {journal} {Class. Quant. Grav.}\ }\textbf {\bibinfo {volume}
  {37}},\ \bibinfo {pages} {075008} (\bibinfo {year} {2020})},\ \Eprint
  {http://arxiv.org/abs/1909.02143} {arXiv:1909.02143 [gr-qc]} \BibitemShut
  {NoStop}%
\bibitem [{\citenamefont {{Nagar}}\ \emph {et~al.}(2021)\citenamefont
  {{Nagar}}, \citenamefont {{Rettegno}}, \citenamefont {{Gamba}},\ and\
  \citenamefont {{Bernuzzi}}}]{Nagar:2021:EccentricWaveform}%
  \BibitemOpen
  \bibfield  {author} {\bibinfo {author} {\bibfnamefont {A.}~\bibnamefont
  {{Nagar}}}, \bibinfo {author} {\bibfnamefont {P.}~\bibnamefont {{Rettegno}}},
  \bibinfo {author} {\bibfnamefont {R.}~\bibnamefont {{Gamba}}}, \ and\
  \bibinfo {author} {\bibfnamefont {S.}~\bibnamefont {{Bernuzzi}}},\ }\href
  {\doibase 10.1103/PhysRevD.103.064013} {\bibfield  {journal} {\bibinfo
  {journal} {\prd}\ }\textbf {\bibinfo {volume} {103}},\ \bibinfo {eid}
  {064013} (\bibinfo {year} {2021})},\ \Eprint
  {http://arxiv.org/abs/2009.12857} {arXiv:2009.12857 [gr-qc]} \BibitemShut
  {NoStop}%
\bibitem [{\citenamefont {{Kozai}}(1962)}]{Kozai}%
  \BibitemOpen
  \bibfield  {author} {\bibinfo {author} {\bibfnamefont {Y.}~\bibnamefont
  {{Kozai}}},\ }\href {\doibase 10.1086/108790} {\bibfield  {journal} {\bibinfo
   {journal} {\aj}\ }\textbf {\bibinfo {volume} {67}},\ \bibinfo {pages} {591}
  (\bibinfo {year} {1962})}\BibitemShut {NoStop}%
\bibitem [{\citenamefont {Lidov}(1962)}]{Lidov}%
  \BibitemOpen
  \bibfield  {author} {\bibinfo {author} {\bibfnamefont {M.}~\bibnamefont
  {Lidov}},\ }\href {\doibase https://doi.org/10.1016/0032-0633(62)90129-0}
  {\bibfield  {journal} {\bibinfo  {journal} {Planetary and Space Science}\
  }\textbf {\bibinfo {volume} {9}},\ \bibinfo {pages} {719} (\bibinfo {year}
  {1962})}\BibitemShut {NoStop}%
\bibitem [{\citenamefont {Naoz}\ \emph {et~al.}(2013)\citenamefont {Naoz},
  \citenamefont {Kocsis}, \citenamefont {Loeb},\ and\ \citenamefont
  {Yunes}}]{Naoz:2012bx}%
  \BibitemOpen
  \bibfield  {author} {\bibinfo {author} {\bibfnamefont {S.}~\bibnamefont
  {Naoz}}, \bibinfo {author} {\bibfnamefont {B.}~\bibnamefont {Kocsis}},
  \bibinfo {author} {\bibfnamefont {A.}~\bibnamefont {Loeb}}, \ and\ \bibinfo
  {author} {\bibfnamefont {N.}~\bibnamefont {Yunes}},\ }\href {\doibase
  10.1088/0004-637X/773/2/187} {\bibfield  {journal} {\bibinfo  {journal}
  {Astrophys. J.}\ }\textbf {\bibinfo {volume} {773}},\ \bibinfo {pages} {187}
  (\bibinfo {year} {2013})},\ \Eprint {http://arxiv.org/abs/1206.4316}
  {arXiv:1206.4316 [astro-ph.SR]} \BibitemShut {NoStop}%
\bibitem [{\citenamefont {Naoz}(2016)}]{Naoz:2016}%
  \BibitemOpen
  \bibfield  {author} {\bibinfo {author} {\bibfnamefont {S.}~\bibnamefont
  {Naoz}},\ }\href {\doibase 10.1146/annurev-astro-081915-023315} {\bibfield
  {journal} {\bibinfo  {journal} {Annual Review of Astronomy and Astrophysics}\
  }\textbf {\bibinfo {volume} {54}},\ \bibinfo {pages} {441} (\bibinfo {year}
  {2016})},\ \Eprint
  {http://arxiv.org/abs/https://doi.org/10.1146/annurev-astro-081915-023315}
  {https://doi.org/10.1146/annurev-astro-081915-023315} \BibitemShut {NoStop}%
\bibitem [{\citenamefont {Silsbee}\ and\ \citenamefont
  {Tremaine}(2017)}]{Silsbee16}%
  \BibitemOpen
  \bibfield  {author} {\bibinfo {author} {\bibfnamefont {K.}~\bibnamefont
  {Silsbee}}\ and\ \bibinfo {author} {\bibfnamefont {S.}~\bibnamefont
  {Tremaine}},\ }\href {\doibase 10.3847/1538-4357/aa5729} {\bibfield
  {journal} {\bibinfo  {journal} {Astrophys. J.}\ }\textbf {\bibinfo {volume}
  {836}},\ \bibinfo {pages} {39} (\bibinfo {year} {2017})},\ \Eprint
  {http://arxiv.org/abs/1608.07642} {arXiv:1608.07642 [astro-ph.HE]}
  \BibitemShut {NoStop}%
\bibitem [{\citenamefont {Antonini}\ \emph {et~al.}(2017)\citenamefont
  {Antonini}, \citenamefont {Toonen},\ and\ \citenamefont
  {Hamers}}]{Antonini17}%
  \BibitemOpen
  \bibfield  {author} {\bibinfo {author} {\bibfnamefont {F.}~\bibnamefont
  {Antonini}}, \bibinfo {author} {\bibfnamefont {S.}~\bibnamefont {Toonen}}, \
  and\ \bibinfo {author} {\bibfnamefont {A.~S.}\ \bibnamefont {Hamers}},\
  }\href {\doibase 10.3847/1538-4357/aa6f5e} {\bibfield  {journal} {\bibinfo
  {journal} {Astrophys. J.}\ }\textbf {\bibinfo {volume} {841}},\ \bibinfo
  {pages} {77} (\bibinfo {year} {2017})},\ \Eprint
  {http://arxiv.org/abs/1703.06614} {arXiv:1703.06614} \BibitemShut {NoStop}%
\bibitem [{\citenamefont {{Liu}}\ and\ \citenamefont {{Lai}}(2017)}]{LiuLai17}%
  \BibitemOpen
  \bibfield  {author} {\bibinfo {author} {\bibfnamefont {B.}~\bibnamefont
  {{Liu}}}\ and\ \bibinfo {author} {\bibfnamefont {D.}~\bibnamefont {{Lai}}},\
  }\href {\doibase 10.3847/2041-8213/aa8727} {\bibfield  {journal} {\bibinfo
  {journal} {Astrophys. J. Lett.}\ }\textbf {\bibinfo {volume} {846}},\
  \bibinfo {eid} {L11} (\bibinfo {year} {2017})},\ \Eprint
  {http://arxiv.org/abs/1706.02309} {arXiv:1706.02309 [astro-ph.HE]}
  \BibitemShut {NoStop}%
\bibitem [{\citenamefont {Randall}\ and\ \citenamefont
  {Xianyu}(2018)}]{Randall18}%
  \BibitemOpen
  \bibfield  {author} {\bibinfo {author} {\bibfnamefont {L.}~\bibnamefont
  {Randall}}\ and\ \bibinfo {author} {\bibfnamefont {Z.-Z.}\ \bibnamefont
  {Xianyu}},\ }\href {\doibase 10.3847/1538-4357/aad7fe} {\bibfield  {journal}
  {\bibinfo  {journal} {Astrophys. J.}\ }\textbf {\bibinfo {volume} {864}},\
  \bibinfo {pages} {134} (\bibinfo {year} {2018})},\ \Eprint
  {http://arxiv.org/abs/1802.05718} {arXiv:1802.05718} \BibitemShut {NoStop}%
\bibitem [{\citenamefont {{Liu}}\ \emph {et~al.}(2019)\citenamefont {{Liu}},
  \citenamefont {{Lai}},\ and\ \citenamefont {{Wang}}}]{Liu19}%
  \BibitemOpen
  \bibfield  {author} {\bibinfo {author} {\bibfnamefont {B.}~\bibnamefont
  {{Liu}}}, \bibinfo {author} {\bibfnamefont {D.}~\bibnamefont {{Lai}}}, \ and\
  \bibinfo {author} {\bibfnamefont {Y.-H.}\ \bibnamefont {{Wang}}},\ }\href
  {\doibase 10.3847/1538-4357/ab2dfb} {\bibfield  {journal} {\bibinfo
  {journal} {Astrophys. J. Lett.}\ }\textbf {\bibinfo {volume} {881}},\
  \bibinfo {eid} {41} (\bibinfo {year} {2019})},\ \Eprint
  {http://arxiv.org/abs/1905.00427} {arXiv:1905.00427 [astro-ph.HE]}
  \BibitemShut {NoStop}%
\bibitem [{\citenamefont {{Li}}\ \emph {et~al.}(2022)\citenamefont {{Li}},
  \citenamefont {{Lai}},\ and\ \citenamefont {{Rodet}}}]{Li:2022:AGNBBH}%
  \BibitemOpen
  \bibfield  {author} {\bibinfo {author} {\bibfnamefont {J.}~\bibnamefont
  {{Li}}}, \bibinfo {author} {\bibfnamefont {D.}~\bibnamefont {{Lai}}}, \ and\
  \bibinfo {author} {\bibfnamefont {L.}~\bibnamefont {{Rodet}}},\ }\href
  {\doibase 10.3847/1538-4357/ac7c0d} {\bibfield  {journal} {\bibinfo
  {journal} {\apj}\ }\textbf {\bibinfo {volume} {934}},\ \bibinfo {eid} {154}
  (\bibinfo {year} {2022})},\ \Eprint {http://arxiv.org/abs/2203.05584}
  {arXiv:2203.05584 [astro-ph.HE]} \BibitemShut {NoStop}%
\bibitem [{\citenamefont {{Gond{\'a}n}}(2023)}]{Gondan:2022:AGN}%
  \BibitemOpen
  \bibfield  {author} {\bibinfo {author} {\bibfnamefont {L.}~\bibnamefont
  {{Gond{\'a}n}}},\ }\href {\doibase 10.1093/mnras/stac3612} {\bibfield
  {journal} {\bibinfo  {journal} {Mon. Not. Roy. Astron. Soc.}\ }\textbf
  {\bibinfo {volume} {519}},\ \bibinfo {pages} {1856} (\bibinfo {year}
  {2023})},\ \Eprint {http://arxiv.org/abs/2210.02975} {arXiv:2210.02975
  [astro-ph.HE]} \BibitemShut {NoStop}%
\bibitem [{\citenamefont {{Deme}}\ \emph {et~al.}(2020)\citenamefont {{Deme}},
  \citenamefont {{Hoang}}, \citenamefont {{Naoz}},\ and\ \citenamefont
  {{Kocsis}}}]{Deme:2022:IMBH}%
  \BibitemOpen
  \bibfield  {author} {\bibinfo {author} {\bibfnamefont {B.}~\bibnamefont
  {{Deme}}}, \bibinfo {author} {\bibfnamefont {B.-M.}\ \bibnamefont {{Hoang}}},
  \bibinfo {author} {\bibfnamefont {S.}~\bibnamefont {{Naoz}}}, \ and\ \bibinfo
  {author} {\bibfnamefont {B.}~\bibnamefont {{Kocsis}}},\ }\href {\doibase
  10.3847/1538-4357/abafa3} {\bibfield  {journal} {\bibinfo  {journal} {\apj}\
  }\textbf {\bibinfo {volume} {901}},\ \bibinfo {eid} {125} (\bibinfo {year}
  {2020})},\ \Eprint {http://arxiv.org/abs/2005.03677} {arXiv:2005.03677
  [astro-ph.HE]} \BibitemShut {NoStop}%
\bibitem [{\citenamefont {Samsing}\ and\ \citenamefont
  {D'Orazio}(2018)}]{Samsing:2018isx}%
  \BibitemOpen
  \bibfield  {author} {\bibinfo {author} {\bibfnamefont {J.}~\bibnamefont
  {Samsing}}\ and\ \bibinfo {author} {\bibfnamefont {D.~J.}\ \bibnamefont
  {D'Orazio}},\ }\href {\doibase 10.1093/mnras/sty2334} {\bibfield  {journal}
  {\bibinfo  {journal} {Mon. Not. Roy. Astron. Soc.}\ }\textbf {\bibinfo
  {volume} {481}},\ \bibinfo {pages} {5445} (\bibinfo {year} {2018})},\ \Eprint
  {http://arxiv.org/abs/1804.06519} {arXiv:1804.06519 [astro-ph.HE]}
  \BibitemShut {NoStop}%
\bibitem [{\citenamefont {Samsing}\ \emph
  {et~al.}(2018{\natexlab{b}})\citenamefont {Samsing}, \citenamefont
  {D'Orazio}, \citenamefont {Askar},\ and\ \citenamefont
  {Giersz}}]{Samsing:2018ykz}%
  \BibitemOpen
  \bibfield  {author} {\bibinfo {author} {\bibfnamefont {J.}~\bibnamefont
  {Samsing}}, \bibinfo {author} {\bibfnamefont {D.~J.}\ \bibnamefont
  {D'Orazio}}, \bibinfo {author} {\bibfnamefont {A.}~\bibnamefont {Askar}}, \
  and\ \bibinfo {author} {\bibfnamefont {M.}~\bibnamefont {Giersz}},\
  }\href@noop {} {\  (\bibinfo {year} {2018}{\natexlab{b}})},\ \Eprint
  {http://arxiv.org/abs/1802.08654} {arXiv:1802.08654 [astro-ph.HE]}
  \BibitemShut {NoStop}%
\bibitem [{\citenamefont {{Martinez}}\ \emph {et~al.}(2020)\citenamefont
  {{Martinez}}, \citenamefont {{Fragione}}, \citenamefont {{Kremer}},
  \citenamefont {{Chatterjee}}, \citenamefont {{Rodriguez}}, \citenamefont
  {{Samsing}}, \citenamefont {{Ye}}, \citenamefont {{Weatherford}},
  \citenamefont {{Zevin}}, \citenamefont {{Naoz}},\ and\ \citenamefont
  {{Rasio}}}]{Martinez:2020}%
  \BibitemOpen
  \bibfield  {author} {\bibinfo {author} {\bibfnamefont {M.~A.~S.}\
  \bibnamefont {{Martinez}}}, \bibinfo {author} {\bibfnamefont
  {G.}~\bibnamefont {{Fragione}}}, \bibinfo {author} {\bibfnamefont
  {K.}~\bibnamefont {{Kremer}}}, \bibinfo {author} {\bibfnamefont
  {S.}~\bibnamefont {{Chatterjee}}}, \bibinfo {author} {\bibfnamefont {C.~L.}\
  \bibnamefont {{Rodriguez}}}, \bibinfo {author} {\bibfnamefont
  {J.}~\bibnamefont {{Samsing}}}, \bibinfo {author} {\bibfnamefont {C.~S.}\
  \bibnamefont {{Ye}}}, \bibinfo {author} {\bibfnamefont {N.~C.}\ \bibnamefont
  {{Weatherford}}}, \bibinfo {author} {\bibfnamefont {M.}~\bibnamefont
  {{Zevin}}}, \bibinfo {author} {\bibfnamefont {S.}~\bibnamefont {{Naoz}}}, \
  and\ \bibinfo {author} {\bibfnamefont {F.~A.}\ \bibnamefont {{Rasio}}},\
  }\href {\doibase 10.3847/1538-4357/abba25} {\bibfield  {journal} {\bibinfo
  {journal} {Astrophys. J.}\ }\textbf {\bibinfo {volume} {903}},\ \bibinfo
  {eid} {67} (\bibinfo {year} {2020})},\ \Eprint
  {http://arxiv.org/abs/2009.08468} {arXiv:2009.08468 [astro-ph.GA]}
  \BibitemShut {NoStop}%
\bibitem [{\citenamefont {{Yu}}\ and\ \citenamefont
  {{Chen}}(2021)}]{YuChen:2021:SMBHinference}%
  \BibitemOpen
  \bibfield  {author} {\bibinfo {author} {\bibfnamefont {H.}~\bibnamefont
  {{Yu}}}\ and\ \bibinfo {author} {\bibfnamefont {Y.}~\bibnamefont {{Chen}}},\
  }\href {\doibase 10.1103/PhysRevLett.126.021101} {\bibfield  {journal}
  {\bibinfo  {journal} {\prl}\ }\textbf {\bibinfo {volume} {126}},\ \bibinfo
  {eid} {021101} (\bibinfo {year} {2021})},\ \Eprint
  {http://arxiv.org/abs/2009.02579} {arXiv:2009.02579 [gr-qc]} \BibitemShut
  {NoStop}%
\bibitem [{\citenamefont {Samsing}(2022)}]{Johan-doppler}%
  \BibitemOpen
  \bibfield  {author} {\bibinfo {author} {\bibfnamefont {J.}~\bibnamefont
  {Samsing}},\ }\href@noop {} {\bibfield  {journal} {\bibinfo  {journal} {{in
  preparation}}\ } (\bibinfo {year} {2022})}\BibitemShut {NoStop}%
\bibitem [{\citenamefont {Chandramouli}\ and\ \citenamefont
  {Yunes}(2022)}]{Chandramouli:2021kts}%
  \BibitemOpen
  \bibfield  {author} {\bibinfo {author} {\bibfnamefont {R.~S.}\ \bibnamefont
  {Chandramouli}}\ and\ \bibinfo {author} {\bibfnamefont {N.}~\bibnamefont
  {Yunes}},\ }\href {\doibase 10.1103/PhysRevD.105.064009} {\bibfield
  {journal} {\bibinfo  {journal} {Phys. Rev. D}\ }\textbf {\bibinfo {volume}
  {105}},\ \bibinfo {pages} {064009} (\bibinfo {year} {2022})},\ \Eprint
  {http://arxiv.org/abs/2107.00741} {arXiv:2107.00741 [gr-qc]} \BibitemShut
  {NoStop}%
\bibitem [{\citenamefont {{Xuan}}\ \emph {et~al.}(2023)\citenamefont {{Xuan}},
  \citenamefont {{Naoz}},\ and\ \citenamefont {{Chen}}}]{Xuan:2022qkw}%
  \BibitemOpen
  \bibfield  {author} {\bibinfo {author} {\bibfnamefont {Z.}~\bibnamefont
  {{Xuan}}}, \bibinfo {author} {\bibfnamefont {S.}~\bibnamefont {{Naoz}}}, \
  and\ \bibinfo {author} {\bibfnamefont {X.}~\bibnamefont {{Chen}}},\ }\href
  {\doibase 10.1103/PhysRevD.107.043009} {\bibfield  {journal} {\bibinfo
  {journal} {Phys. Rev. D.}\ }\textbf {\bibinfo {volume} {107}},\ \bibinfo
  {eid} {043009} (\bibinfo {year} {2023})},\ \Eprint
  {http://arxiv.org/abs/2210.03129} {arXiv:2210.03129 [astro-ph.HE]}
  \BibitemShut {NoStop}%
\bibitem [{\citenamefont {Toubiana}\ \emph {et~al.}(2021)\citenamefont
  {Toubiana} \emph {et~al.}}]{Toubiana:2020drf}%
  \BibitemOpen
  \bibfield  {author} {\bibinfo {author} {\bibfnamefont {A.}~\bibnamefont
  {Toubiana}} \emph {et~al.},\ }\href {\doibase 10.1103/PhysRevLett.126.101105}
  {\bibfield  {journal} {\bibinfo  {journal} {Phys. Rev. Lett.}\ }\textbf
  {\bibinfo {volume} {126}},\ \bibinfo {pages} {101105} (\bibinfo {year}
  {2021})},\ \Eprint {http://arxiv.org/abs/2010.06056} {arXiv:2010.06056
  [astro-ph.HE]} \BibitemShut {NoStop}%
\bibitem [{\citenamefont {{Inayoshi}}\ \emph {et~al.}(2017)\citenamefont
  {{Inayoshi}}, \citenamefont {{Tamanini}}, \citenamefont {{Caprini}},\ and\
  \citenamefont {{Haiman}}}]{Inayoshi:2017:LISAThreeBody}%
  \BibitemOpen
  \bibfield  {author} {\bibinfo {author} {\bibfnamefont {K.}~\bibnamefont
  {{Inayoshi}}}, \bibinfo {author} {\bibfnamefont {N.}~\bibnamefont
  {{Tamanini}}}, \bibinfo {author} {\bibfnamefont {C.}~\bibnamefont
  {{Caprini}}}, \ and\ \bibinfo {author} {\bibfnamefont {Z.}~\bibnamefont
  {{Haiman}}},\ }\href {\doibase 10.1103/PhysRevD.96.063014} {\bibfield
  {journal} {\bibinfo  {journal} {\prd}\ }\textbf {\bibinfo {volume} {96}},\
  \bibinfo {eid} {063014} (\bibinfo {year} {2017})},\ \Eprint
  {http://arxiv.org/abs/1702.06529} {arXiv:1702.06529 [astro-ph.HE]}
  \BibitemShut {NoStop}%
\bibitem [{\citenamefont {{Randall}}\ and\ \citenamefont
  {{Xianyu}}(2019)}]{RandallXianyu:2019:BBHProbeTertiaryWithLISA}%
  \BibitemOpen
  \bibfield  {author} {\bibinfo {author} {\bibfnamefont {L.}~\bibnamefont
  {{Randall}}}\ and\ \bibinfo {author} {\bibfnamefont {Z.-Z.}\ \bibnamefont
  {{Xianyu}}},\ }\href {\doibase 10.3847/1538-4357/ab20c6} {\bibfield
  {journal} {\bibinfo  {journal} {apj}\ }\textbf {\bibinfo {volume} {878}},\
  \bibinfo {eid} {75} (\bibinfo {year} {2019})},\ \Eprint
  {http://arxiv.org/abs/1805.05335} {arXiv:1805.05335 [gr-qc]} \BibitemShut
  {NoStop}%
\bibitem [{\citenamefont {{Meiron}}\ \emph {et~al.}(2017)\citenamefont
  {{Meiron}}, \citenamefont {{Kocsis}},\ and\ \citenamefont
  {{Loeb}}}]{Meiron:2017:LISALIGOThreeBody}%
  \BibitemOpen
  \bibfield  {author} {\bibinfo {author} {\bibfnamefont {Y.}~\bibnamefont
  {{Meiron}}}, \bibinfo {author} {\bibfnamefont {B.}~\bibnamefont {{Kocsis}}},
  \ and\ \bibinfo {author} {\bibfnamefont {A.}~\bibnamefont {{Loeb}}},\ }\href
  {\doibase 10.3847/1538-4357/834/2/200} {\bibfield  {journal} {\bibinfo
  {journal} {\apj}\ }\textbf {\bibinfo {volume} {834}},\ \bibinfo {eid} {200}
  (\bibinfo {year} {2017})},\ \Eprint {http://arxiv.org/abs/1604.02148}
  {arXiv:1604.02148 [astro-ph.HE]} \BibitemShut {NoStop}%
\bibitem [{\citenamefont {Klimenko}\ \emph {et~al.}(2008)\citenamefont
  {Klimenko}, \citenamefont {Yakushin}, \citenamefont {Mercer},\ and\
  \citenamefont {Mitselmakher}}]{Klimenko:cWB:2008}%
  \BibitemOpen
  \bibfield  {author} {\bibinfo {author} {\bibfnamefont {S.}~\bibnamefont
  {Klimenko}}, \bibinfo {author} {\bibfnamefont {I.}~\bibnamefont {Yakushin}},
  \bibinfo {author} {\bibfnamefont {A.}~\bibnamefont {Mercer}}, \ and\ \bibinfo
  {author} {\bibfnamefont {G.}~\bibnamefont {Mitselmakher}},\ }\href {\doibase
  10.1088/0264-9381/25/11/114029} {\bibfield  {journal} {\bibinfo  {journal}
  {Class. Quant. Grav.}\ }\textbf {\bibinfo {volume} {25}},\ \bibinfo {pages}
  {114029} (\bibinfo {year} {2008})},\ \Eprint {http://arxiv.org/abs/0802.3232}
  {arXiv:0802.3232 [gr-qc]} \BibitemShut {NoStop}%
\bibitem [{\citenamefont {{Robinet}}\ \emph {et~al.}(2020)\citenamefont
  {{Robinet}}, \citenamefont {{Arnaud}}, \citenamefont {{Leroy}}, \citenamefont
  {{Lundgren}}, \citenamefont {{Macleod}},\ and\ \citenamefont
  {{McIver}}}]{Robinet:2020:Omicron}%
  \BibitemOpen
  \bibfield  {author} {\bibinfo {author} {\bibfnamefont {F.}~\bibnamefont
  {{Robinet}}}, \bibinfo {author} {\bibfnamefont {N.}~\bibnamefont {{Arnaud}}},
  \bibinfo {author} {\bibfnamefont {N.}~\bibnamefont {{Leroy}}}, \bibinfo
  {author} {\bibfnamefont {A.}~\bibnamefont {{Lundgren}}}, \bibinfo {author}
  {\bibfnamefont {D.}~\bibnamefont {{Macleod}}}, \ and\ \bibinfo {author}
  {\bibfnamefont {J.}~\bibnamefont {{McIver}}},\ }\href {\doibase
  10.1016/j.softx.2020.100620} {\bibfield  {journal} {\bibinfo  {journal}
  {SoftwareX}\ }\textbf {\bibinfo {volume} {12}},\ \bibinfo {eid} {100620}
  (\bibinfo {year} {2020})},\ \Eprint {http://arxiv.org/abs/2007.11374}
  {arXiv:2007.11374 [astro-ph.IM]} \BibitemShut {NoStop}%
\bibitem [{\citenamefont {Tai}\ \emph {et~al.}(2014)\citenamefont {Tai},
  \citenamefont {McWilliams},\ and\ \citenamefont {Pretorius}}]{Tai:2014bfa}%
  \BibitemOpen
  \bibfield  {author} {\bibinfo {author} {\bibfnamefont {K.~S.}\ \bibnamefont
  {Tai}}, \bibinfo {author} {\bibfnamefont {S.~T.}\ \bibnamefont {McWilliams}},
  \ and\ \bibinfo {author} {\bibfnamefont {F.}~\bibnamefont {Pretorius}},\
  }\href {\doibase 10.1103/PhysRevD.90.103001} {\bibfield  {journal} {\bibinfo
  {journal} {Phys. Rev. D}\ }\textbf {\bibinfo {volume} {90}},\ \bibinfo
  {pages} {103001} (\bibinfo {year} {2014})},\ \Eprint
  {http://arxiv.org/abs/1403.7754} {arXiv:1403.7754 [gr-qc]} \BibitemShut
  {NoStop}%
\bibitem [{\citenamefont {Cornish}\ and\ \citenamefont
  {Littenberg}(2015)}]{Cornish:2014kda}%
  \BibitemOpen
  \bibfield  {author} {\bibinfo {author} {\bibfnamefont {N.~J.}\ \bibnamefont
  {Cornish}}\ and\ \bibinfo {author} {\bibfnamefont {T.~B.}\ \bibnamefont
  {Littenberg}},\ }\href {\doibase 10.1088/0264-9381/32/13/135012} {\bibfield
  {journal} {\bibinfo  {journal} {Class. Quant. Grav.}\ }\textbf {\bibinfo
  {volume} {32}},\ \bibinfo {pages} {135012} (\bibinfo {year} {2015})},\
  \Eprint {http://arxiv.org/abs/1410.3835} {arXiv:1410.3835 [gr-qc]}
  \BibitemShut {NoStop}%
\bibitem [{\citenamefont {Cornish}\ \emph {et~al.}(2021)\citenamefont
  {Cornish}, \citenamefont {Littenberg}, \citenamefont {B\'ecsy}, \citenamefont
  {Chatziioannou}, \citenamefont {Clark}, \citenamefont {Ghonge},\ and\
  \citenamefont {Millhouse}}]{Cornish:2020dwh}%
  \BibitemOpen
  \bibfield  {author} {\bibinfo {author} {\bibfnamefont {N.~J.}\ \bibnamefont
  {Cornish}}, \bibinfo {author} {\bibfnamefont {T.~B.}\ \bibnamefont
  {Littenberg}}, \bibinfo {author} {\bibfnamefont {B.}~\bibnamefont {B\'ecsy}},
  \bibinfo {author} {\bibfnamefont {K.}~\bibnamefont {Chatziioannou}}, \bibinfo
  {author} {\bibfnamefont {J.~A.}\ \bibnamefont {Clark}}, \bibinfo {author}
  {\bibfnamefont {S.}~\bibnamefont {Ghonge}}, \ and\ \bibinfo {author}
  {\bibfnamefont {M.}~\bibnamefont {Millhouse}},\ }\href {\doibase
  10.1103/PhysRevD.103.044006} {\bibfield  {journal} {\bibinfo  {journal}
  {Phys. Rev. D}\ }\textbf {\bibinfo {volume} {103}},\ \bibinfo {pages}
  {044006} (\bibinfo {year} {2021})},\ \Eprint
  {http://arxiv.org/abs/2011.09494} {arXiv:2011.09494 [gr-qc]} \BibitemShut
  {NoStop}%
\bibitem [{\citenamefont {{Cheeseboro}}\ and\ \citenamefont
  {{Baker}}(2021)}]{CheeseboroBaker:2021:EccentricBurstSearch}%
  \BibitemOpen
  \bibfield  {author} {\bibinfo {author} {\bibfnamefont {B.~D.}\ \bibnamefont
  {{Cheeseboro}}}\ and\ \bibinfo {author} {\bibfnamefont {P.~T.}\ \bibnamefont
  {{Baker}}},\ }\href {\doibase 10.1103/PhysRevD.104.104016} {\bibfield
  {journal} {\bibinfo  {journal} {\prd}\ }\textbf {\bibinfo {volume} {104}},\
  \bibinfo {eid} {104016} (\bibinfo {year} {2021})},\ \Eprint
  {http://arxiv.org/abs/2108.01050} {arXiv:2108.01050 [gr-qc]} \BibitemShut
  {NoStop}%
\bibitem [{\citenamefont {Loutrel}\ \emph {et~al.}(2014)\citenamefont
  {Loutrel}, \citenamefont {Yunes},\ and\ \citenamefont
  {Pretorius}}]{Loutrel:2014vja}%
  \BibitemOpen
  \bibfield  {author} {\bibinfo {author} {\bibfnamefont {N.}~\bibnamefont
  {Loutrel}}, \bibinfo {author} {\bibfnamefont {N.}~\bibnamefont {Yunes}}, \
  and\ \bibinfo {author} {\bibfnamefont {F.}~\bibnamefont {Pretorius}},\ }\href
  {\doibase 10.1103/PhysRevD.90.104010} {\bibfield  {journal} {\bibinfo
  {journal} {Phys. Rev. D}\ }\textbf {\bibinfo {volume} {90}},\ \bibinfo
  {pages} {104010} (\bibinfo {year} {2014})},\ \Eprint
  {http://arxiv.org/abs/1404.0092} {arXiv:1404.0092 [gr-qc]} \BibitemShut
  {NoStop}%
\bibitem [{\citenamefont {Loutrel}\ and\ \citenamefont
  {Yunes}(2017)}]{Loutrel:2017fgu}%
  \BibitemOpen
  \bibfield  {author} {\bibinfo {author} {\bibfnamefont {N.}~\bibnamefont
  {Loutrel}}\ and\ \bibinfo {author} {\bibfnamefont {N.}~\bibnamefont
  {Yunes}},\ }\href {\doibase 10.1088/1361-6382/aa7449} {\bibfield  {journal}
  {\bibinfo  {journal} {Class. Quant. Grav.}\ }\textbf {\bibinfo {volume}
  {34}},\ \bibinfo {pages} {135011} (\bibinfo {year} {2017})},\ \Eprint
  {http://arxiv.org/abs/1702.01818} {arXiv:1702.01818 [gr-qc]} \BibitemShut
  {NoStop}%
\bibitem [{\citenamefont {Arredondo}\ and\ \citenamefont
  {Loutrel}(2021)}]{Arredondo:2021rdt}%
  \BibitemOpen
  \bibfield  {author} {\bibinfo {author} {\bibfnamefont {J.~N.}\ \bibnamefont
  {Arredondo}}\ and\ \bibinfo {author} {\bibfnamefont {N.}~\bibnamefont
  {Loutrel}},\ }\href {\doibase 10.1088/1361-6382/ac1083} {\bibfield  {journal}
  {\bibinfo  {journal} {Class. Quant. Grav.}\ }\textbf {\bibinfo {volume}
  {38}},\ \bibinfo {pages} {165001} (\bibinfo {year} {2021})},\ \Eprint
  {http://arxiv.org/abs/2101.10963} {arXiv:2101.10963 [gr-qc]} \BibitemShut
  {NoStop}%
\bibitem [{\citenamefont {Poisson}\ and\ \citenamefont
  {Will}(2014)}]{PoissonWill}%
  \BibitemOpen
  \bibfield  {author} {\bibinfo {author} {\bibfnamefont {E.}~\bibnamefont
  {Poisson}}\ and\ \bibinfo {author} {\bibfnamefont {C.}~\bibnamefont {Will}},\
  }\href@noop {} {\emph {\bibinfo {title} {{Gravity: Newtonian, Post-Newtonian,
  Relativistic}}}}\ (\bibinfo  {publisher} {Cambridge University Press},\
  \bibinfo {address} {Cambridge, UK},\ \bibinfo {year} {2014})\BibitemShut
  {NoStop}%
\bibitem [{\citenamefont {Lincoln}\ and\ \citenamefont
  {Will}(1990)}]{LincolnWill}%
  \BibitemOpen
  \bibfield  {author} {\bibinfo {author} {\bibfnamefont {C.~W.}\ \bibnamefont
  {Lincoln}}\ and\ \bibinfo {author} {\bibfnamefont {C.~M.}\ \bibnamefont
  {Will}},\ }\href {\doibase 10.1103/PhysRevD.42.1123} {\bibfield  {journal}
  {\bibinfo  {journal} {Phys. Rev. D}\ }\textbf {\bibinfo {volume} {42}},\
  \bibinfo {pages} {1123} (\bibinfo {year} {1990})}\BibitemShut {NoStop}%
\bibitem [{\citenamefont {Mora}\ and\ \citenamefont
  {Will}(2004)}]{Mora:2003wt}%
  \BibitemOpen
  \bibfield  {author} {\bibinfo {author} {\bibfnamefont {T.}~\bibnamefont
  {Mora}}\ and\ \bibinfo {author} {\bibfnamefont {C.~M.}\ \bibnamefont
  {Will}},\ }\href {\doibase 10.1103/PhysRevD.71.129901} {\bibfield  {journal}
  {\bibinfo  {journal} {Phys. Rev. D}\ }\textbf {\bibinfo {volume} {69}},\
  \bibinfo {pages} {104021} (\bibinfo {year} {2004})},\ \bibinfo {note}
  {[Erratum: Phys.Rev.D 71, 129901 (2005)]},\ \Eprint
  {http://arxiv.org/abs/gr-qc/0312082} {arXiv:gr-qc/0312082} \BibitemShut
  {NoStop}%
\bibitem [{\citenamefont {Will}\ and\ \citenamefont
  {Maitra}(2017)}]{Will:2016pgm}%
  \BibitemOpen
  \bibfield  {author} {\bibinfo {author} {\bibfnamefont {C.~M.}\ \bibnamefont
  {Will}}\ and\ \bibinfo {author} {\bibfnamefont {M.}~\bibnamefont {Maitra}},\
  }\href {\doibase 10.1103/PhysRevD.95.064003} {\bibfield  {journal} {\bibinfo
  {journal} {Phys. Rev. D}\ }\textbf {\bibinfo {volume} {95}},\ \bibinfo
  {pages} {064003} (\bibinfo {year} {2017})},\ \Eprint
  {http://arxiv.org/abs/1611.06931} {arXiv:1611.06931 [gr-qc]} \BibitemShut
  {NoStop}%
\bibitem [{\citenamefont {Konigsdorffer}\ and\ \citenamefont
  {Gopakumar}(2006)}]{Konigsdorffer:2006zt}%
  \BibitemOpen
  \bibfield  {author} {\bibinfo {author} {\bibfnamefont {C.}~\bibnamefont
  {Konigsdorffer}}\ and\ \bibinfo {author} {\bibfnamefont {A.}~\bibnamefont
  {Gopakumar}},\ }\href {\doibase 10.1103/PhysRevD.73.124012} {\bibfield
  {journal} {\bibinfo  {journal} {Phys. Rev. D}\ }\textbf {\bibinfo {volume}
  {73}},\ \bibinfo {pages} {124012} (\bibinfo {year} {2006})},\ \Eprint
  {http://arxiv.org/abs/gr-qc/0603056} {arXiv:gr-qc/0603056} \BibitemShut
  {NoStop}%
\bibitem [{\citenamefont {Loutrel}\ \emph {et~al.}(2019)\citenamefont
  {Loutrel}, \citenamefont {Liebersbach}, \citenamefont {Yunes},\ and\
  \citenamefont {Cornish}}]{Loutrel:2018ydu}%
  \BibitemOpen
  \bibfield  {author} {\bibinfo {author} {\bibfnamefont {N.}~\bibnamefont
  {Loutrel}}, \bibinfo {author} {\bibfnamefont {S.}~\bibnamefont
  {Liebersbach}}, \bibinfo {author} {\bibfnamefont {N.}~\bibnamefont {Yunes}},
  \ and\ \bibinfo {author} {\bibfnamefont {N.}~\bibnamefont {Cornish}},\ }\href
  {\doibase 10.1088/1361-6382/aaf2a9} {\bibfield  {journal} {\bibinfo
  {journal} {Class. Quant. Grav.}\ }\textbf {\bibinfo {volume} {36}},\ \bibinfo
  {pages} {025004} (\bibinfo {year} {2019})},\ \Eprint
  {http://arxiv.org/abs/1810.03521} {arXiv:1810.03521 [gr-qc]} \BibitemShut
  {NoStop}%
\bibitem [{\citenamefont {Pound}\ and\ \citenamefont
  {Poisson}(2008)}]{Pound:2007th}%
  \BibitemOpen
  \bibfield  {author} {\bibinfo {author} {\bibfnamefont {A.}~\bibnamefont
  {Pound}}\ and\ \bibinfo {author} {\bibfnamefont {E.}~\bibnamefont
  {Poisson}},\ }\href {\doibase 10.1103/PhysRevD.77.044013} {\bibfield
  {journal} {\bibinfo  {journal} {Phys. Rev. D}\ }\textbf {\bibinfo {volume}
  {77}},\ \bibinfo {pages} {044013} (\bibinfo {year} {2008})},\ \Eprint
  {http://arxiv.org/abs/0708.3033} {arXiv:0708.3033 [gr-qc]} \BibitemShut
  {NoStop}%
\bibitem [{\citenamefont {{McKernan}}\ \emph {et~al.}(2018)\citenamefont
  {{McKernan}}, \citenamefont {{Ford}}, \citenamefont {{Bellovary}},
  \citenamefont {{Leigh}}, \citenamefont {{Haiman}}, \citenamefont {{Kocsis}},
  \citenamefont {{Lyra}}, \citenamefont {{Mac Low}}, \citenamefont {{Metzger}},
  \citenamefont {{O'Dowd}}, \citenamefont {{Endlich}},\ and\ \citenamefont
  {{Rosen}}}]{McKernan:2018:AGNLIGO}%
  \BibitemOpen
  \bibfield  {author} {\bibinfo {author} {\bibfnamefont {B.}~\bibnamefont
  {{McKernan}}}, \bibinfo {author} {\bibfnamefont {K.~E.~S.}\ \bibnamefont
  {{Ford}}}, \bibinfo {author} {\bibfnamefont {J.}~\bibnamefont {{Bellovary}}},
  \bibinfo {author} {\bibfnamefont {N.~W.~C.}\ \bibnamefont {{Leigh}}},
  \bibinfo {author} {\bibfnamefont {Z.}~\bibnamefont {{Haiman}}}, \bibinfo
  {author} {\bibfnamefont {B.}~\bibnamefont {{Kocsis}}}, \bibinfo {author}
  {\bibfnamefont {W.}~\bibnamefont {{Lyra}}}, \bibinfo {author} {\bibfnamefont
  {M.~M.}\ \bibnamefont {{Mac Low}}}, \bibinfo {author} {\bibfnamefont
  {B.}~\bibnamefont {{Metzger}}}, \bibinfo {author} {\bibfnamefont
  {M.}~\bibnamefont {{O'Dowd}}}, \bibinfo {author} {\bibfnamefont
  {S.}~\bibnamefont {{Endlich}}}, \ and\ \bibinfo {author} {\bibfnamefont
  {D.~J.}\ \bibnamefont {{Rosen}}},\ }\href {\doibase 10.3847/1538-4357/aadae5}
  {\bibfield  {journal} {\bibinfo  {journal} {Astrophys. J.}\ }\textbf
  {\bibinfo {volume} {866}},\ \bibinfo {eid} {66} (\bibinfo {year} {2018})},\
  \Eprint {http://arxiv.org/abs/1702.07818} {arXiv:1702.07818 [astro-ph.HE]}
  \BibitemShut {NoStop}%
\bibitem [{\citenamefont {{McKernan}}\ \emph {et~al.}(2020)\citenamefont
  {{McKernan}}, \citenamefont {{Ford}}, \citenamefont {{O'Shaugnessy}},\ and\
  \citenamefont {{Wysocki}}}]{McKernan:2020:AGNLIGO2}%
  \BibitemOpen
  \bibfield  {author} {\bibinfo {author} {\bibfnamefont {B.}~\bibnamefont
  {{McKernan}}}, \bibinfo {author} {\bibfnamefont {K.~E.~S.}\ \bibnamefont
  {{Ford}}}, \bibinfo {author} {\bibfnamefont {R.}~\bibnamefont
  {{O'Shaugnessy}}}, \ and\ \bibinfo {author} {\bibfnamefont {D.}~\bibnamefont
  {{Wysocki}}},\ }\href {\doibase 10.1093/mnras/staa740} {\bibfield  {journal}
  {\bibinfo  {journal} {Mon. Not. Roy. Astron. Soc.}\ }\textbf {\bibinfo
  {volume} {494}},\ \bibinfo {pages} {1203} (\bibinfo {year} {2020})},\ \Eprint
  {http://arxiv.org/abs/1907.04356} {arXiv:1907.04356 [astro-ph.HE]}
  \BibitemShut {NoStop}%
\bibitem [{\citenamefont {{Secunda}}\ \emph {et~al.}(2019)\citenamefont
  {{Secunda}}, \citenamefont {{Bellovary}}, \citenamefont {{Mac Low}},
  \citenamefont {{Ford}}, \citenamefont {{McKernan}}, \citenamefont {{Leigh}},
  \citenamefont {{Lyra}},\ and\ \citenamefont
  {{S{\'a}ndor}}}]{Secunda:2019:AGNMigration}%
  \BibitemOpen
  \bibfield  {author} {\bibinfo {author} {\bibfnamefont {A.}~\bibnamefont
  {{Secunda}}}, \bibinfo {author} {\bibfnamefont {J.}~\bibnamefont
  {{Bellovary}}}, \bibinfo {author} {\bibfnamefont {M.-M.}\ \bibnamefont {{Mac
  Low}}}, \bibinfo {author} {\bibfnamefont {K.~E.~S.}\ \bibnamefont {{Ford}}},
  \bibinfo {author} {\bibfnamefont {B.}~\bibnamefont {{McKernan}}}, \bibinfo
  {author} {\bibfnamefont {N.~W.~C.}\ \bibnamefont {{Leigh}}}, \bibinfo
  {author} {\bibfnamefont {W.}~\bibnamefont {{Lyra}}}, \ and\ \bibinfo {author}
  {\bibfnamefont {Z.}~\bibnamefont {{S{\'a}ndor}}},\ }\href {\doibase
  10.3847/1538-4357/ab20ca} {\bibfield  {journal} {\bibinfo  {journal}
  {Astrophys. J.}\ }\textbf {\bibinfo {volume} {878}},\ \bibinfo {eid} {85}
  (\bibinfo {year} {2019})},\ \Eprint {http://arxiv.org/abs/1807.02859}
  {arXiv:1807.02859 [astro-ph.HE]} \BibitemShut {NoStop}%
\bibitem [{\citenamefont {Bender}\ and\ \citenamefont {Orszag}(1999)}]{Bender}%
  \BibitemOpen
  \bibfield  {author} {\bibinfo {author} {\bibfnamefont {C.~M.}\ \bibnamefont
  {Bender}}\ and\ \bibinfo {author} {\bibfnamefont {S.}~\bibnamefont
  {Orszag}},\ }\href {\doibase 10.1007/978-1-4757-3069-2} {\emph {\bibinfo
  {title} {{Advanced Mathematical Methods for Scientists and Engineers I:
  Asymptotic Methods and Perturbation Theory}}}}\ (\bibinfo  {publisher}
  {Springer},\ \bibinfo {address} {New York, NY},\ \bibinfo {year}
  {1999})\BibitemShut {NoStop}%
\bibitem [{\citenamefont {Edwards}\ \emph {et~al.}(2006)\citenamefont
  {Edwards}, \citenamefont {Hobbs},\ and\ \citenamefont
  {Manchester}}]{Edwards:2006zg}%
  \BibitemOpen
  \bibfield  {author} {\bibinfo {author} {\bibfnamefont {R.~T.}\ \bibnamefont
  {Edwards}}, \bibinfo {author} {\bibfnamefont {G.~B.}\ \bibnamefont {Hobbs}},
  \ and\ \bibinfo {author} {\bibfnamefont {R.~N.}\ \bibnamefont {Manchester}},\
  }\href {\doibase 10.1111/j.1365-2966.2006.10870.x} {\bibfield  {journal}
  {\bibinfo  {journal} {Mon. Not. Roy. Astron. Soc.}\ }\textbf {\bibinfo
  {volume} {372}},\ \bibinfo {pages} {1549} (\bibinfo {year} {2006})},\ \Eprint
  {http://arxiv.org/abs/astro-ph/0607664} {arXiv:astro-ph/0607664} \BibitemShut
  {NoStop}%
\bibitem [{\citenamefont {{Manchester}}(2017)}]{Manchester:2017:PulsarTiming}%
  \BibitemOpen
  \bibfield  {author} {\bibinfo {author} {\bibfnamefont {R.~N.}\ \bibnamefont
  {{Manchester}}},\ }in\ \href {\doibase 10.1088/1742-6596/932/1/012002} {\emph
  {\bibinfo {booktitle} {Journal of Physics Conference Series}}},\ \bibinfo
  {series} {Journal of Physics Conference Series}, Vol.\ \bibinfo {volume}
  {932}\ (\bibinfo {year} {2017})\ p.\ \bibinfo {pages} {012002},\ \Eprint
  {http://arxiv.org/abs/1801.04318} {arXiv:1801.04318 [astro-ph.HE]}
  \BibitemShut {NoStop}%
\bibitem [{\citenamefont {{Verbiest}}\ \emph {et~al.}(2021)\citenamefont
  {{Verbiest}}, \citenamefont {{Os{\l}owski}},\ and\ \citenamefont
  {{Burke-Spolaor}}}]{Verbiest:2021:PulsarTiming}%
  \BibitemOpen
  \bibfield  {author} {\bibinfo {author} {\bibfnamefont {J.~P.~W.}\
  \bibnamefont {{Verbiest}}}, \bibinfo {author} {\bibfnamefont
  {S.}~\bibnamefont {{Os{\l}owski}}}, \ and\ \bibinfo {author} {\bibfnamefont
  {S.}~\bibnamefont {{Burke-Spolaor}}},\ }in\ \href {\doibase
  10.1007/978-981-15-4702-7_4-1} {\emph {\bibinfo {booktitle} {Handbook of
  Gravitational Wave Astronomy}}}\ (\bibinfo {year} {2021})\ p.~\bibinfo
  {pages} {4}\BibitemShut {NoStop}%
\bibitem [{\citenamefont {{Thrane}}\ and\ \citenamefont
  {{Talbot}}(2019)}]{ThraneTalbot:BayesianInference:2019}%
  \BibitemOpen
  \bibfield  {author} {\bibinfo {author} {\bibfnamefont {E.}~\bibnamefont
  {{Thrane}}}\ and\ \bibinfo {author} {\bibfnamefont {C.}~\bibnamefont
  {{Talbot}}},\ }\href {\doibase 10.1017/pasa.2019.2} {\bibfield  {journal}
  {\bibinfo  {journal} {Publications of the Astronomical Society of Australia}\
  }\textbf {\bibinfo {volume} {36}},\ \bibinfo {eid} {e010} (\bibinfo {year}
  {2019})},\ \Eprint {http://arxiv.org/abs/1809.02293} {arXiv:1809.02293
  [astro-ph.IM]} \BibitemShut {NoStop}%
\bibitem [{\citenamefont {{Abbott}}\ \emph {et~al.}(2023)\citenamefont
  {{Abbott}}, \citenamefont {{Abbott}}, \citenamefont {{Acernese}},
  \citenamefont {{Ackley}}, \citenamefont {{Adams}}, \citenamefont
  {{Adhikari}}, \citenamefont {{Adhikari}}, \citenamefont {{Adya}},
  \citenamefont {{Affeldt}}, \citenamefont {{Agarwal}} \emph
  {et~al.}}]{LVK:2021:Population3}%
  \BibitemOpen
  \bibfield  {author} {\bibinfo {author} {\bibfnamefont {R.}~\bibnamefont
  {{Abbott}}}, \bibinfo {author} {\bibfnamefont {T.~D.}\ \bibnamefont
  {{Abbott}}}, \bibinfo {author} {\bibfnamefont {F.}~\bibnamefont
  {{Acernese}}}, \bibinfo {author} {\bibfnamefont {K.}~\bibnamefont
  {{Ackley}}}, \bibinfo {author} {\bibfnamefont {C.}~\bibnamefont {{Adams}}},
  \bibinfo {author} {\bibfnamefont {N.}~\bibnamefont {{Adhikari}}}, \bibinfo
  {author} {\bibfnamefont {R.~X.}\ \bibnamefont {{Adhikari}}}, \bibinfo
  {author} {\bibfnamefont {V.~B.}\ \bibnamefont {{Adya}}}, \bibinfo {author}
  {\bibfnamefont {C.}~\bibnamefont {{Affeldt}}}, \bibinfo {author}
  {\bibfnamefont {D.}~\bibnamefont {{Agarwal}}},  \emph {et~al.},\ }\href
  {\doibase 10.1103/PhysRevX.13.011048} {\bibfield  {journal} {\bibinfo
  {journal} {Phys. Rev. X}\ }\textbf {\bibinfo {volume} {13}},\ \bibinfo {eid}
  {011048} (\bibinfo {year} {2023})},\ \Eprint
  {http://arxiv.org/abs/2111.03634} {arXiv:2111.03634 [astro-ph.HE]}
  \BibitemShut {NoStop}%
\bibitem [{\citenamefont {{Drago}}\ \emph {et~al.}(2021)\citenamefont
  {{Drago}}, \citenamefont {{Klimenko}}, \citenamefont {{Lazzaro}},
  \citenamefont {{Milotti}}, \citenamefont {{Mitselmakher}}, \citenamefont
  {{Necula}}, \citenamefont {{O'Brian}}, \citenamefont {{Prodi}}, \citenamefont
  {{Salemi}}, \citenamefont {{Szczepanczyk}}, \citenamefont {{Tiwari}},
  \citenamefont {{Tiwari}}, \citenamefont {{Gayathri}}, \citenamefont
  {{Vedovato}},\ and\ \citenamefont {{Yakushin}}}]{Drago:cWB:2020}%
  \BibitemOpen
  \bibfield  {author} {\bibinfo {author} {\bibfnamefont {M.}~\bibnamefont
  {{Drago}}}, \bibinfo {author} {\bibfnamefont {S.}~\bibnamefont {{Klimenko}}},
  \bibinfo {author} {\bibfnamefont {C.}~\bibnamefont {{Lazzaro}}}, \bibinfo
  {author} {\bibfnamefont {E.}~\bibnamefont {{Milotti}}}, \bibinfo {author}
  {\bibfnamefont {G.}~\bibnamefont {{Mitselmakher}}}, \bibinfo {author}
  {\bibfnamefont {V.}~\bibnamefont {{Necula}}}, \bibinfo {author}
  {\bibfnamefont {B.}~\bibnamefont {{O'Brian}}}, \bibinfo {author}
  {\bibfnamefont {G.~A.}\ \bibnamefont {{Prodi}}}, \bibinfo {author}
  {\bibfnamefont {F.}~\bibnamefont {{Salemi}}}, \bibinfo {author}
  {\bibfnamefont {M.}~\bibnamefont {{Szczepanczyk}}}, \bibinfo {author}
  {\bibfnamefont {S.}~\bibnamefont {{Tiwari}}}, \bibinfo {author}
  {\bibfnamefont {V.}~\bibnamefont {{Tiwari}}}, \bibinfo {author}
  {\bibfnamefont {V.}~\bibnamefont {{Gayathri}}}, \bibinfo {author}
  {\bibfnamefont {G.}~\bibnamefont {{Vedovato}}}, \ and\ \bibinfo {author}
  {\bibfnamefont {I.}~\bibnamefont {{Yakushin}}},\ }\href {\doibase
  10.1016/j.softx.2021.100678} {\bibfield  {journal} {\bibinfo  {journal}
  {SoftwareX}\ }\textbf {\bibinfo {volume} {14}},\ \bibinfo {eid} {100678}
  (\bibinfo {year} {2021})},\ \Eprint {http://arxiv.org/abs/2006.12604}
  {arXiv:2006.12604 [gr-qc]} \BibitemShut {NoStop}%
\bibitem [{\citenamefont {{Martel}}\ and\ \citenamefont
  {{Poisson}}(1999)}]{MartelPoisson:1999:EccentricSignalLoss}%
  \BibitemOpen
  \bibfield  {author} {\bibinfo {author} {\bibfnamefont {K.}~\bibnamefont
  {{Martel}}}\ and\ \bibinfo {author} {\bibfnamefont {E.}~\bibnamefont
  {{Poisson}}},\ }\href {\doibase 10.1103/PhysRevD.60.124008} {\bibfield
  {journal} {\bibinfo  {journal} {\prd}\ }\textbf {\bibinfo {volume} {60}},\
  \bibinfo {eid} {124008} (\bibinfo {year} {1999})},\ \Eprint
  {http://arxiv.org/abs/gr-qc/9907006} {arXiv:gr-qc/9907006 [gr-qc]}
  \BibitemShut {NoStop}%
\bibitem [{\citenamefont {{Brown}}\ and\ \citenamefont
  {{Zimmerman}}(2010)}]{BrownZimmerman:2010:EccentricSignalLoss}%
  \BibitemOpen
  \bibfield  {author} {\bibinfo {author} {\bibfnamefont {D.~A.}\ \bibnamefont
  {{Brown}}}\ and\ \bibinfo {author} {\bibfnamefont {P.~J.}\ \bibnamefont
  {{Zimmerman}}},\ }\href {\doibase 10.1103/PhysRevD.81.024007} {\bibfield
  {journal} {\bibinfo  {journal} {\prd}\ }\textbf {\bibinfo {volume} {81}},\
  \bibinfo {eid} {024007} (\bibinfo {year} {2010})},\ \Eprint
  {http://arxiv.org/abs/0909.0066} {arXiv:0909.0066 [gr-qc]} \BibitemShut
  {NoStop}%
\bibitem [{\citenamefont {{Ramos-Buades}}\ \emph {et~al.}(2020)\citenamefont
  {{Ramos-Buades}}, \citenamefont {{Tiwari}}, \citenamefont {{Haney}},\ and\
  \citenamefont {{Husa}}}]{RamosBuades:2020:EccentricSearchEfficiency}%
  \BibitemOpen
  \bibfield  {author} {\bibinfo {author} {\bibfnamefont {A.}~\bibnamefont
  {{Ramos-Buades}}}, \bibinfo {author} {\bibfnamefont {S.}~\bibnamefont
  {{Tiwari}}}, \bibinfo {author} {\bibfnamefont {M.}~\bibnamefont {{Haney}}}, \
  and\ \bibinfo {author} {\bibfnamefont {S.}~\bibnamefont {{Husa}}},\ }\href
  {\doibase 10.1103/PhysRevD.102.043005} {\bibfield  {journal} {\bibinfo
  {journal} {\prd}\ }\textbf {\bibinfo {volume} {102}},\ \bibinfo {eid}
  {043005} (\bibinfo {year} {2020})},\ \Eprint
  {http://arxiv.org/abs/2005.14016} {arXiv:2005.14016 [gr-qc]} \BibitemShut
  {NoStop}%
\bibitem [{\citenamefont {{Gamba}}\ \emph {et~al.}(2021)\citenamefont
  {{Gamba}}, \citenamefont {{Breschi}}, \citenamefont {{Carullo}},
  \citenamefont {{Rettegno}}, \citenamefont {{Albanesi}}, \citenamefont
  {{Bernuzzi}},\ and\ \citenamefont {{Nagar}}}]{Gamba:2021:GW190521}%
  \BibitemOpen
  \bibfield  {author} {\bibinfo {author} {\bibfnamefont {R.}~\bibnamefont
  {{Gamba}}}, \bibinfo {author} {\bibfnamefont {M.}~\bibnamefont {{Breschi}}},
  \bibinfo {author} {\bibfnamefont {G.}~\bibnamefont {{Carullo}}}, \bibinfo
  {author} {\bibfnamefont {P.}~\bibnamefont {{Rettegno}}}, \bibinfo {author}
  {\bibfnamefont {S.}~\bibnamefont {{Albanesi}}}, \bibinfo {author}
  {\bibfnamefont {S.}~\bibnamefont {{Bernuzzi}}}, \ and\ \bibinfo {author}
  {\bibfnamefont {A.}~\bibnamefont {{Nagar}}},\ }\href@noop {} {\bibfield
  {journal} {\bibinfo  {journal} {arXiv e-prints}\ ,\ \bibinfo {eid}
  {arXiv:2106.05575}} (\bibinfo {year} {2021})},\ \Eprint
  {http://arxiv.org/abs/2106.05575} {arXiv:2106.05575 [gr-qc]} \BibitemShut
  {NoStop}%
\bibitem [{\citenamefont {{Tiwari}}\ \emph {et~al.}(2019)\citenamefont
  {{Tiwari}}, \citenamefont {{Gopakumar}}, \citenamefont {{Haney}},\ and\
  \citenamefont {{Hemantakumar}}}]{Tiwari:2019:EccentricWaveform}%
  \BibitemOpen
  \bibfield  {author} {\bibinfo {author} {\bibfnamefont {S.}~\bibnamefont
  {{Tiwari}}}, \bibinfo {author} {\bibfnamefont {A.}~\bibnamefont
  {{Gopakumar}}}, \bibinfo {author} {\bibfnamefont {M.}~\bibnamefont
  {{Haney}}}, \ and\ \bibinfo {author} {\bibfnamefont {P.}~\bibnamefont
  {{Hemantakumar}}},\ }\href {\doibase 10.1103/PhysRevD.99.124008} {\bibfield
  {journal} {\bibinfo  {journal} {\prd}\ }\textbf {\bibinfo {volume} {99}},\
  \bibinfo {eid} {124008} (\bibinfo {year} {2019})},\ \Eprint
  {http://arxiv.org/abs/1905.07956} {arXiv:1905.07956 [gr-qc]} \BibitemShut
  {NoStop}%
\bibitem [{\citenamefont {{Islam}}\ \emph {et~al.}(2021)\citenamefont
  {{Islam}}, \citenamefont {{Varma}}, \citenamefont {{Lodman}}, \citenamefont
  {{Field}}, \citenamefont {{Khanna}}, \citenamefont {{Scheel}}, \citenamefont
  {{Pfeiffer}}, \citenamefont {{Gerosa}},\ and\ \citenamefont
  {{Kidder}}}]{Islam:2021:EccentricWaveform}%
  \BibitemOpen
  \bibfield  {author} {\bibinfo {author} {\bibfnamefont {T.}~\bibnamefont
  {{Islam}}}, \bibinfo {author} {\bibfnamefont {V.}~\bibnamefont {{Varma}}},
  \bibinfo {author} {\bibfnamefont {J.}~\bibnamefont {{Lodman}}}, \bibinfo
  {author} {\bibfnamefont {S.~E.}\ \bibnamefont {{Field}}}, \bibinfo {author}
  {\bibfnamefont {G.}~\bibnamefont {{Khanna}}}, \bibinfo {author}
  {\bibfnamefont {M.~A.}\ \bibnamefont {{Scheel}}}, \bibinfo {author}
  {\bibfnamefont {H.~P.}\ \bibnamefont {{Pfeiffer}}}, \bibinfo {author}
  {\bibfnamefont {D.}~\bibnamefont {{Gerosa}}}, \ and\ \bibinfo {author}
  {\bibfnamefont {L.~E.}\ \bibnamefont {{Kidder}}},\ }\href {\doibase
  10.1103/PhysRevD.103.064022} {\bibfield  {journal} {\bibinfo  {journal}
  {\prd}\ }\textbf {\bibinfo {volume} {103}},\ \bibinfo {eid} {064022}
  (\bibinfo {year} {2021})},\ \Eprint {http://arxiv.org/abs/2101.11798}
  {arXiv:2101.11798 [gr-qc]} \BibitemShut {NoStop}%
\bibitem [{\citenamefont {{Setyawati}}\ and\ \citenamefont
  {{Ohme}}(2021)}]{SetyawatiOhme:2021:EccentricWaveform}%
  \BibitemOpen
  \bibfield  {author} {\bibinfo {author} {\bibfnamefont {Y.}~\bibnamefont
  {{Setyawati}}}\ and\ \bibinfo {author} {\bibfnamefont {F.}~\bibnamefont
  {{Ohme}}},\ }\href {\doibase 10.1103/PhysRevD.103.124011} {\bibfield
  {journal} {\bibinfo  {journal} {\prd}\ }\textbf {\bibinfo {volume} {103}},\
  \bibinfo {eid} {124011} (\bibinfo {year} {2021})},\ \Eprint
  {http://arxiv.org/abs/2101.11033} {arXiv:2101.11033 [gr-qc]} \BibitemShut
  {NoStop}%
\bibitem [{\citenamefont {{Liu}}\ \emph {et~al.}(2022)\citenamefont {{Liu}},
  \citenamefont {{Cao}},\ and\ \citenamefont
  {{Zhu}}}]{Liu:2022:EccentricWaveform}%
  \BibitemOpen
  \bibfield  {author} {\bibinfo {author} {\bibfnamefont {X.}~\bibnamefont
  {{Liu}}}, \bibinfo {author} {\bibfnamefont {Z.}~\bibnamefont {{Cao}}}, \ and\
  \bibinfo {author} {\bibfnamefont {Z.-H.}\ \bibnamefont {{Zhu}}},\ }\href
  {\doibase 10.1088/1361-6382/ac4119} {\bibfield  {journal} {\bibinfo
  {journal} {Classical and Quantum Gravity}\ }\textbf {\bibinfo {volume}
  {39}},\ \bibinfo {eid} {035009} (\bibinfo {year} {2022})},\ \Eprint
  {http://arxiv.org/abs/2102.08614} {arXiv:2102.08614 [gr-qc]} \BibitemShut
  {NoStop}%
\bibitem [{\citenamefont {{Ramos-Buades}}\ \emph {et~al.}(2022)\citenamefont
  {{Ramos-Buades}}, \citenamefont {{Buonanno}}, \citenamefont {{Khalil}},\ and\
  \citenamefont {{Ossokine}}}]{RamosBuades:2022:EccentricWaveform}%
  \BibitemOpen
  \bibfield  {author} {\bibinfo {author} {\bibfnamefont {A.}~\bibnamefont
  {{Ramos-Buades}}}, \bibinfo {author} {\bibfnamefont {A.}~\bibnamefont
  {{Buonanno}}}, \bibinfo {author} {\bibfnamefont {M.}~\bibnamefont
  {{Khalil}}}, \ and\ \bibinfo {author} {\bibfnamefont {S.}~\bibnamefont
  {{Ossokine}}},\ }\href {\doibase 10.1103/PhysRevD.105.044035} {\bibfield
  {journal} {\bibinfo  {journal} {\prd}\ }\textbf {\bibinfo {volume} {105}},\
  \bibinfo {eid} {044035} (\bibinfo {year} {2022})},\ \Eprint
  {http://arxiv.org/abs/2112.06952} {arXiv:2112.06952 [gr-qc]} \BibitemShut
  {NoStop}%
\bibitem [{\citenamefont {{Cho}}\ \emph {et~al.}(2022)\citenamefont {{Cho}},
  \citenamefont {{Tanay}}, \citenamefont {{Gopakumar}},\ and\ \citenamefont
  {{Lee}}}]{Cho:2022:EccentricWaveform}%
  \BibitemOpen
  \bibfield  {author} {\bibinfo {author} {\bibfnamefont {G.}~\bibnamefont
  {{Cho}}}, \bibinfo {author} {\bibfnamefont {S.}~\bibnamefont {{Tanay}}},
  \bibinfo {author} {\bibfnamefont {A.}~\bibnamefont {{Gopakumar}}}, \ and\
  \bibinfo {author} {\bibfnamefont {H.~M.}\ \bibnamefont {{Lee}}},\ }\href
  {\doibase 10.1103/PhysRevD.105.064010} {\bibfield  {journal} {\bibinfo
  {journal} {\prd}\ }\textbf {\bibinfo {volume} {105}},\ \bibinfo {eid}
  {064010} (\bibinfo {year} {2022})},\ \Eprint
  {http://arxiv.org/abs/2110.09608} {arXiv:2110.09608 [gr-qc]} \BibitemShut
  {NoStop}%
\bibitem [{\citenamefont {Moore}\ and\ \citenamefont
  {Yunes}(2019)}]{Moore:2019xkm}%
  \BibitemOpen
  \bibfield  {author} {\bibinfo {author} {\bibfnamefont {B.}~\bibnamefont
  {Moore}}\ and\ \bibinfo {author} {\bibfnamefont {N.}~\bibnamefont {Yunes}},\
  }\href {\doibase 10.1088/1361-6382/ab3778} {\bibfield  {journal} {\bibinfo
  {journal} {Class. Quant. Grav.}\ }\textbf {\bibinfo {volume} {36}},\ \bibinfo
  {pages} {185003} (\bibinfo {year} {2019})},\ \Eprint
  {http://arxiv.org/abs/1903.05203} {arXiv:1903.05203 [gr-qc]} \BibitemShut
  {NoStop}%
\bibitem [{\citenamefont {Romero-Shaw}\ \emph {et~al.}(2022)\citenamefont
  {Romero-Shaw}, \citenamefont {Gerosa},\ and\ \citenamefont
  {Loutrel}}]{Romero-Shaw:2022:EccOrPrecc}%
  \BibitemOpen
  \bibfield  {author} {\bibinfo {author} {\bibfnamefont {I.}~\bibnamefont
  {Romero-Shaw}}, \bibinfo {author} {\bibfnamefont {D.}~\bibnamefont {Gerosa}},
  \ and\ \bibinfo {author} {\bibfnamefont {N.}~\bibnamefont {Loutrel}},\
  }\href@noop {} {\bibfield  {journal} {\bibinfo  {journal} {{in preparation}}\
  } (\bibinfo {year} {2022})}\BibitemShut {NoStop}%
\bibitem [{\citenamefont {{Chen}}\ \emph {et~al.}(2021)\citenamefont {{Chen}},
  \citenamefont {{Huerta}}, \citenamefont {{Adamo}}, \citenamefont {{Haas}},
  \citenamefont {{O'Shea}}, \citenamefont {{Kumar}},\ and\ \citenamefont
  {{Moore}}}]{Chen:2021:2G3GEccentricProspects}%
  \BibitemOpen
  \bibfield  {author} {\bibinfo {author} {\bibfnamefont {Z.}~\bibnamefont
  {{Chen}}}, \bibinfo {author} {\bibfnamefont {E.~A.}\ \bibnamefont
  {{Huerta}}}, \bibinfo {author} {\bibfnamefont {J.}~\bibnamefont {{Adamo}}},
  \bibinfo {author} {\bibfnamefont {R.}~\bibnamefont {{Haas}}}, \bibinfo
  {author} {\bibfnamefont {E.}~\bibnamefont {{O'Shea}}}, \bibinfo {author}
  {\bibfnamefont {P.}~\bibnamefont {{Kumar}}}, \ and\ \bibinfo {author}
  {\bibfnamefont {C.}~\bibnamefont {{Moore}}},\ }\href {\doibase
  10.1103/PhysRevD.103.084018} {\bibfield  {journal} {\bibinfo  {journal}
  {\prd}\ }\textbf {\bibinfo {volume} {103}},\ \bibinfo {eid} {084018}
  (\bibinfo {year} {2021})},\ \Eprint {http://arxiv.org/abs/2008.03313}
  {arXiv:2008.03313 [gr-qc]} \BibitemShut {NoStop}%
\bibitem [{\citenamefont {{Kocsis}}\ and\ \citenamefont
  {{Levin}}(2012)}]{KocsisLevin:2012:BurstsGN}%
  \BibitemOpen
  \bibfield  {author} {\bibinfo {author} {\bibfnamefont {B.}~\bibnamefont
  {{Kocsis}}}\ and\ \bibinfo {author} {\bibfnamefont {J.}~\bibnamefont
  {{Levin}}},\ }\href {\doibase 10.1103/PhysRevD.85.123005} {\bibfield
  {journal} {\bibinfo  {journal} {\prd}\ }\textbf {\bibinfo {volume} {85}},\
  \bibinfo {eid} {123005} (\bibinfo {year} {2012})},\ \Eprint
  {http://arxiv.org/abs/1109.4170} {arXiv:1109.4170 [astro-ph.CO]} \BibitemShut
  {NoStop}%
\bibitem [{\citenamefont {Abbott}\ \emph {et~al.}(2016)\citenamefont {Abbott}
  \emph {et~al.}}]{Abbott:2017:Exploring3G}%
  \BibitemOpen
  \bibfield  {author} {\bibinfo {author} {\bibfnamefont {B.}~\bibnamefont
  {Abbott}} \emph {et~al.},\ }\href {\doibase 10.1088/1361-6382/aa51f4}
  {\bibfield  {journal} {\bibinfo  {journal} {Classical and Quantum Gravity}\
  }\textbf {\bibinfo {volume} {34}} (\bibinfo {year} {2016}),\
  10.1088/1361-6382/aa51f4}\BibitemShut {NoStop}%
\bibitem [{\citenamefont {{Abbott}}\ \emph
  {et~al.}(2020{\natexlab{b}})\citenamefont {{Abbott}} \emph
  {et~al.}}]{Abbott:2020:Noise}%
  \BibitemOpen
  \bibfield  {author} {\bibinfo {author} {\bibfnamefont {B.~P.}\ \bibnamefont
  {{Abbott}}} \emph {et~al.},\ }\href {\doibase 10.1088/1361-6382/ab685e}
  {\bibfield  {journal} {\bibinfo  {journal} {Classical and Quantum Gravity}\
  }\textbf {\bibinfo {volume} {37}},\ \bibinfo {eid} {055002} (\bibinfo {year}
  {2020}{\natexlab{b}})},\ \Eprint {http://arxiv.org/abs/1908.11170}
  {arXiv:1908.11170 [gr-qc]} \BibitemShut {NoStop}%
\bibitem [{\citenamefont {{Talbot}}\ and\ \citenamefont
  {{Thrane}}(2020)}]{TalbotThrane:2020:UncertainPSD}%
  \BibitemOpen
  \bibfield  {author} {\bibinfo {author} {\bibfnamefont {C.}~\bibnamefont
  {{Talbot}}}\ and\ \bibinfo {author} {\bibfnamefont {E.}~\bibnamefont
  {{Thrane}}},\ }\href {\doibase 10.1103/PhysRevResearch.2.043298} {\bibfield
  {journal} {\bibinfo  {journal} {Physical Review Research}\ }\textbf {\bibinfo
  {volume} {2}},\ \bibinfo {eid} {043298} (\bibinfo {year} {2020})},\ \Eprint
  {http://arxiv.org/abs/2006.05292} {arXiv:2006.05292 [astro-ph.IM]}
  \BibitemShut {NoStop}%
\bibitem [{\citenamefont {Ashton}\ \emph {et~al.}(2019)\citenamefont {Ashton}
  \emph {et~al.}}]{bilby}%
  \BibitemOpen
  \bibfield  {author} {\bibinfo {author} {\bibfnamefont {G.}~\bibnamefont
  {Ashton}} \emph {et~al.},\ }\href {\doibase 10.3847/1538-4365/ab06fc}
  {\bibfield  {journal} {\bibinfo  {journal} {Astrophys. J. Suppl.}\ }\textbf
  {\bibinfo {volume} {241}},\ \bibinfo {pages} {27} (\bibinfo {year} {2019})},\
  \Eprint {http://arxiv.org/abs/1811.02042} {arXiv:1811.02042 [astro-ph.IM]}
  \BibitemShut {NoStop}%
\bibitem [{\citenamefont {Romero-Shaw}\ \emph {et~al.}(2020)\citenamefont
  {Romero-Shaw} \emph {et~al.}}]{Romero-Shaw:2020:Bilby}%
  \BibitemOpen
  \bibfield  {author} {\bibinfo {author} {\bibfnamefont {I.~M.}\ \bibnamefont
  {Romero-Shaw}} \emph {et~al.},\ }\href {\doibase 10.1093/mnras/staa2850}
  {\bibfield  {journal} {\bibinfo  {journal} {Mon. Not. Roy. Astron. Soc.}\
  }\textbf {\bibinfo {volume} {499}},\ \bibinfo {pages} {3295} (\bibinfo {year}
  {2020})},\ \Eprint {http://arxiv.org/abs/2006.00714} {arXiv:2006.00714
  [astro-ph.IM]} \BibitemShut {NoStop}%
\bibitem [{\citenamefont {{Hild}}\ \emph {et~al.}(2008)\citenamefont {{Hild}},
  \citenamefont {{Chelkowski}},\ and\ \citenamefont
  {{Freise}}}]{Hild:2008:ETB}%
  \BibitemOpen
  \bibfield  {author} {\bibinfo {author} {\bibfnamefont {S.}~\bibnamefont
  {{Hild}}}, \bibinfo {author} {\bibfnamefont {S.}~\bibnamefont
  {{Chelkowski}}}, \ and\ \bibinfo {author} {\bibfnamefont {A.}~\bibnamefont
  {{Freise}}},\ }\href@noop {} {\bibfield  {journal} {\bibinfo  {journal}
  {arXiv e-prints}\ ,\ \bibinfo {eid} {arXiv:0810.0604}} (\bibinfo {year}
  {2008})},\ \Eprint {http://arxiv.org/abs/0810.0604} {arXiv:0810.0604 [gr-qc]}
  \BibitemShut {NoStop}%
\bibitem [{\citenamefont {{Ananna}}\ \emph {et~al.}(2020)\citenamefont
  {{Ananna}}, \citenamefont {{Urry}}, \citenamefont {{Treister}}, \citenamefont
  {{Hickox}}, \citenamefont {{Shankar}}, \citenamefont {{Ricci}}, \citenamefont
  {{Cappelluti}}, \citenamefont {{Marchesi}},\ and\ \citenamefont
  {{Turner}}}]{Ananna:2020:AGNSpin}%
  \BibitemOpen
  \bibfield  {author} {\bibinfo {author} {\bibfnamefont {T.~T.}\ \bibnamefont
  {{Ananna}}}, \bibinfo {author} {\bibfnamefont {C.~M.}\ \bibnamefont
  {{Urry}}}, \bibinfo {author} {\bibfnamefont {E.}~\bibnamefont {{Treister}}},
  \bibinfo {author} {\bibfnamefont {R.~C.}\ \bibnamefont {{Hickox}}}, \bibinfo
  {author} {\bibfnamefont {F.}~\bibnamefont {{Shankar}}}, \bibinfo {author}
  {\bibfnamefont {C.}~\bibnamefont {{Ricci}}}, \bibinfo {author} {\bibfnamefont
  {N.}~\bibnamefont {{Cappelluti}}}, \bibinfo {author} {\bibfnamefont
  {S.}~\bibnamefont {{Marchesi}}}, \ and\ \bibinfo {author} {\bibfnamefont
  {T.~J.}\ \bibnamefont {{Turner}}},\ }\href {\doibase
  10.3847/1538-4357/abb815} {\bibfield  {journal} {\bibinfo  {journal}
  {Astrophys. J.}\ }\textbf {\bibinfo {volume} {903}},\ \bibinfo {eid} {85}
  (\bibinfo {year} {2020})},\ \Eprint {http://arxiv.org/abs/2009.07711}
  {arXiv:2009.07711 [astro-ph.HE]} \BibitemShut {NoStop}%
\bibitem [{\citenamefont {{Bardeen}}\ \emph {et~al.}(1972)\citenamefont
  {{Bardeen}}, \citenamefont {{Press}},\ and\ \citenamefont
  {{Teukolsky}}}]{Bardeen:1972:KerrBH}%
  \BibitemOpen
  \bibfield  {author} {\bibinfo {author} {\bibfnamefont {J.~M.}\ \bibnamefont
  {{Bardeen}}}, \bibinfo {author} {\bibfnamefont {W.~H.}\ \bibnamefont
  {{Press}}}, \ and\ \bibinfo {author} {\bibfnamefont {S.~A.}\ \bibnamefont
  {{Teukolsky}}},\ }\href {\doibase 10.1086/151796} {\bibfield  {journal}
  {\bibinfo  {journal} {Astrophys. J.}\ }\textbf {\bibinfo {volume} {178}},\
  \bibinfo {pages} {347} (\bibinfo {year} {1972})}\BibitemShut {NoStop}%
\bibitem [{\citenamefont {{Fang}}\ \emph {et~al.}(2019)\citenamefont {{Fang}},
  \citenamefont {{Chen}},\ and\ \citenamefont {{Huang}}}]{Fang:2019:SMBHSpin}%
  \BibitemOpen
  \bibfield  {author} {\bibinfo {author} {\bibfnamefont {Y.}~\bibnamefont
  {{Fang}}}, \bibinfo {author} {\bibfnamefont {X.}~\bibnamefont {{Chen}}}, \
  and\ \bibinfo {author} {\bibfnamefont {Q.-G.}\ \bibnamefont {{Huang}}},\
  }\href {\doibase 10.3847/1538-4357/ab510e} {\bibfield  {journal} {\bibinfo
  {journal} {\apj}\ }\textbf {\bibinfo {volume} {887}},\ \bibinfo {eid} {210}
  (\bibinfo {year} {2019})},\ \Eprint {http://arxiv.org/abs/1908.01443}
  {arXiv:1908.01443 [astro-ph.HE]} \BibitemShut {NoStop}%
\bibitem [{\citenamefont {{Nasim}}\ \emph {et~al.}(2022)\citenamefont
  {{Nasim}}, \citenamefont {{Fabj}}, \citenamefont {{Caban}}, \citenamefont
  {{Secunda}}, \citenamefont {{Ford}}, \citenamefont {{McKernan}},
  \citenamefont {{Bellovary}}, \citenamefont {{Leigh}},\ and\ \citenamefont
  {{Lyra}}}]{Nasim:2022:AGNalignplane}%
  \BibitemOpen
  \bibfield  {author} {\bibinfo {author} {\bibfnamefont {S.~S.}\ \bibnamefont
  {{Nasim}}}, \bibinfo {author} {\bibfnamefont {G.}~\bibnamefont {{Fabj}}},
  \bibinfo {author} {\bibfnamefont {F.}~\bibnamefont {{Caban}}}, \bibinfo
  {author} {\bibfnamefont {A.}~\bibnamefont {{Secunda}}}, \bibinfo {author}
  {\bibfnamefont {K.~E.~S.}\ \bibnamefont {{Ford}}}, \bibinfo {author}
  {\bibfnamefont {B.}~\bibnamefont {{McKernan}}}, \bibinfo {author}
  {\bibfnamefont {J.~M.}\ \bibnamefont {{Bellovary}}}, \bibinfo {author}
  {\bibfnamefont {N.~W.~C.}\ \bibnamefont {{Leigh}}}, \ and\ \bibinfo {author}
  {\bibfnamefont {W.}~\bibnamefont {{Lyra}}},\ }\href@noop {} {\bibfield
  {journal} {\bibinfo  {journal} {arXiv e-prints}\ ,\ \bibinfo {eid}
  {arXiv:2207.09540}} (\bibinfo {year} {2022})},\ \Eprint
  {http://arxiv.org/abs/2207.09540} {arXiv:2207.09540 [astro-ph.GA]}
  \BibitemShut {NoStop}%
\bibitem [{\citenamefont {{Cho}}(2022)}]{Cho:2022:AccuracyET}%
  \BibitemOpen
  \bibfield  {author} {\bibinfo {author} {\bibfnamefont {H.-S.}\ \bibnamefont
  {{Cho}}},\ }\href {\doibase 10.1088/1361-6382/ac5b31} {\bibfield  {journal}
  {\bibinfo  {journal} {Classical and Quantum Gravity}\ }\textbf {\bibinfo
  {volume} {39}},\ \bibinfo {eid} {085006} (\bibinfo {year} {2022})},\ \Eprint
  {http://arxiv.org/abs/2202.10858} {arXiv:2202.10858 [gr-qc]} \BibitemShut
  {NoStop}%
\bibitem [{\citenamefont {Pieroni}\ \emph {et~al.}(2022)\citenamefont
  {Pieroni}, \citenamefont {Ricciardone},\ and\ \citenamefont
  {Barausse}}]{Pieroni:2022:AccuracyETCE}%
  \BibitemOpen
  \bibfield  {author} {\bibinfo {author} {\bibfnamefont {M.}~\bibnamefont
  {Pieroni}}, \bibinfo {author} {\bibfnamefont {A.}~\bibnamefont
  {Ricciardone}}, \ and\ \bibinfo {author} {\bibfnamefont {E.}~\bibnamefont
  {Barausse}},\ }\href {\doibase 10.1038/s41598-022-19540-7} {\bibfield
  {journal} {\bibinfo  {journal} {Scientific Reports}\ }\textbf {\bibinfo
  {volume} {12}},\ \bibinfo {pages} {17940} (\bibinfo {year}
  {2022})}\BibitemShut {NoStop}%
\bibitem [{\citenamefont {Gralla}\ \emph {et~al.}(2016)\citenamefont {Gralla},
  \citenamefont {Hughes},\ and\ \citenamefont {Warburton}}]{Gralla:2016qfw}%
  \BibitemOpen
  \bibfield  {author} {\bibinfo {author} {\bibfnamefont {S.~E.}\ \bibnamefont
  {Gralla}}, \bibinfo {author} {\bibfnamefont {S.~A.}\ \bibnamefont {Hughes}},
  \ and\ \bibinfo {author} {\bibfnamefont {N.}~\bibnamefont {Warburton}},\
  }\href {\doibase 10.1088/0264-9381/33/15/155002} {\bibfield  {journal}
  {\bibinfo  {journal} {Class. Quant. Grav.}\ }\textbf {\bibinfo {volume}
  {33}},\ \bibinfo {pages} {155002} (\bibinfo {year} {2016})},\ \bibinfo {note}
  {[Erratum: Class.Quant.Grav. 37, 109501 (2020)]},\ \Eprint
  {http://arxiv.org/abs/1603.01221} {arXiv:1603.01221 [gr-qc]} \BibitemShut
  {NoStop}%
\bibitem [{\citenamefont {Hill}(1878{\natexlab{a}})}]{Hill:1878a}%
  \BibitemOpen
  \bibfield  {author} {\bibinfo {author} {\bibfnamefont {G.~W.}\ \bibnamefont
  {Hill}},\ }\href {http://www.jstor.org/stable/2369430} {\bibfield  {journal}
  {\bibinfo  {journal} {American Journal of Mathematics}\ }\textbf {\bibinfo
  {volume} {1}},\ \bibinfo {pages} {5} (\bibinfo {year}
  {1878}{\natexlab{a}})}\BibitemShut {NoStop}%
\bibitem [{\citenamefont {Hill}(1878{\natexlab{b}})}]{Hill:1878b}%
  \BibitemOpen
  \bibfield  {author} {\bibinfo {author} {\bibfnamefont {G.~W.}\ \bibnamefont
  {Hill}},\ }\href {http://www.jstor.org/stable/2369304} {\bibfield  {journal}
  {\bibinfo  {journal} {American Journal of Mathematics}\ }\textbf {\bibinfo
  {volume} {1}},\ \bibinfo {pages} {129} (\bibinfo {year}
  {1878}{\natexlab{b}})}\BibitemShut {NoStop}%
\bibitem [{\citenamefont {Hill}(1878{\natexlab{c}})}]{Hill:1878c}%
  \BibitemOpen
  \bibfield  {author} {\bibinfo {author} {\bibfnamefont {G.~W.}\ \bibnamefont
  {Hill}},\ }\href {http://www.jstor.org/stable/2369313} {\bibfield  {journal}
  {\bibinfo  {journal} {American Journal of Mathematics}\ }\textbf {\bibinfo
  {volume} {1}},\ \bibinfo {pages} {245} (\bibinfo {year}
  {1878}{\natexlab{c}})}\BibitemShut {NoStop}%
\bibitem [{\citenamefont {Zare}(1976)}]{Zare:1976a}%
  \BibitemOpen
  \bibfield  {author} {\bibinfo {author} {\bibfnamefont {K.}~\bibnamefont
  {Zare}},\ }\href {\doibase 10.1007/BF01247133} {\bibfield  {journal}
  {\bibinfo  {journal} {Celestial mechanics}\ }\textbf {\bibinfo {volume}
  {14}},\ \bibinfo {pages} {73} (\bibinfo {year} {1976})}\BibitemShut {NoStop}%
\bibitem [{\citenamefont {Zare}(1977)}]{Zare:1977}%
  \BibitemOpen
  \bibfield  {author} {\bibinfo {author} {\bibfnamefont {K.}~\bibnamefont
  {Zare}},\ }\href {\doibase 10.1007/BF01235726} {\bibfield  {journal}
  {\bibinfo  {journal} {Celestial mechanics}\ }\textbf {\bibinfo {volume}
  {16}},\ \bibinfo {pages} {35} (\bibinfo {year} {1977})}\BibitemShut {NoStop}%
\bibitem [{\citenamefont {{Mardling}}\ and\ \citenamefont
  {{Aarseth}}(2001)}]{Mardling:2001}%
  \BibitemOpen
  \bibfield  {author} {\bibinfo {author} {\bibfnamefont {R.~A.}\ \bibnamefont
  {{Mardling}}}\ and\ \bibinfo {author} {\bibfnamefont {S.~J.}\ \bibnamefont
  {{Aarseth}}},\ }\href {\doibase 10.1046/j.1365-8711.2001.03974.x} {\bibfield
  {journal} {\bibinfo  {journal} {Mon. Not. Roy. Astron. Soc.}\ }\textbf
  {\bibinfo {volume} {321}},\ \bibinfo {pages} {398} (\bibinfo {year}
  {2001})}\BibitemShut {NoStop}%
\bibitem [{\citenamefont {Georgakarakos}(2008)}]{Georga:2008}%
  \BibitemOpen
  \bibfield  {author} {\bibinfo {author} {\bibfnamefont {N.}~\bibnamefont
  {Georgakarakos}},\ }\href {\doibase 10.1007/s10569-007-9109-2} {\bibfield
  {journal} {\bibinfo  {journal} {Celestial Mechanics and Dynamical Astronomy}\
  }\textbf {\bibinfo {volume} {100}},\ \bibinfo {pages} {151} (\bibinfo {year}
  {2008})}\BibitemShut {NoStop}%
\end{thebibliography}%
\bibliographystyle{apsrev4-1}
\end{document}